\documentclass[preprint,longabstract]{aastex}

\usepackage{macros}

\begin{document}

\title{Spectropolarimetry with the Allen Telescope Array:  Faraday Rotation toward Bright Polarized Radio Galaxies}
\shorttitle{ATA Spectropolarimetry}
\shortauthors{Law et al.}

\author{C. J. Law\altaffilmark{1}, B. M. Gaensler\altaffilmark{2}, G. C. Bower\altaffilmark{1}, D. C. Backer\altaffilmark{1}, A. Bauermeister\altaffilmark{1}, S. Croft\altaffilmark{1}, R. Forster\altaffilmark{1}, C. Gutierrez-Kraybill\altaffilmark{1}, L. Harvey-Smith\altaffilmark{2,3}, C. Heiles\altaffilmark{1}, C. Hull\altaffilmark{1}, G. Keating\altaffilmark{1}, D. MacMahon\altaffilmark{1}, D. Whysong\altaffilmark{1}, P. K. G. Williams\altaffilmark{1}, M. Wright\altaffilmark{1}}
\altaffiltext{1}{Radio Astronomy Lab, University of California, Berkeley, CA 94720, USA; claw@astro.berkeley.edu}
\altaffiltext{2}{Sydney Institute for Astronomy, School of Physics A29, The University of Sydney, NSW 2006, Australia}
\altaffiltext{3}{CSIRO Astronomy and Space Science, Australia Telescope National Facility, PO Box 76, Epping NSW 1710, Australia}

\begin{abstract}
We have observed 37 bright, polarized radio sources with the Allen Telescope Array (ATA) to present a novel analysis of their Faraday rotation properties.  Each source was observed during the commissioning phase with 2 to 4 100-MHz bands at frequencies ranging from 1 to 2 GHz.  These observations demonstrate how the continuous frequency coverage of the ATA's log-periodic receiver can be applied to the study of Faraday rotation measures (RMs).  We use RM synthesis to show that wide-bandwidth data can find multiple RM components toward a single source.  Roughly a quarter of the sources studied have extra RM components with high confidence (brighter than $\approx$\ 40 mJy), when observing with a RM resolution of roughly 100 rad m$^{-2}$.  These extra components contribute 10\%--70\% of the total polarized flux.  This is the first time multiple RM components have been identified in a large sample of point sources.  For our observing configuration, these extra RM components bias the measurement of the peak RM by 10--15 rad m$^{-2}$;  more generally, the peak RM cannot be determined more precisely than the RM beam size.  Comparing our 1--2 GHz RM spectra to VLBA polarimetric maps shows both techniques can identify complicated Faraday structures in the sources.  However, the RM values and fractional polarization are generally smaller at lower frequencies than in the higher-frequency VLBA maps.  With a few exceptions, the RMs from this work are consistent with that of earlier, narrow-bandwidth, all-sky surveys.  This work also describes the polarimetry calibration procedure and that on-axis ATA observations of linear polarization can be calibrated to an accuracy of 0.2\% of Stokes I.  Future research directions include studying the time-dependent RM structure in Active Galactic Nuclei (AGNs) and enabling accurate, wide-area RM surveys to test models of Galactic and extragalactic magnetic fields.
\end{abstract}


\section{Introduction}
Radio waves encode information not only about their origin, but about their entire path of propagation.  One way this information is encoded is through Faraday rotation, the frequency-dependent rotation of the radiation polarization angle caused by dispersion in a magnetized plasma \citep{f44,b66}.  Observationally, Faraday rotation is parameterized by the rotation measure (RM):
\begin{equation}
\label{dtdl}
\rm{RM} = \Delta \theta/\Delta(\lambda^2),
\end{equation}
\noindent where $\theta$ is typically measured in radians and $\lambda$ in meters.  The plasma dispersion law predicts RM induced by propagation along a line as:
\begin{equation}
\label{phi}
\rm{RM} = 0.81 \int_{0}^{d} \! n_e \, \mathbf{B} \, \cdot d\mathbf{l} \, \rm{rad} \, \rm{m}^{-2},
\end{equation}
\noindent where $n_e$ is in cm$^{-3}$ and $\mathbf{B}$ is in $\mu$G, and $d$\ is the distance to the source \citep{b66}.  Thus, measurements of Faraday rotation constrain physical conditions critical to understanding a wide variety of problems.

Measurements of RM have expanded our knowledge of magnetic fields in our own Galaxy and in other galaxies.  Several efforts have been made to compile RM along lines of sight over large areas of the sky \citep{s81,b03}.  Finding patterns in these RM values have been used to constrain Galactic magnetic structure and turbulence \citep{h89,h06,h08}.  Similar studies of other galaxies constrain models for the amplification of galactic magnetic fields \citep{g05}.  Observations of individual AGNs and the massive black hole in the Galactic center have been used to constrain the geometry of their magnetic fields and depolarization \citep{b99,z05}.

Recently, Taylor et al.\ (2009;  hereafter ``T09'')\defcitealias{t09}{T09}, have expanded the number of sources with measured RM by nearly two orders of magnitude.  This was done by reanalyzing the 1.4 GHz NRAO VLA Sky Survey \citep[NVSS; ][]{c98}, producing RMs for 37,543 radio sources located throughout the sky north of declination --40\sdeg.  Each source was observed in two bands, which, assuming Equation \ref{dtdl}, can be used to estimate RM.  The high density and large coverage of the \citetalias{t09} sample make it very effective at statistically measuring the structure and strength of the magnetic fields.

However, while the \citetalias{t09} RMs are undoubtedly precise, it is not clear that they are accurate.  \citet{b66} first noted that the Faraday rotation does not require the polarization angle to change as $\lambda^2$.  When the emitting and Faraday-rotating media are mixed or multiple sources with different RM are spatially unresolved, one observes complicated changes in the polarization angle with wavelength \citep{g84}.  These kinds of changes, combined with the possibility of $n\pi$ ambiguities in measuring polarization angle, make it difficult to measure RM robustly.  \citet{b05} noted that discrete sampling of the Stokes vector in $\lambda^2$-space constrains different kinds of Faraday structures.  They introduce the concept of ``rotation measure synthesis'', which takes advantage of the mathematical similarity between how aperture synthesis is done with multiple antennas and how RM is measured by sampling multiple wavelengths.  The output from this technique is a ``RM spectrum'' showing the amount of polarized brightness as a function of its Faraday rotation \citep{h09,mao10}.

In parallel with this algorithmic development has been the technical development of wide-bandwidth radio receivers and digital signal processing.  The Allen Telescope Array (ATA), a radio interferometer in northern California, is commissioning these and many other new technologies \citep{w09}.  One strength of the ATA design is its log-periodic receiver, which gives it continuous access to frequencies from 0.5 to 10 GHz \citep{w10}.  The array and receiver design are optimized for large surveys \citep{c10}, which could be very powerful in the study of cosmic magnetism.

The coincidence of these new algorithms and technologies for wide-bandwidth polarimetry inspired us to conduct a commissioning survey with the ATA.  The first goal of this survey is to test the polarimetry capabilities of the ATA telescope design.  Second, comparing the RM measured by the ATA and by \citetalias{t09} can test for biases associated with narrow bandwidth and coarse frequency resolution of the NVSS data.  If the \citetalias{t09} RM measurements are trustworthy, it will give confidence in their application to measuring large-scale RM structure.  Finally, the wide bandwidth and fine frequency resolution of the ATA demonstrates the power of RM synthesis to study complicated RM spectra.  The paper begins with a description of the observations in \S \ref{obs}.  The data reduction, including a detailed description of the ATA polarimetry calibration process, is described in \S \ref{red}.  The analysis of the RM spectra is given in \S \ref{ana}, and implications of this work are given in \S \ref{dis}.

\section{Observations}
\label{obs}
\begin{figure}[tbp]
\includegraphics[width=\textwidth]{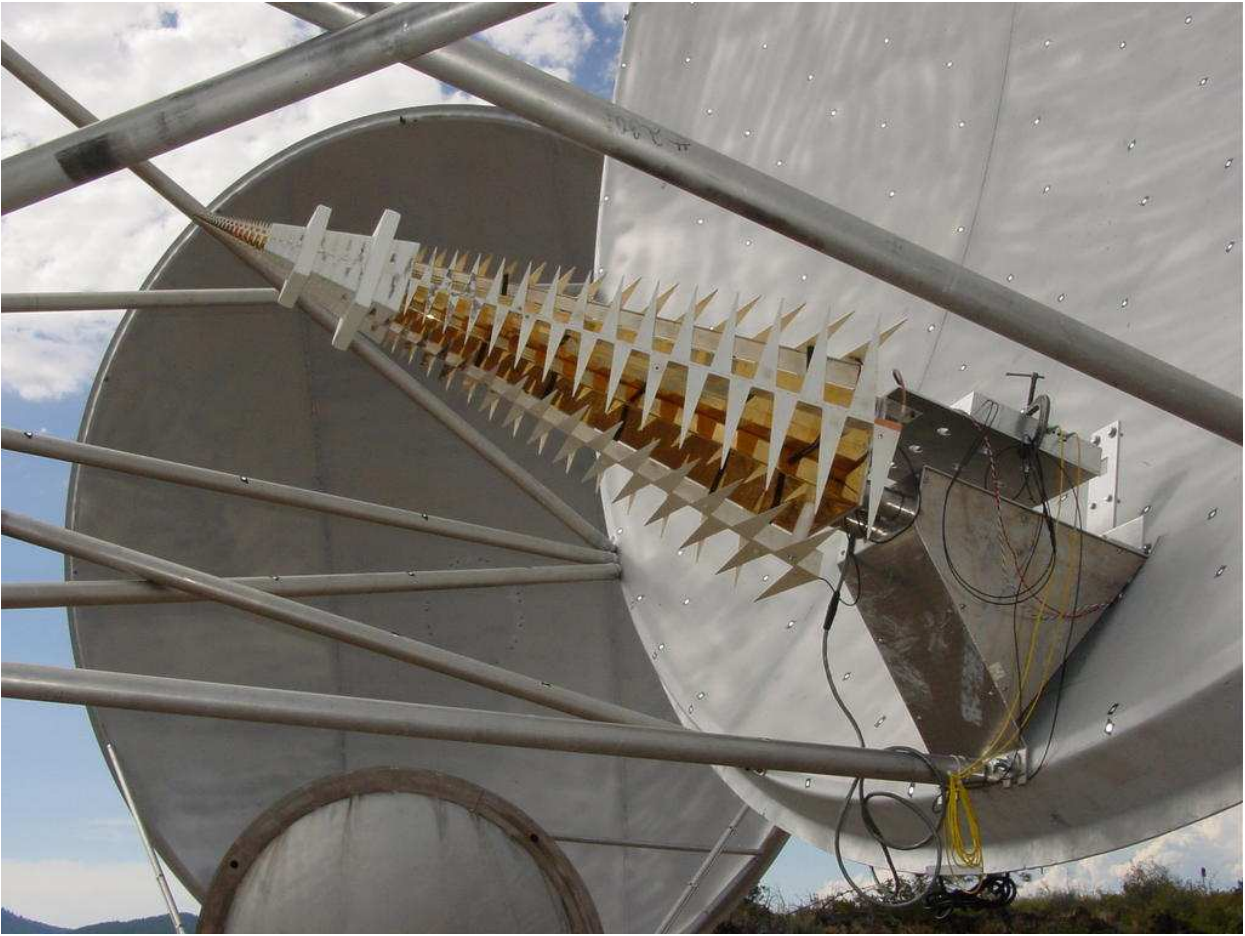}
\caption{The log-periodic receiver of the ATA.  This picture shows the open antenna/receiver configuration used during commissioning.  In the current configuration, the c-clamp is removed, the receiver is enclosed from below by a shroud, and from above by a radio-transparent radome.  \label{atafeed}}
\end{figure}

The ATA is currently composed of 42 6.1-m dishes with an offset Gregorian optical design on an alt-az mount.  The dishes are distributed to produce a Gaussian distribution of baselines with a maximum length of 300 m (1500$\lambda$\ at 1.4 GHz).  Typical observations will have a resolution of 150\arcsec\ at the zenith and a primary beam FWHM of $2\ddeg5$\ at 1.4 GHz.  As shown in Fig. \ref{atafeed}, the receiver has a pyramid shape with a log-periodic pattern, giving it sensitivity from 0.5 to 10 GHz \citep{w10}.  The entire bandpass is sent as an analog signal over $\sim$100m of optical fiber into the on-site signal processing room.  In the signal processing room, the signal is mixed, filtered, and digitized before being passed to the backends.  

For our experiment, the signal was processed simultaneously by two correlators tuned to different frequencies.  Each correlator takes 32 dual-linear polarization antenna inputs and produces 1024-channel spectra covering 100 MHz.  Each visibility cross product ($xx$, $xy$, $yx$, $yy$) was integrated for 10 seconds.  All fields were observed with the correlators tuned to 1.43 and 2.01 GHz, but several fields were also observed with correlators tuned to 1.0 and 1.8 GHz.  In both cases, the focus was set to the higher frequency, which gives the array good sensitivity in both bands \citep{w09}.  These frequencies were chosen as a rough ``minimum redundant spacing'' in $\lambda^2$-space \citep{h07}, which is important for creating good RM spectra.  The frequencies were also selected such that some portion of their bands were clear of radio frequency interference (RFI).

With the goals of testing \citetalias{t09} and demonstrating the power of RM synthesis, we sought a target list of a few tens of sources.  We started with the catalog of polarized sources presented by \citetalias{t09} from the 1.4 GHz NVSS data.  We built a list of 46 sources by selecting sources with linearly-polarized brightness greater than 200 mJy beam$^{-1}$\ and declinations greater than $-30$\sdeg.  This list excludes two sources in \citetalias{t09} that were not real sources (sidelobes of Cas A).  Filtering on polarized brightness produced a source list with a range of locations and RM that are relatively easy to detect.  Note that the target selection criteria include our calibrator, 3C 286.  We included 3C 286 as a target in our survey, which we refer to as J133108+303032, as a test of our calibration procedure.  Calibration is derived from specific observations of 3C 286 that exclude J133108+303032.  Observations of J133108+303032 are treated as any other target, so we can use errors seen in J133108+303032 to estimate calibration errors toward all targets.

The survey was conducted over three separate days, as summarized in Table \ref{observing}.  In total, 44 fields were observed over 34.5 hours at two or four frequencies.  The typical spatial resolution at the zenith is shown for each frequency.  The schedule was designed to observe each target with two, 7-minute scans at different hour angles to improve image quality;  the actual observations range from one to six scans.  On day 1, 41 fields were observed, eight of which had data collection problems that led us to conservatively reject them.  Another two of the fields (centered on J195904+393805, J200259+410150) were empty and are likely to be sidelobes of Cassiopeia A in the NVSS catalog.  On day 2, we observed eight new fields at 1.43 and 2.01 GHz and 11 fields from day 1 at 1.0 and 1.8 GHz.  Of these, three and two fields at 1.0 and 1.8 GHz, respectively, were too short and had \emph{uv} coverage too poor to be useful.  On day 3, the target list was a subset of 14 fields from day 2, observed only at 1.0 and 1.8 GHz.  Of this list, one field was rejected for having poor data quality.  After these data quality checks, we had 37 fields with at least two 100-MHz bands, as shown in Table \ref{fields}.

\begin{deluxetable}{lccccccc}
\tablecaption{Summary of Observations \label{observing}}
\setlength{\tabcolsep}{0.01in}
\tablewidth{0pt}
\tablehead{
\colhead{Date} & \colhead{Start (UT)} & \colhead{Length (hrs)} & \multicolumn{4}{c}{Resolution (arcsec)} & \colhead{Fields} \\
& & & \colhead{1.0 GHz} & \colhead{1.43 GHz} & \colhead{1.8 GHz} & \colhead{2.01 GHz} &  
}
\startdata
19-20 Sep.\ 2009 & 19   & 20    & $\ldots$ & $250\times120$ & $\ldots$ & $180\times90$ & 41 \\
27 Nov.\ 2009     & 15   &  9    & $360\times170$ & $250\times120$ & $200\times100$ & $180\times90$ & 19 \\
4 Dec.\ 2009      & 18.5 &  5.5  & $360\times170$ & $\ldots$ & $200\times100$ & $\ldots$ & 14 \\
\enddata
\end{deluxetable}

\begin{deluxetable}{lccccc}
\tablecaption{Fields Observed in ATA Polarimetry Survey \label{fields}}
\tablewidth{0pt}
\tablehead{
\colhead{T09 Source} & \colhead{Common Name} & \multicolumn{4}{c}{Integration time (minutes)} \\
 & & \colhead{1.0 GHz} & \colhead{1.43 GHz} & \colhead{1.8 GHz} & \colhead{2.0 GHz} 
}
\startdata
J005558+682218 &	3C 27 & $\ldots$ & 20 & $\ldots$ & 20 \\
J005734--012258 &	3C 29 & $\ldots$ & 20 & $\ldots$ & 21 \\
J010850+131831 &	3C 33 & $\ldots$ & 20 & $\ldots$ & 20 \\
J012644+331309 &	3C 41 & $\ldots$ & 28 & $\ldots$ & 28  \\
J022248+861851 &	3C 61.1 & $\ldots$ & 28 & $\ldots$ & 28  \\
J030824+040639 &	3C 78 & $\ldots$ & 28 & $\ldots$ & 28 \\
J035232--071104 &	3C 94 & $\ldots$ & 14 & $\ldots$ & 18 \\
J052109+163822 &	3C 138 & $\ldots$ & 28 & $\ldots$ & 28 \\
J063633--204233 &	     & $\ldots$ & 21 & $\ldots$ & 20 \\
J074948+555421 &	4C 56.16 & 7 & 13 & 7 & 13 \\
J084124+705341 &	4C 71.07 & 20 & 34 & 40 & 34 \\ 
J094752+072517 &	3C 227   & $\ldots$ & 18 & $\ldots$ & 17 \\ 
J104244+120331 &	3C 245 & 7 & 11 & 7 & 11 \\
J113007--144927 &	PKSJ1130--1449 & $\ldots$ & 10 & $\ldots$ & 10 \\
J122906+020305 &        3C 273 &	7 & 11 & 14 & 10 \\
J123039+121758 &	M87 (jet) & 21 & 7 & 36 & 7 \\
J123522+212018 &               & $\ldots$ & 14 & $\ldots$ & 14 \\
J125611--054720 &	3C 279 & 14 & $\ldots$ & 14 & 11 \\
J133108+303032 &        3C 286 &	21 & 7 & 31 & 7 \\
J153150+240243 &        3C 321 &	19 & 14 & 35 & 14 \\
J160231+015748 &        3C 327 &	$\ldots$ & 22 & $\ldots$ & 24 \\
J160939+655652 &        3C 330 &	$\ldots$ & 22 & $\ldots$ & 14 \\
J162803+274136 &        3C 341 &	$\ldots$ & 7 & $\ldots$ & 7 \\
J164258+394837 &        3C 345 &	$\ldots$ & 7 & $\ldots$ & 7 \\
J165111+045919\tablenotemark{a} &       &	19 & 20 & 37 & 20 \\
J165112+045917\tablenotemark{a} &       &	$\ldots$ & 20 & $\ldots$ & 20 \\
J172025--005852 &        3C 353 &	7 & 14 & 42 & 14 \\
J184226+794517 &        3C 390.3 & 19 & 7 & 38 & 7 \\
J192451--291431 &        PKS1921--293 &	20 & 20 & 20 & 17 \\
J194114--152431 &               & $\ldots$ & 21 & $\ldots$ & 20 \\
J201713+334546 &        4C 33.50 & $\ldots$ & 20 & $\ldots$ & 20 \\
J211636--205551 &        PKS J2116--2055 & 31 & 13 & 37 & 13 \\
J212344+250410\tablenotemark{b} & 3C 433 & $\ldots$ & 20 & $\ldots$ & 20 \\
J212345+250448\tablenotemark{b} & 3C 433 & $\ldots$ & 14 & $\ldots$ & 14 \\
J222547--045701 &        3C 446 &	$\ldots$ & 13 & $\ldots$ & 13 \\
J225357+160853 &        3C 454.3 & $\ldots$ & 20 & $\ldots$ & 20 \\
J231956--272713 & PKS 2317--27 &	$\ldots$ & 14 & $\ldots$ & 13 \\
\enddata
\tablenotetext{a}{J165111+045919 and J165112+045917 are unresolved by ATA}
\tablenotetext{b}{J212344+250410 and J212345+250448 are unresolved by ATA}
\end{deluxetable}

\clearpage

\section{Data Reduction}
\label{red}

\subsection{Flagging}
The visibility data in each band were edited to remove RFI and poorly-performing antennas.  Since the array was being commissioned during this survey, some receivers were being refurbished and were not available.  System health tests revealed a few antennas with large $xy$\ beam squint \citep{h01} that we flagged entirely.  Initial flagging removed 100 channels from each edge of the 1024-channel band, where sensitivity is low;  this removes roughly 20\% of usable channels.  Next, the ATA automated flagging routine, RAPID \citep{k09}, was run on the data.  RAPID iteratively removed antennas, times, and frequency channels with large-amplitude deviations from their means.  Daytime observations are sometimes affected by flux from the sun falling in distant sidelobes;  for simplicity, we manually removed all baselines with \emph{uv} distance less than 200$\lambda$.  Finally, a round of manual flagging was done by visually inspecting the data as (time, frequency) plots \citep[described in][]{w10}.  The number of antennas remaining after flagging changed with frequency and time, but typically data from about 21 antennas were used in the final analysis.  The result of all flagging reduced final sensitivity by roughly 50\%.

Table \ref{fields} shows the total integration time for each field and frequency after flagging.  The total time on a field ranges from 7 to 42 minutes, with a typical duration of 18 minutes.  This produced 15.1 hours of dual-polarization data on the target fields.  The combined useful integration time for all sources and calibrators is about 18 hours, which is about 50\% of the total scheduled time.

\subsection{Calibration}
The correlator outputs a complex visibility for each polarization, frequency, time, and baseline.  The observed visibilities change with the distribution of sources in the field and the response of the telescope, both of which can change with polarization, frequency, time, and baseline.  A general expression of how an astrophysical signal is observed as visibilities is given by the ``measurement equation'' formalism \citep{h96}.  The measurement equation treats the emitted signal as a vector that is modified by a series of matrix operations before being detected as visibilities.  The equation can be written as $\vec{v} = \mathbf{J} \vec{e}$, where $\vec{e}$ is the emitted signal as a Stokes vector (I, Q, U, V), $\mathbf{J}$ is the Jones matrix that describes the transformation of that vector during propagation and detection, and $\vec{v}$ is a vector of the observed complex visibilities ($v_{xx}$, $v_{xy}$, $v_{yx}$, $v_{yy}$).

The measurement equation formalism is useful because we can use the tools of linear algebra to simplify effects of signal propagation and detection.  In the case of the dual-linear feeds used by the ATA, the Jones matrix can be written as an ordinary product of several matrices:  

\begin{equation}
\mathbf{J} = \mathbf{G}*\mathbf{D}*\mathbf{C}*\mathbf{P}*\mathbf{S}
\end{equation}

\noindent where $\mathbf{G}$ represents the antenna gain, $\mathbf{D}$ is the ``leakage'' of the two feed polarization into one another, $\mathbf{C}$ is the rotation of the feed relative to the mount, $\mathbf{P}$ is the rotation of the feed relative to the sky (the parallactic angle), and $\mathbf{S}$ transforms the Stokes vector into the $xy$ coordinate system.  The latter three of these terms ($\mathbf{C}$, $\mathbf{P}$, $\mathbf{S}$) are geometric transformations of the electromagnetic wave and have been derived \citep{h96}.  The feed is oriented such that the horizontal polarization is $x$\ and the vertical polarization is $y$.  By rotating the Stokes vector into the antenna coordinate system (applying $\mathbf{C}$ and $\mathbf{P}$), we can see how the rotated Stokes vector appears as visibilities modified by the antenna gains and leakages.  Expanding these equations and dropping second-order terms produces the following \citep{s96}:

\begin{equation}
\label{meq}
\left(
\begin{array}{c}
  v_{xx} \\
  v_{xy} \\
  v_{yx} \\
  v_{yy}
\end{array}
\right)
\approx
\frac{1}{2}
\left(
\begin{array}{c c c c}
  g_{xx}  &  g_{xx}  &  0  &  0 \\
  g_{xy} (d_{1x} - d^{*}_{2y})  &  0  &  g_{xy}  &  i g_{xy} \\
  -g_{yx} (d_{1y} - d^{*}_{2x})  &  0  &  g_{yx}  &  -i g_{yx} \\
  g_{yy}  &  -g_{yy}  &  0  &  0 
\end{array}
\right)
\left(
\begin{array}{c}
  I_{rot} \\
  Q_{rot} \\
  U_{rot} \\
  V_{rot}
\end{array}
\right)
\end{equation}

\noindent where $g_{xx,xy,yx,yy}$ are the products of the gains of any two antenna polarizations, and $d_{1x,1y,\cdots}$ are the complex leakages of $y$\ into $x$\ for antenna 1, of $x$\ into $y$\ for antenna 1, etc..  These are the parameters solved during calibration.

In general, the gains and leakages can change with time, frequency\footnote{Traditionally, the frequency dependence of the gain is treated separately as the ``bandpass'';  here we include frequency dependence to define a more general gain.}, antenna, and polarization.  The quasar 3C 286 is an excellent calibrator for complex gains and leakages since it is bright (12--17 Jy from 1--2 GHz), isolated, unresolved by the ATA, strongly linearly-polarized ($\sim10$\%), and widely used as a flux standard.  Furthermore, 3C 286 has RM $\approx0$ rad m$^{-2}$ \citep{r83}, which makes it very easy to identify frequency-dependent calibration problems.  As a result, each observing period was scattered with repeated, on-axis observations of 3C 286.  On day 1, 3C 286 was not up during the entire observation, so 3C 138 was used to extend the time- and antenna-dependent gain calibration in time;  3C 286 was always used exclusively for leakage and frequency-dependent gain calibration.  The frequency dependence of total intensity of 3C 286 is based on the flux scale of \citet{b77} as parameterized by \citet{o94}.  The linear polarization of 3C 286 is assumed to follow the VLA model\footnote{See \url{http://www.vla.nrao.edu/astro/calib/manual/polcal.html}; R. Perley \& N. Killeen, private communication}.

The data were calibrated using MIRIAD \citep[polarimetry calibration described in][]{s91}.  To account for the frequency dependence of the leakages, the data were split into frequency segments of 5 channels;  most analysis is done with these $\sim$0.5 MHz segments.  Next, frequency-dependent gain calibration was done with MFCAL with a calibration interval of 1 hour, the typical time scale of phase variations for the ATA.  Finally, frequency-independent gains and leakages were calculated with GPCAL with a calibration interval of 1 hour.  Equation \ref{meq} has four equations and eight unknowns, so a single measurement of the calibrator is not enough to solve both gains and leakages uniquely.  One can use the change in parallactic angle to effectively provide multiple sets of ($v_{xx}$, $v_{xy}$, $v_{yx}$, $v_{yy}$) and (I$_{rot}$, Q$_{rot}$, U$_{rot}$, V$_{rot}$) for which there is one solution to the (time-dependent) gains and (constant) leakages.  For our observations, GPCAL converged to a solution for an observation of 3C 286 over a range of parallactic angle greater than a few tens of degrees.  Due to a scheduling error, observations of 3C 286 on day 3 did not have enough parallactic angle coverage to solve for the leakages ($\Delta\theta\approx10$\sdeg).  The day 3 gain and leakage solutions converged after more aggressively cutting antennas with known large ($>20$\%) leakage and using the leakages measured on day 2 as an initial guess for day 3.

The final gain solution had an amplitude that changed by less than 10\% and gain phase scatter of less than 10\sdeg\ within each day and band.  After calibrating 3C 286, the gain and leakage solutions for each band, day, and frequency segment were applied to their corresponding target field data.  The time-dependent gain solutions were applied to the target fields by linearly interpolating in time.

The calibration and visualization of the data taught us several things.  First, a gain amplitude calibration error is an important source of error in Stokes Q and U.  This is the well-known challenge of measuring linear polarization with linear feeds (i.e., Stokes Q is formed by differencing two large numbers, the $xx$\ and $yy$\ visibilities).  This kind of error limits us to using data within 1 hour of a gain calibration solution.  Second, Stokes I can also mix with Q and U if one doesn't solve for the $xy$\ phase difference for all antennas.  In this work, we know the Stokes parameters for 3C 286 and can solve for the absolute $xy$\ phase, which we assume does not change within a day.  Third, not solving for the absolute leakages of all antennas will cause Stokes Q and U to mix, potentially biasing the measurement of RM.  Fortunately, all of these parameters can be solved for with observations over a range of parallactic angle of a strongly linearly-polarized source like 3C 286.

\begin{figure}[tbp]
\centering
\plottwo{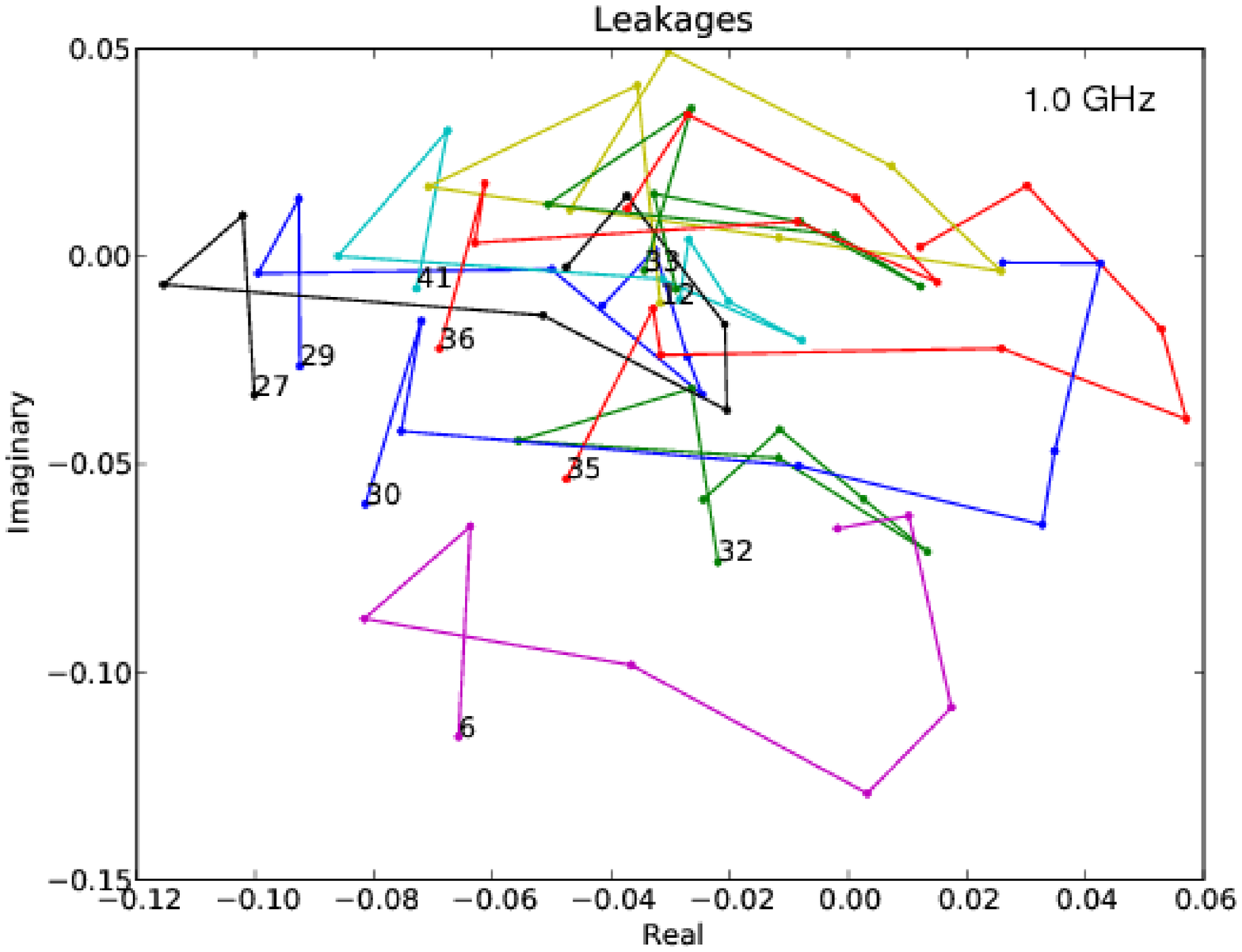}{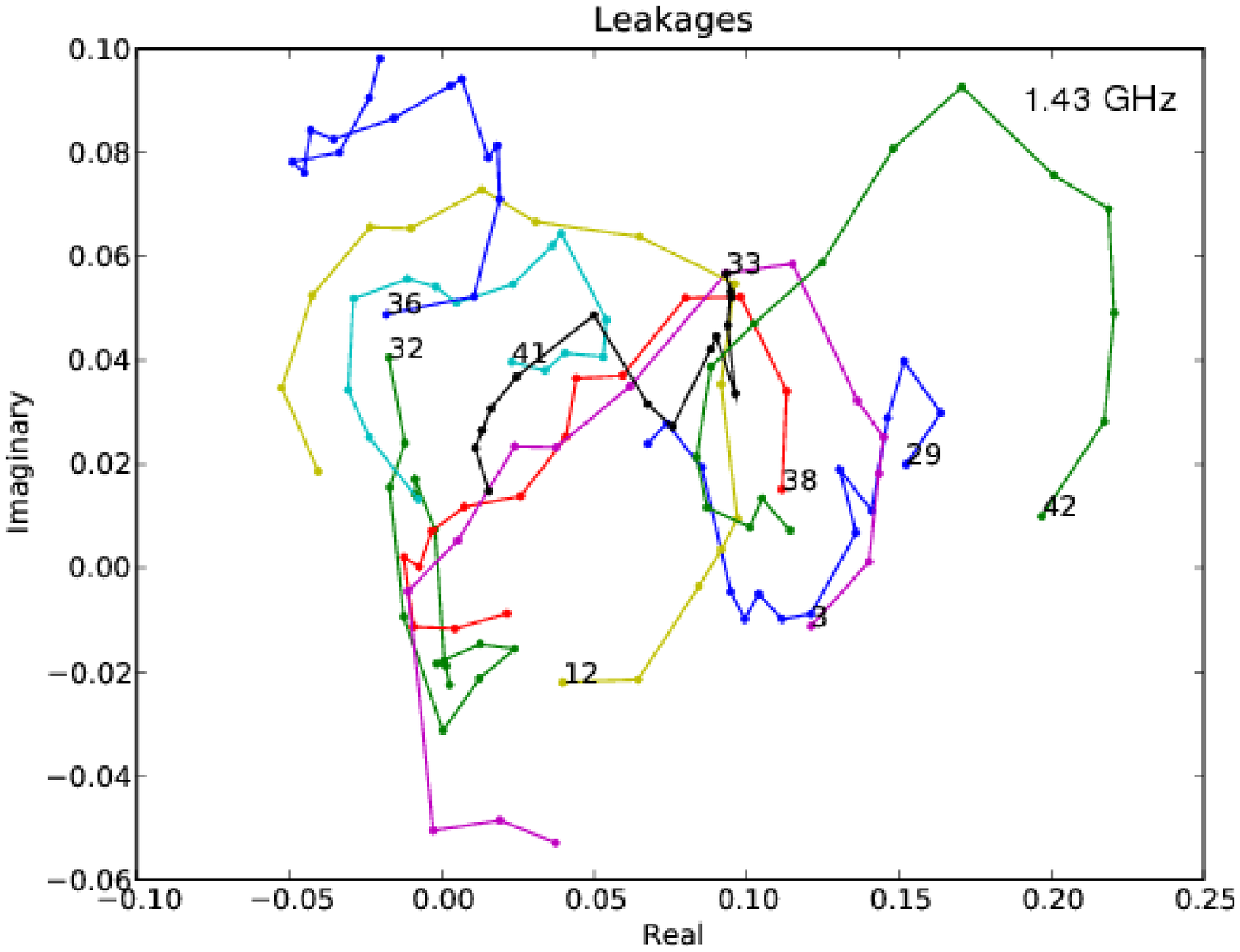}
\plottwo{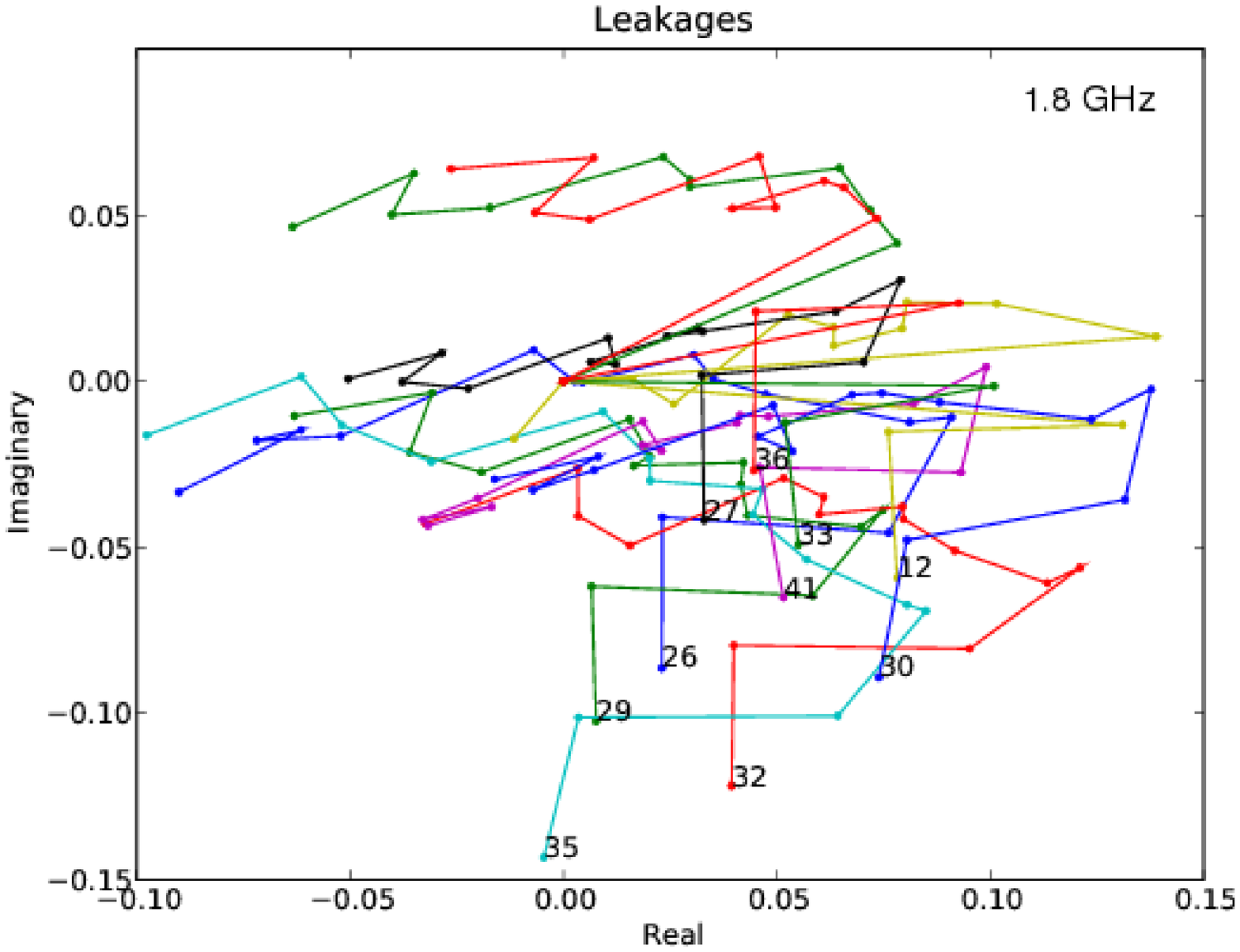}{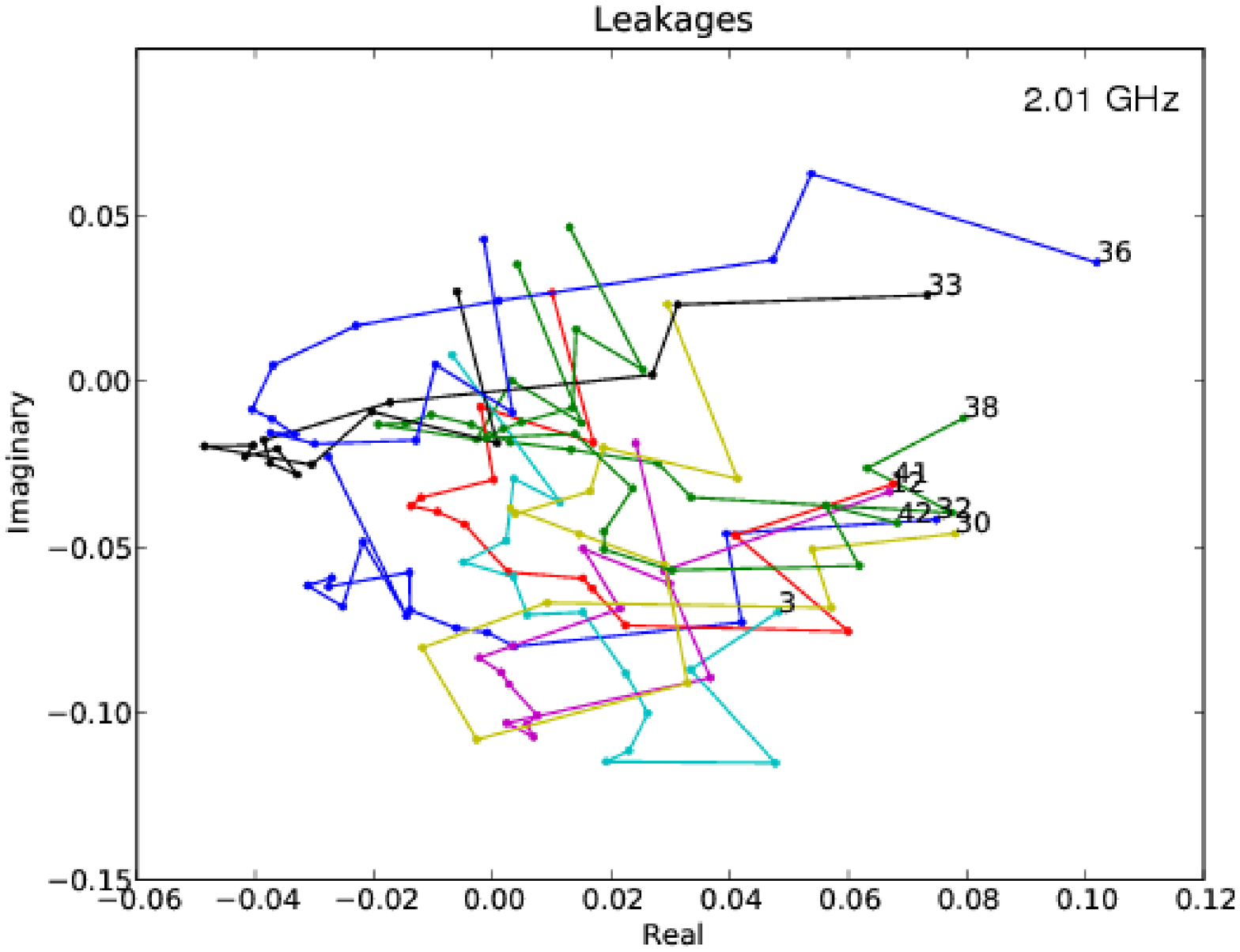}
\caption{Sample of the frequency-dependent polarization leakage of several antennas.  Each dot shows the complex leakage in a 5 MHz segment of the band;  later analysis uses a resolution of 0.5 MHz.  The lowest-frequency, 5-MHz segment is labeled with the antenna number.  For clarity, only half of the antennas are shown.  The plots show different bands:  1.0 GHz (top left), 1.43 GHz (top right), 1.8 GHz (bottom left), and 2.01 GHz (bottom right).  \label{leakages}}
\end{figure}

Figure \ref{leakages} plots the $x$-polarization leakages for different bands, based on observations of 3C 286.  The typical magnitude of the leakages is less than 10\%, with a few antennas ranging up to 20\%.  The leakage changes in frequency by about 3\% per 10 MHz with a tendency to make loops in (real, imaginary) space.  Comparing repeated observations of 3C 286 shows that the leakages change by less than 5\% over a time scale of a month.  Observations with 3C 286 located off-axis find false Stokes (Q, U) signals typically less than 1\% of Stokes I within a third of the half-power point of the primary beam ($0\ddeg3$ at 2.01 GHz).  Applying calibration derived on-axis is likely reliable for the offsets up to $0\ddeg1$\ considered here.  However, since the errors for off-axis sources are not explicitly measured, results for these sources should be treated cautiously.  The behavior of leakage in the ATA is discussed in more detail elsewhere \citep{polcalmemo1,polcalmemo2}.

\subsection{Calibrated Data Quality}
\label{dataqual}
\begin{figure}[tbp]
\includegraphics[width=0.6\textwidth]{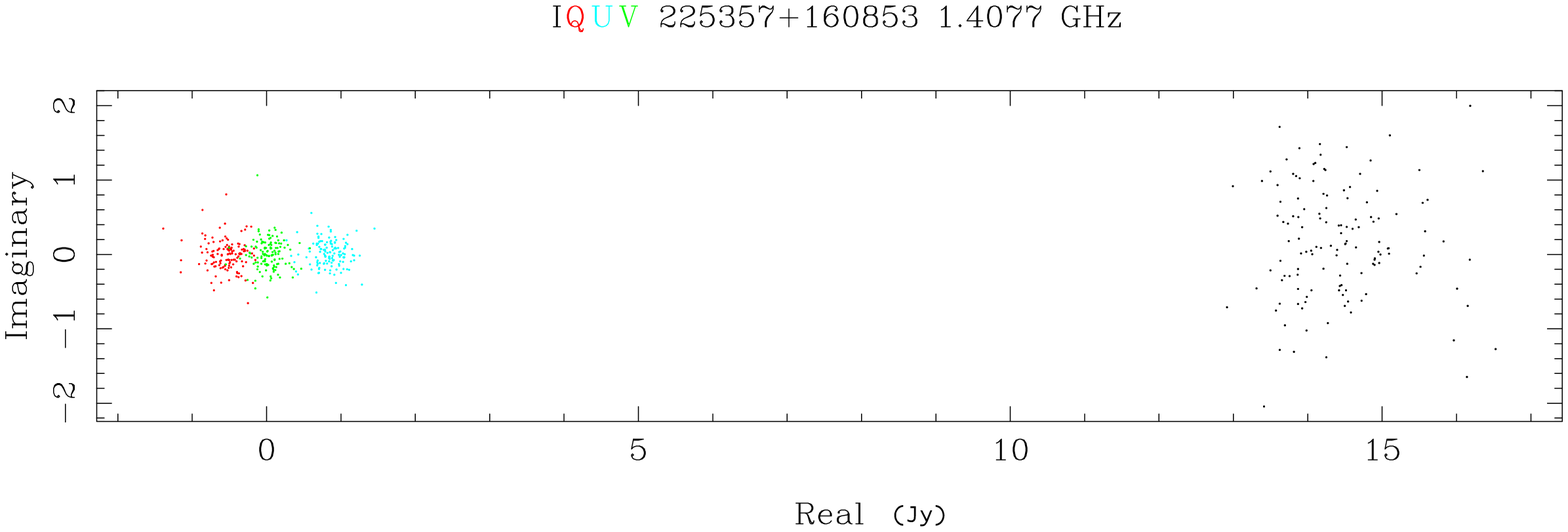}

\vspace{-2cm}

\includegraphics[width=0.6\textwidth]{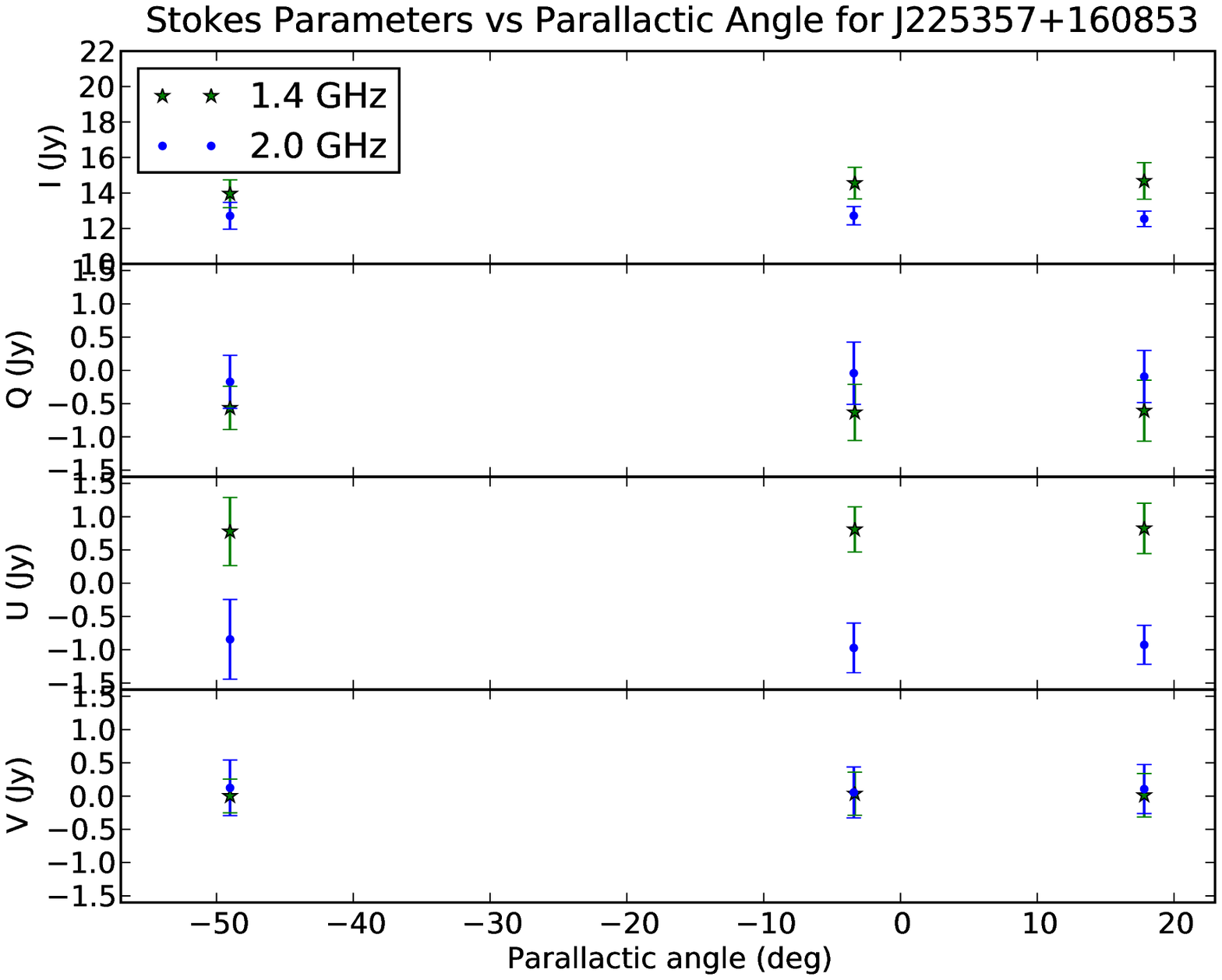}
\caption{\emph{Top:} Visibilities in (real, imaginary) space for J225357+160853 at 1.4 GHz after applying the 3C 286 calibration solution.  Visibilities were averaged over a 0.5 MHz segment and 20 minutes of integration time.  Stokes I has larger scatter because there is an unpolarized, $\sim2$\ Jy source off axis. \emph{Bottom:}  The Stokes I, Q, U, and V of J225357+160853 plotted as a function of parallactic angle at two frequencies.  Stokes parameters from typical 0.5 MHz segments in the 1.43 and 2.01 GHz bands are shown.  The mean and standard deviation in Stokes parameters are measured from the distribution of values over all baselines and channels in the segment.  The plots show that there are no significant changes in the Stokes parameters with parallactic angle.  This plot also shows that the daily variation in the ionosphere is not significant.  The data points with parallactic angle of 50\sdeg\ corresponds to sunset at the ATA, which should have the largest RM contribution from the ionosphere, but shows little deviation from the other points. \label{calquality}}
\end{figure}

As a first-order check of data quality we inspected the calibrated Stokes parameters of simple sources.  Figure \ref{calquality} shows three plots that demonstrate the calibration quality at 1.43 GHz toward J225357+160853, a bright, compact source seen over a wide range of parallactic angle.  The calibrated visibilities for this compact source in (real, imaginary) space have Gaussian distributions centered at the expected flux densities.  Plots of the Stokes parameters as a function of parallactic angle indicate leakage calibration errors, since those kinds of changes move with the feed as it rotates on the sky.  The plot shows a change in Stokes Q, U, and V with parallactic angle by at most 0.1 Jy (less than 1\% of Stokes I).  Errors of this scale should be expected in any 0.5 MHz segment;  as described in \S \ref{imaging}, the error analysis accounts for this by measuring the noise in Stokes V images.

The ultimate test of the data quality was to measure the Stokes Q, U and RM of 3C 286 when treated as a calibrator and as a target.  As described earlier, we observed 3C 286 for calibration, which required long observations over a range of parallactic angle.  But we also observed 3C 286 briefly as a target, sometimes called J133108+303032 (see Table \ref{fields}).  The general calibration solution is derived only from observations of 3C 286 as a calibrator, which is transferred to J133108+303032 and all other targets.  This approach allows us to quantify two kinds of error.  First, when treating 3C 286 as a calibrator, the calibration method will ideally produce the fluxes in the model of 3C 286.  In practice, there are deviations that indicate some aspect of the ATA calibration model that is not correct (ionospheric Faraday rotation, unflagged RFI, second-order leakage corrections, etc.).  Second, when treating 3C 286 as a target, any additional errors can be attributed to the application of calibration solutions to different times and observing conditions.  Thus, measuring the RM of our calibrator tests the validity of our calibration model and estimates systematic error in the RM measured in the target fields.  

Table \ref{3c286rm} shows the errors in observed polarization fraction and RM for 3C 286 when treated as a calibrator and as a target.  The deviation of the calibrated data from the model tells us about the calibration applied to targets in the survey.  Each row shows a combination of data from different days or bands that have calibration applied to them.  For example, data from days 2 and 3 are merged in two cases because the same targets were observed on those days in those bands.  In contrast, day 1 and 2 observations at 1.43 and 2.01 GHz are treated separately because different targets were observed on both of these days.  The last two lines of Table \ref{3c286rm} are the most relevant, since they show errors measured for 3C 286 for combinations of bands used to generate the final Stokes Q and U spectra used to generate the RM spectra.  

\begin{deluxetable}{l|cc|cc}
\tablecaption{Calibration Error toward 3C 286 Treated as Calibrator and Target \label{3c286rm}}
\tablewidth{0pt}
\tablehead{
\colhead{Day, Frequency} & \multicolumn{2}{c}{3C 286 as Calibrator} & \multicolumn{2}{c}{3C 286 as Target} \\
 & \colhead{$|p_{\rm{err}}|$} & \colhead{RM} & \colhead{$|p_{\rm{err}}|$} & \colhead{RM} \\
\colhead{---, (GHz)} & \colhead{(\% I)} & \colhead{(rad m$^{-2}$)} & \colhead{(\% I)} & \colhead{(rad m$^{-2}$)}
}
\startdata
Day 1, 1.43 & $0.26\pm0.02$ & $0.05\pm0.33$ & $0.16\pm0.06$ & $-3.22 \pm1.12$ \\
Day 1, 2.01 & $0.18\pm0.03$ & $-0.09\pm2.30$ & $0.15\pm0.26$ & $92.5\pm29.12$\tablenotemark{a} \\
Day 2, 1.43 & $0.30\pm0.05$ & $-0.82\pm1.10$ & $\ldots$ & $\ldots$ \\
Day 2, 2.01 & $0.21\pm0.08$ & $1.32\pm5.17$ & $\ldots$ & $\ldots$ \\
Day 2+3, 1.0 & $0.27\pm0.02$ & $-1.18\pm0.27$ & $0.65\pm0.10$ & $-2.79\pm1.36$ \\
Day 2+3, 1.8 & $0.07\pm0.01$ & $1.73\pm0.24$ & $0.42\pm0.08$ & $0.75\pm3.79$ \\ \hline
Day 1, 1.43--2.01 & $0.20\pm0.02$ & $0.02\pm0.08$ & $0.15\pm0.08$ &  $-0.76\pm0.34$ \\ 
All\tablenotemark{b} & $0.08\pm0.01$ & $0.02\pm0.02$ & $0.20\pm0.05$ & $0.56\pm0.08$ 
\enddata
\tablenotetext{a}{Flagging reduced the bandwidth and RM constraint significantly for all 2.01 GHz observations on Day 1.}
\tablenotetext{b}{Day 1 data at 1.43 and 2.01 GHz is combined with day 2 and day 3 data at 1.0 and 1.8 GHz in this measurement.  There is no significant difference in the result if we use day 2 data at 1.43 and 2.01 GHz instead.}
\end{deluxetable}

All error quantities are measured from observed Stokes Q and U values as a function of frequency.  The change in the polarization angle as a function of $\lambda^2$\ is used to measure RM (in practice, RM synthesis is used;  see \S \ref{rmsynth}).  The observed Stokes parameters are compared to the expected frequency-dependence of 3C 286 to measure fractional polarization error, $p_{err}$.  The absolute polarization error is the difference between the modeled and observed polarization vector averaged over all frequencies.  The error in this quantity is the standard deviation in the error vector for all channels in the band.  This analysis shows that the error in the polarization vector expected toward targets is roughly 0.2\% of Stokes I, equivalent to an absolute position angle error of $\sim1$\sdeg.  The RM error toward 3C 286, when treated as a calibrator, is generally consistent with the measured noise.  The RM measured toward 3C 286, when treated as a target, has a slightly larger deviation from RM $=0$, suggesting that systematic effects are present at the $\sim2\sigma$ level.  The RM measured between two and four bands (relevant for all target fields) is more reliable, but shows that the RM of the target fields are not more accurate than $\sim0.6$\ rad m$^{-2}$.  A second polarization calibrator, 3C 138 (J052109+163822), was also calibrated as a ordinary target field and has a RM consistent with its known value (shown in \S \ref{rmcompsec}).

The accuracy of the RM measurement is consistent with the variation expected from the ionosphere, which is typically less than 1 rad m$^{-2}$ \citep{m05,b08,mao10}.  Ionospheric Faraday rotation may exceed this limit at sunrise and sunset, which spans about three hours of observing on day 1 and 1 hour of observing on day 2.  Figure \ref{calquality} shows how the Stokes parameters change for one source during sunset, when ionospheric RM is expected to be largest.  The constancy of Stokes Q and U in that case limits RM variation due to the ionosphere to less than 2 rad m$^{-2}$.

\subsection{Imaging}
\label{imaging}
After applying calibration solutions to target fields, each frequency segment was imaged in all four Stokes parameters.  Imaging parameters at each frequency were chosen to maximize image fidelity for measuring peak flux density.  In general, images were not forced to the same beam size, since most sources are unresolved by the ATA.  As described in \S \ref{polsrc}, only one field (J123039+121758) was imaged with a fixed beam size because its RM spectra showed some beam-size dependence.  Section \ref{polsrc} discusses two other sources, J172025–005852 and J184226+794517, that are doubles only marginally resolved at 1.0 GHz.

Visibilities for the target fields were Fourier transformed with a ``robustness'' of 0, which slightly downweights long baselines to improve image fidelity.  Images at all wavelengths had pixel size of 10\arcsec\ and 1024 pixels on a side, giving a field of view of $2\ddeg8$.  The pixels overresolve the synthesized beam (ranging from 107\arcsec\ to 215\arcsec\ from 2.0 to 1.0 GHz) and the images cover most of the primary beam at all wavelengths \citep[ranging from $1\ddeg8$ to $3\ddeg6$ from 2.0 to 1.0 GHz; ][]{h10}, helping make robust images.  The region within 5 to 10\arcmin\ of the center of each image was ``cleaned'' by iteratively fitting and removing the brightest sources down to a threshold of 4.5 times the Stokes I image theoretical noise level.  Inspection of the images showed no obvious signs of cleaning problems.  All subsequent analysis is done on a restored image formed by convolving the best-fit source distribution with an idealized, Gaussian synthesized beam \citep{h74,t01}.

The radiometer equation predicts the Stokes I thermal noise for ATA should be $\sigma = \rm{SEFD}/\sqrt{BW*t}$, where SEFD is the system equivalent flux density, $BW$\ is the bandwidth and $t$\ is the length of the observation.  During these observations, we commonly used 21 dishes with a typical system temperature of 65 K, which translates to SEFD$=450$\ Jy \citep{w09,w10}.  For a typical image of a 5-channel segment, $BW=0.5$\ MHz, $t=7$\ minutes, the ideal sensitivity is 30 mJy beam$^{-1}$.  

In practice, short ATA commissioning observations are dynamic range limited \footnote{Image dynamic range is defined here as the maximum of an image divided by its standard deviation measured off peak.} \citep{c10}.  The dynamic range limit manifests as a featureless, noise-like signal for all Stokes parameters that scales with stokes I.  Our 1.43 GHz Stokes I images have a maximum dynamic range of 400 and a typical value of about 150.  ATA images of Stokes Q, U, V and polarized intensity tend to be less limited by dynamic range and have noise levels within a factor of 2--5 of theoretical expectations.  However, fields with Stokes I brighter than 20 Jy have more noise in polarized intensity.  This sets the ATA dynamic range limit for images of polarized intensity at 3000 relative to Stokes I \footnote{Comparing the polarized intensity noise to the peak polarized intensity is not as meaningful, since some sources are depolarized.}.

\begin{figure}[tbp]
\includegraphics[width=0.4\textwidth,angle=270]{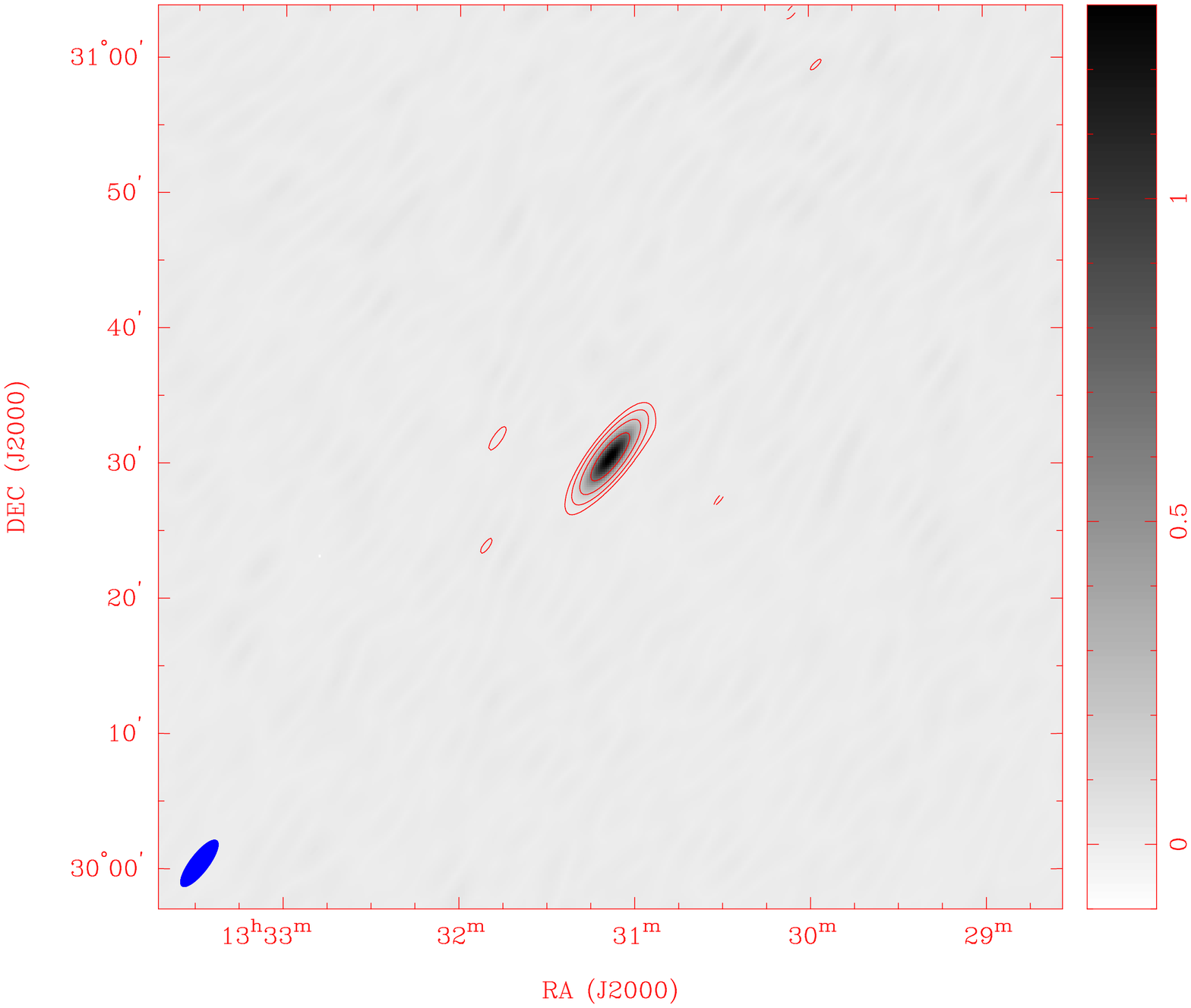}
\includegraphics[width=0.4\textwidth,angle=270]{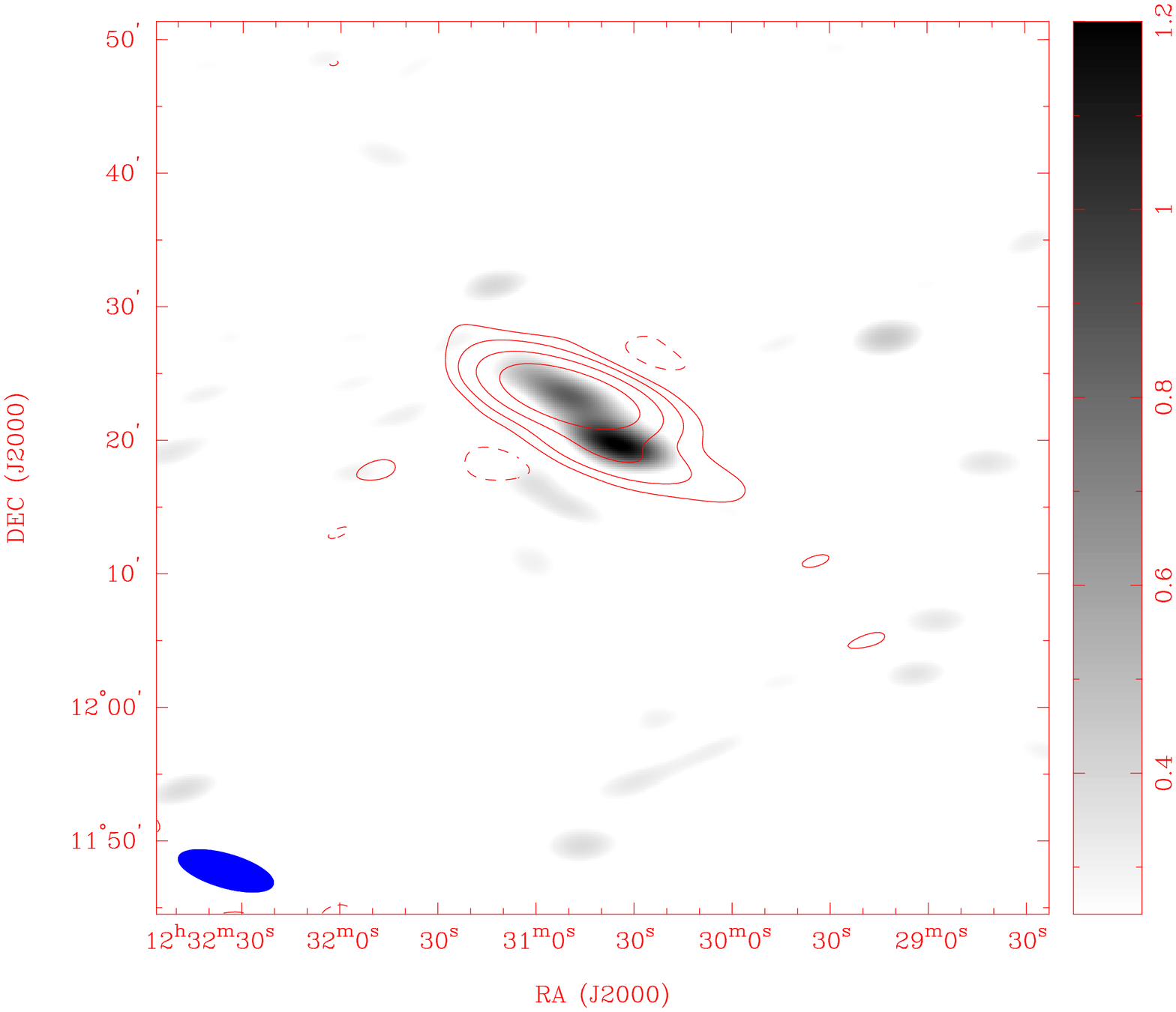}
\caption{Images showing imaging quality for a typical and a challenging field averaged over the 1.43 GHz band on day 1.  \emph{Left:}  Gray scale shows linearly-polarized flux of field J133108+303032 (a.k.a. 3C 286) with the gray scale bar at right given in units of Jy beam$^{-1}$.  Contours show the total intensity at levels of --0.25, 0.25, 0.75, and 2.25, 6.75 Jy beam$^{-1}$.  \emph{Right:}  Polarized and total intensity images of field J123039+121758.  This field contains M87, the brightest and most complex source in this survey, to demonstrate the most challenging field studied in this survey.  Contours show total intensity at levels of --1.3, 1.3, 3.9, 11.7, and 35.1 Jy beam$^{-1}$.  The core of M87 is very bright in Stokes I, but largely depolarized when averaging across the 1.43 GHz band.  \label{images}}
\end{figure}

Examples of the image quality in two fields are shown in Figure \ref{images}.  First we show an image of J133108+303032 to demonstrate the image quality for the calibrator field when it is treated as a target.  As with all fields, the Stokes I image of J133108+303032 is dynamic range limited at a level of roughly 350;  the noise in the polarized intensity image is about two times the ideal, so it does not seem to be dynamic range limited.  The second image shows field J123039+121758, which contains M87 \citep{r96}.  As the brightest and most complex field in our survey, it represents the worst-case scenario for image quality.  The Stokes I image shows negative bowls produced by flagging visibilities with \emph{uv}-distance less than $200\lambda$.  The polarized flux image shows that the M87 core is depolarized and that the jet is the brightest polarized source in the field with a brightness of 1.2 Jy beam$^{-1}$.  The image has a standard deviation of about 55 mJy beam$^{-1}$, which corresponds to a polarization dynamic range of roughly 20 relative to the polarized peak and 2600 relative to the total intensity peak.  So, while the polarized intensity image noise is large, it is largely caused by the extreme brightness of M87.

The presence of calibration errors and a dynamic range limit for the ATA require all image error analysis to be based on the observed noise.  Since these effects and thermal noise in Stokes Q, U, and V scale together, we can use Stokes V to estimate the noise in Stokes Q and U.  True Stokes V emission is generally much less than 1\% of Stokes I \citep{h99}, which is too weak to be detected in our individual 0.5 MHz images.  We inspected dirty images of Stokes V and found that most images were noise-like and suitable for estimating noise in Stokes Q and U.  In some cases, calibration errors produced weak sources in Stokes V at a level less than 1\%, as discussed in \S \ref{dataqual}.  However, the uncleaned Stokes V images tend to spread this false emission and increase the standard deviation of pixel values in the image.  For this reason, we set the error for each image of Stokes Q and U to the standard deviation of pixel values in uncleaned Stokes V image.

\section{Analysis}
\label{ana}
\subsection{Rotation Measure Synthesis}
\label{rmsynth}
Traditionally, the RM was measured by fitting the standard, $\lambda^2$ plasma dispersion law to measurements of $\theta$ at multiple wavelengths.  However, this technique does not account for ambiguity in the rotation of the polarization angle between sampling points in $\lambda^2$-space (known as the ``$n\pi$ ambiguity'').  Also, in many cases the polarization angle does not follow the $\lambda^2$ law, giving misleading results from the angle-fitting technique \citep{b66}.

\citet{b05} derived a more generally valid way of studying RM:  rotation measure synthesis.  First, the technique treats the linear polarization as a complex vector ($\rm{P} = \rm{Q} + i \rm{U}$) rather than a scalar ($\theta$).  Now the polarization angle change is visualized as a rotation of the Stokes vector with frequency.  Second, the concept of rotation measure, which is essentially an observed quantity, is changed to a more general ``Faraday depth'', $\phi$.  In this way, the polarization vector from each Faraday depth, $F(\phi)$, can be integrated to produce the apparent complex polarization surface brightness \citep{b66,b99,b05}:
\begin{equation}
\label{acpsb}
P(\lambda^2) = \int^{\infty}_{-\infty} \! F(\phi) \, e^{2i\phi\lambda^2} \, d\phi.
\end{equation}

The similarity of Equation \ref{acpsb} to a Fourier transform makes it easy to invert and derive the distribution of polarized brightness in Faraday depth-space.  Indeed, the name ``RM synthesis'' was chosen to show similarity of this technique to aperture synthesis \citep{b05}.  The distribution of samples in $\lambda^2$-space is analogous to antennas in physical space;  widely separated points constrain closely spaced RM and compact sources, respectively.  The shortest wavelength determines the largest ``Faraday thickness'' detectable ($\Delta\phi\approx\pi/\lambda_{\rm{min}}^2$).  A Fourier transform of the distribution of points in $\lambda^2$-space shows the ``rotation measure spread function'' (RMSF), which is analogous to a ``dirty beam'' in aperture synthesis.

\begin{figure}[tbp]
\includegraphics[angle=270,width=\textwidth]{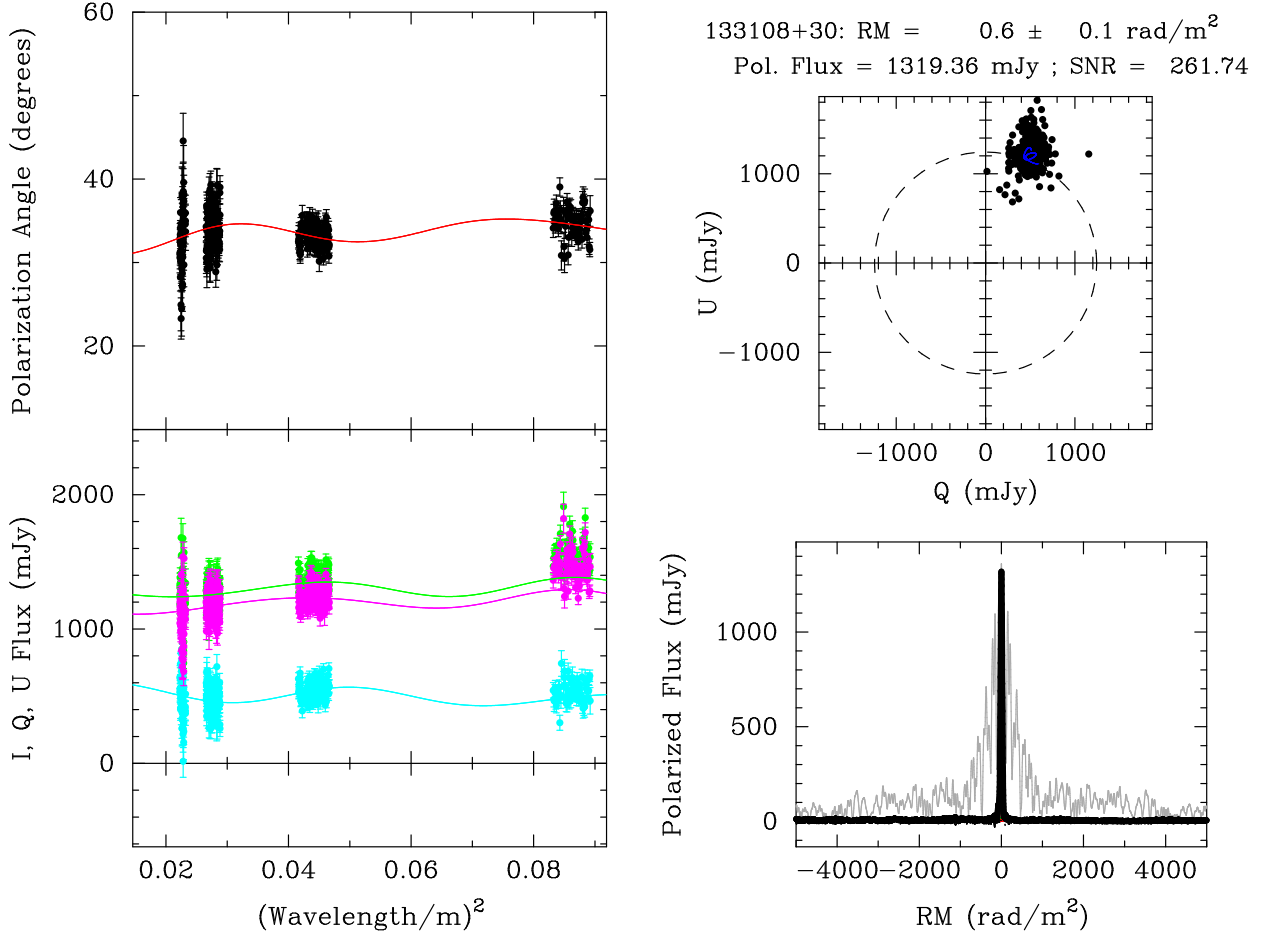}
\caption{An example of the input to and output from the RM synthesis algorithm for all observations of 3C 286, when treated as a target.  Deviations of the RM clean model from the ideal (RM $=0$) should be similar to the typical errors in RM spectra for all targets presented in this survey.  \emph{Top right:} Each point shows the measured Stokes Q and U brightness at one frequency.  The clean model for the brightest 10 clean components is shown with a red line. \emph{Top left:} The points show the measured polarization angle as a function of $\lambda^2$ with a red line showing the clean model.  \emph{Bottom left:}  The green, blue, and purple points show the measured polarized brightness, Stokes Q, and Stokes U, respectively, as a function of $\lambda^2$.  The corresponding colored lines show the clean model.  \emph{Bottom right:} The RM spectrum from $-5000$ to 5000 rad m$^{-2}$.  The thin and thick lines show the dirty and clean RM spectra, respectively.  \label{rmsynthplot}}
\end{figure}

The output from the RM synthesis algorithm is shown in Figure \ref{rmsynthplot}.  This shows the analysis of all data toward 3C 286, when treated as a target (see Table \ref{3c286rm}).  The script takes Stokes (Q, U) and their error (see \S \ref{imaging}) as a function of frequency.  The algorithm applies a Fourier transform to the complex Stokes vectors to generate a dirty RM spectrum.  The RM spectrum is then interactively cleaned by fitting the RMSF to the dirty RM spectrum and subtracting the best fit component \citep[``RM-CLEAN'' algorithm; ][]{h09}.  Since 3C 286 has only one RM component, the thin line in the bottom right plot shows a typical RMSF for an observation with 1--2 GHz data.  

Much as with image cleaning algorithms, RM-CLEAN has many parameters that control the depth and robustness of the final cleaned spectrum.  We developed a two-step cleaning process to avoid undercleaning or confusing RM sidelobes with noise.  The initial RM cleaning threshold is 3 times the mean of the noise of all input Q and U images.  We then measure the standard deviation in the cleaned Q and U spectra for RM larger than 10 times the RM resolution;  this is equivalent to measuring the noise in a cleaned image far from sources near the phase center.  The final RM cleaning was stopped when the peak brightness was less than 3 times this measured noise level.  The final RM spectrum is made by convolving the brightest 10 clean components with a Gaussian of width equal to the central part of the RMSF and adding in residuals left after cleaning.

\subsection{Polarized Source Detections}
\label{polsrc}
Of the 37 fields, most contain only a bright, polarized source at the phase center.  However, since the fields were chosen by a simple cut on 1.4 GHz polarized flux density, some fields are more complicated.  Of the 10 complicated fields, five have a second, polarized source within $0\ddeg1$ of the phase center, where the polarization calibration is believed to be reliable.

The RM synthesis analysis was applied to both the central and off-axis sources.  All complicated fields, including the coordinates of secondary sources, are listed in Table \ref{complexfields}.  The RM synthesis analysis required a Stokes Q and U flux for each source at each frequency.  For the sources at phase center, the fluxes were measured at the central pixel of each image, where the \citetalias{t09} source is known to be.  Two fields, J172025–005852 and J184226+794517, have doubles resolved only at frequencies of 1.43 GHz and higher.  This means that the 1.0-GHz polarized properties of the two sources in each field are mixed.  Since these sources are observed with four bands, this ambiguity means that roughly a quarter of the total polarized flux may be shared between these double sources.

Sources in fields J022248+861851, J063633-204233, and J123039+121758 were at least partially resolved.  Since we measure fluxes in Jy beam$^{-1}$, the frequency dependence of the beam size could systematically affect flux measurements in these cases.  For these three fields, we repeated RM synthesis analysis for these sources based on images with a fixed beam size at all frequencies.  For sources in fields J022248+861851 and J063633-204233 there was no significant change to the RM synthesis results.  In the analysis of field J123039+121758, there was significant difference, so all analysis is done on images with a beam size of 440\arcsec$\times$150\arcsec at a position angle of 74\sdeg.  The sources in this field are discussed in detail in \S \ref{m87}.

\begin{deluxetable}{lccc}
\tablecaption{Fields with Complexity \label{complexfields}}
\tablewidth{0pt}
\tablehead{
\colhead{Field Name(s)} & \multicolumn{2}{c}{Morphology} & \colhead{Second Source} \\
  & Stokes I & Pol. Intensity & \colhead{RA, Dec (J2000)} \\
}
\startdata
J010850+131831       &  double & double & (1:08:55, 13:22:14) \\
J022248+861851       &  extended & extended &  \\
J063633--204233      &  extended & extended &  \\
J123039+121758       &  double & extended   & (12:30:49, 12:23:23) \\
J123522+212018       &  double & double     & (12:35:30, 21:20:48) \\
J160231+015748       &  double & single     & (16:02:19, 1:58:25) \\
J165111/2+045919/7\tablenotemark{1} &  single & single  &  \\
J172025--005852      &  double & double     & (17:20:34, --0:58:43) \\
J184226+794517       &  double & double     & (18:41:50, 79:47:28) \\
J212344/5+250410/48\tablenotemark{1}  & single & single & 
\enddata
\tablenotetext{1}{Double resolved by VLA, unresolved by ATA}
\end{deluxetable}

\subsection{RM Spectra}
\label{rmspectra}
\begin{figure}[tbp]
\includegraphics[width=\textwidth]{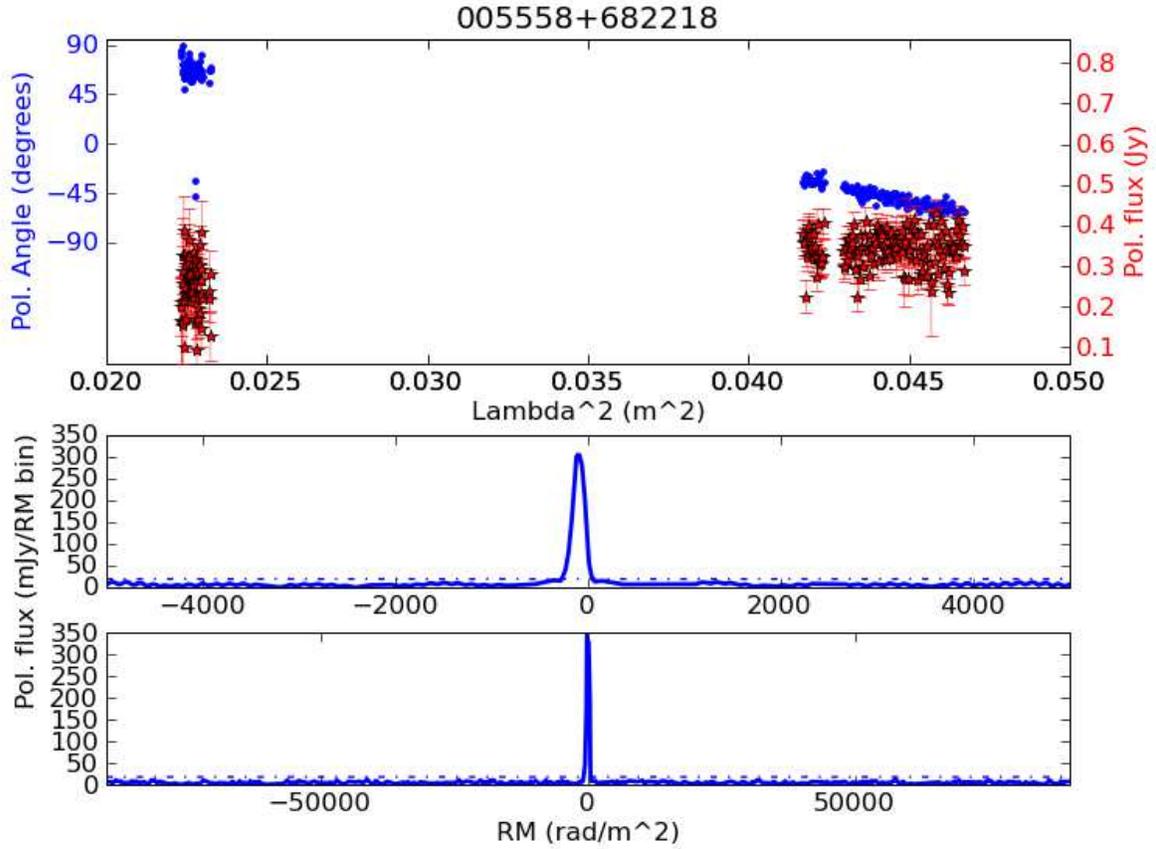}
\caption{RM synthesis summary plots for 3C 27 (J005558+682218).  (\emph{Top:}) Plot showing the polarized flux (red stars) and polarization angle (blue points) as a function of $\lambda^2$. (\emph{Middle:})  Plot showing the RM spectrum for $-5000< RM <5000$\ rad m$^{-2}$, derived from all data for each source.  The dashed line shows 5 times the observed noise in the RM spectrum.  (\emph{Bottom:}) Plot showing the RM spectrum for $-90000< RM <90000$\ rad m$^{-2}$ derived from the 1.43 GHz data only. \label{rmplot1}}
\end{figure}

\begin{figure}[tbp]
\includegraphics[width=\textwidth]{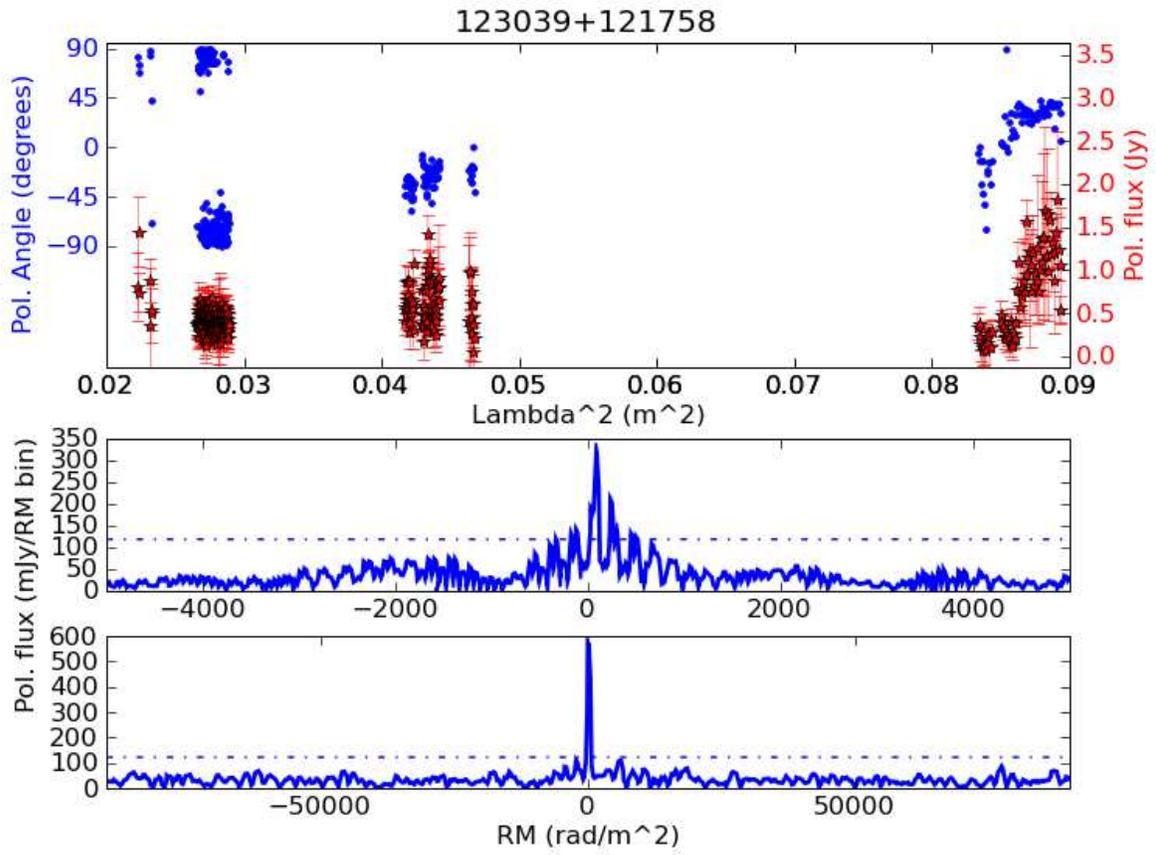}
\caption{RM synthesis summary plots for the M87 jet (J123039+121758), as in Figure \ref{rmplot1}. \label{rmplot2}}
\end{figure}

\begin{figure}[tbp]
\includegraphics[width=\textwidth]{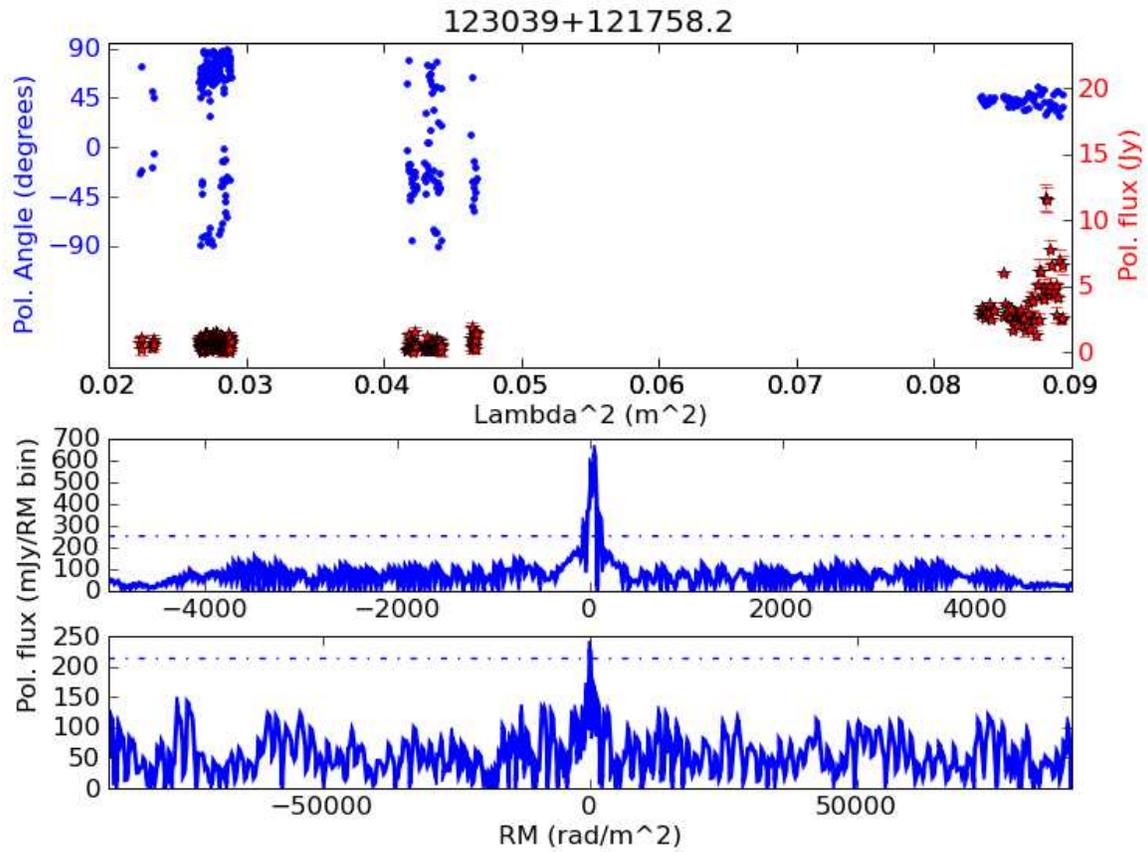}
\caption{RM synthesis summary plots for the M87 core (J123049+122323), as in Figure \ref{rmplot1}. \label{rmplot3}}
\end{figure}

\begin{figure}[tbp]
\includegraphics[width=\textwidth]{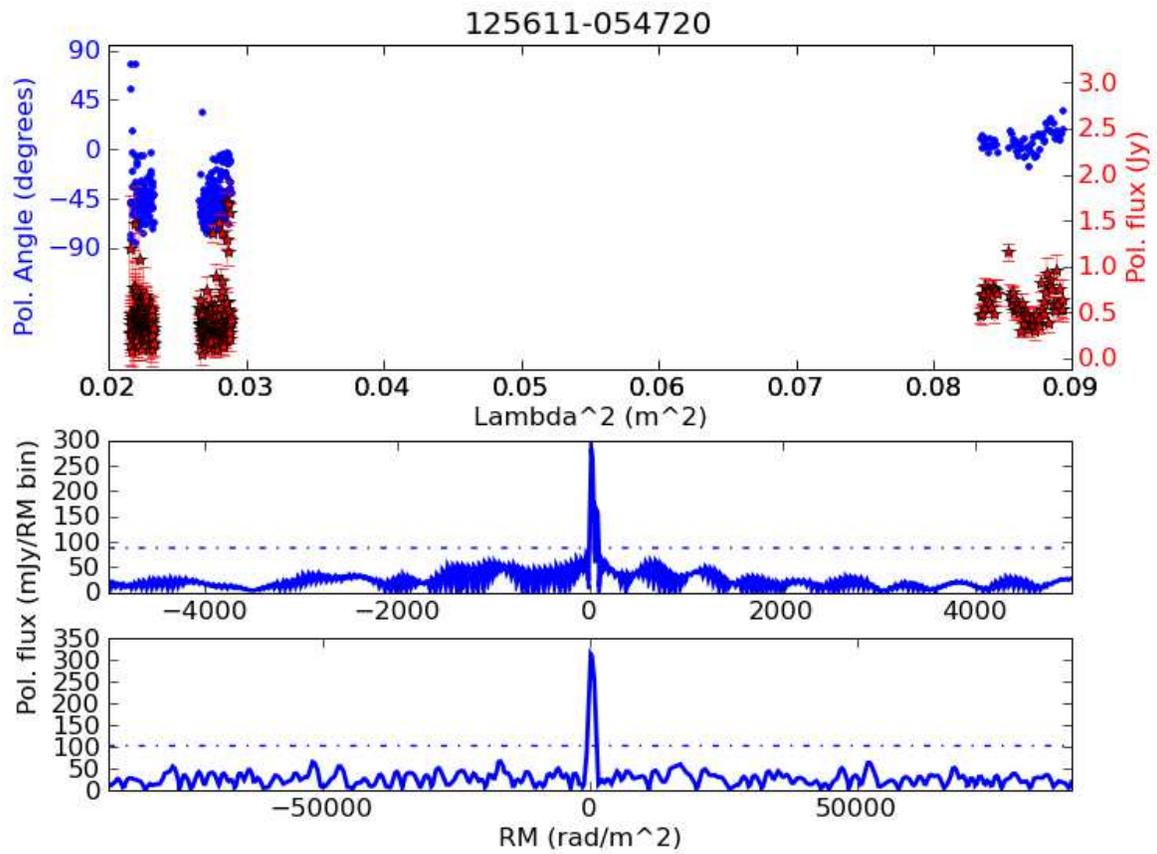}
\caption{RM synthesis summary plots for 3C 279 (J125611--054720), as in Figure \ref{rmplot1}. \label{rmplot6}}
\end{figure}

\begin{figure}[tbp]
\includegraphics[width=\textwidth]{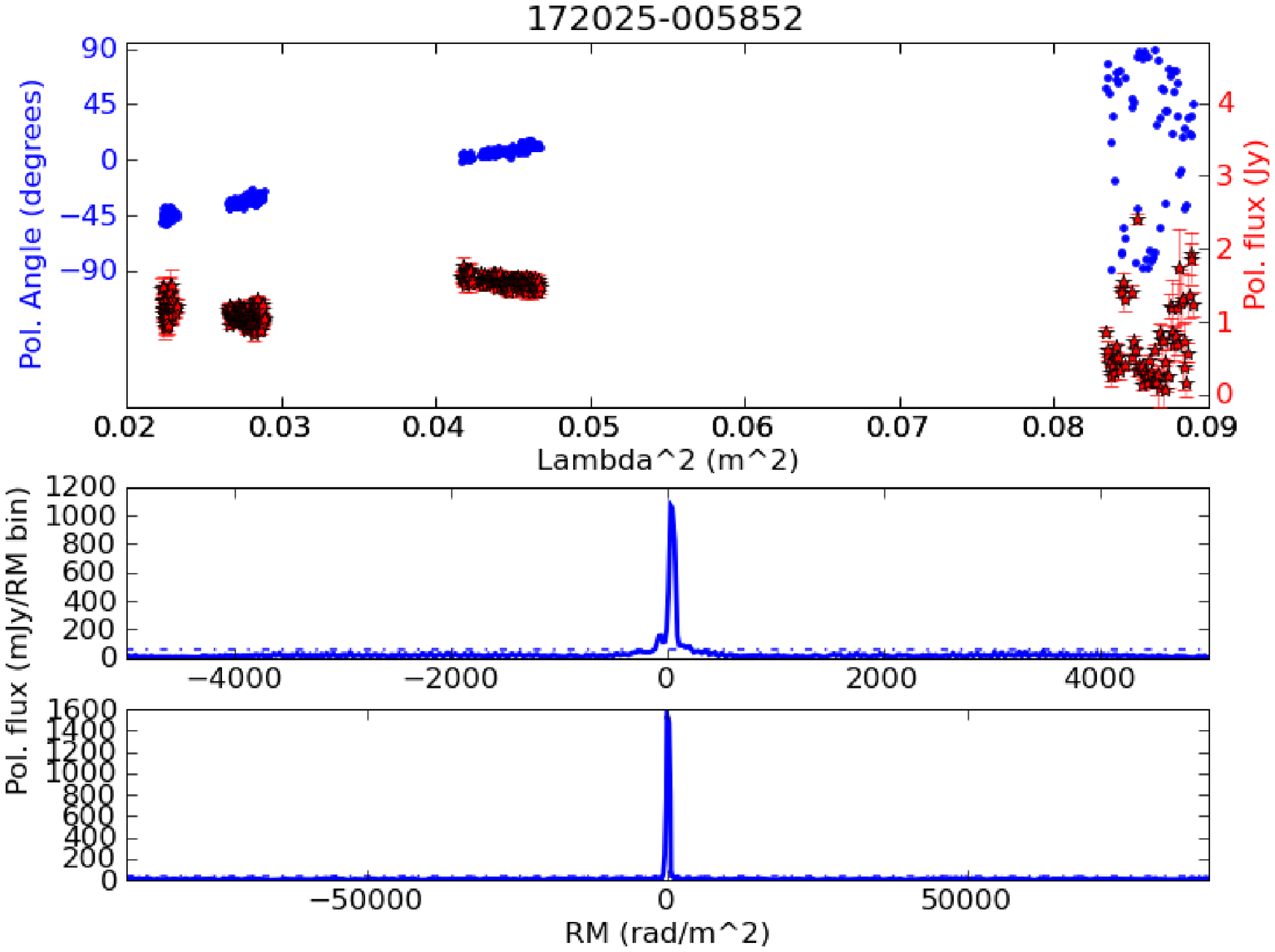}
\caption{RM synthesis summary plots for 3C 353 (J172025--005852), as in Figure \ref{rmplot1}.  Note that J172025--005852 and J172034--005843 are unresolved at 1.0 GHz, so up to a quarter of the polarized flux may be shared between this and Figure \ref{172025.2}.}
\label{172025.1}
\end{figure}

\begin{figure}[tbp]
\includegraphics[width=\textwidth]{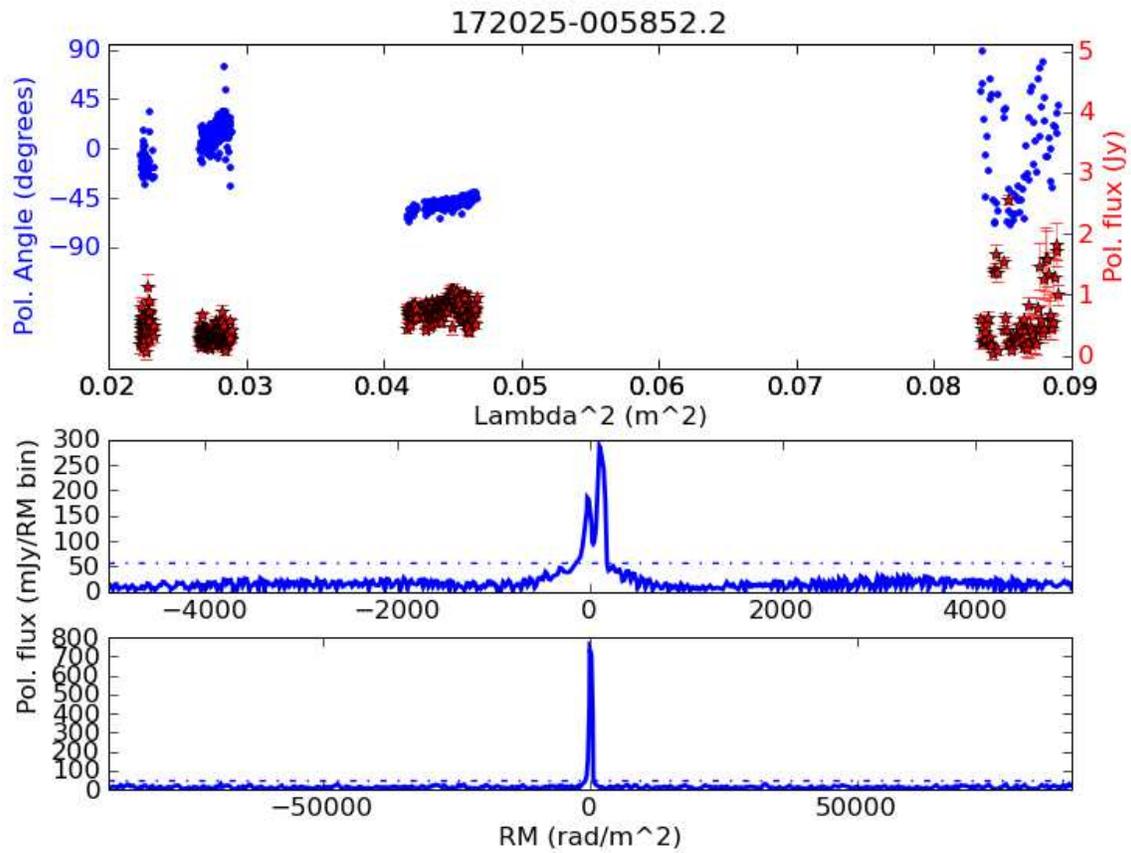}
\caption{RM synthesis summary plots for the secondary source in the field of 3C 353 (J172034--005843), as in Figure \ref{rmplot1}.  Note that J172025--005852 and J172034--005843 are unresolved at 1.0 GHz, so up to a quarter of the polarized flux may be shared between this and Figure \ref{172025.1}.}
\label{172025.2}
\end{figure}

\begin{figure}[tbp]
\includegraphics[width=\textwidth]{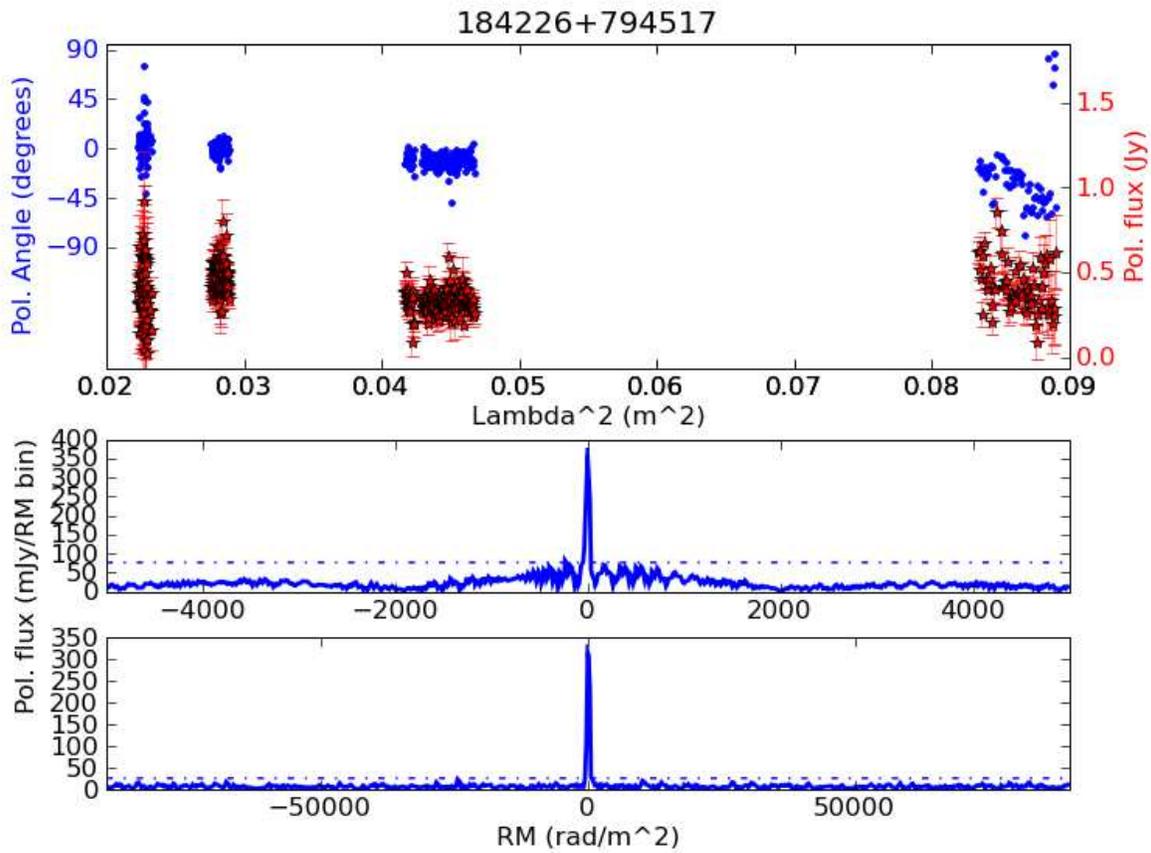}
\caption{RM synthesis summary plots for 3C 390.3 (J184226+794517), as in Figure \ref{rmplot1}.  Note that J184226+794517 and J184150+794728 are unresolved at 1.0 GHz, so up to a quarter of the polarized flux may be shared between this and Figure \ref{184226.2}.
\label{184226.1}}
\end{figure}

\begin{figure}[tbp]
\includegraphics[width=\textwidth]{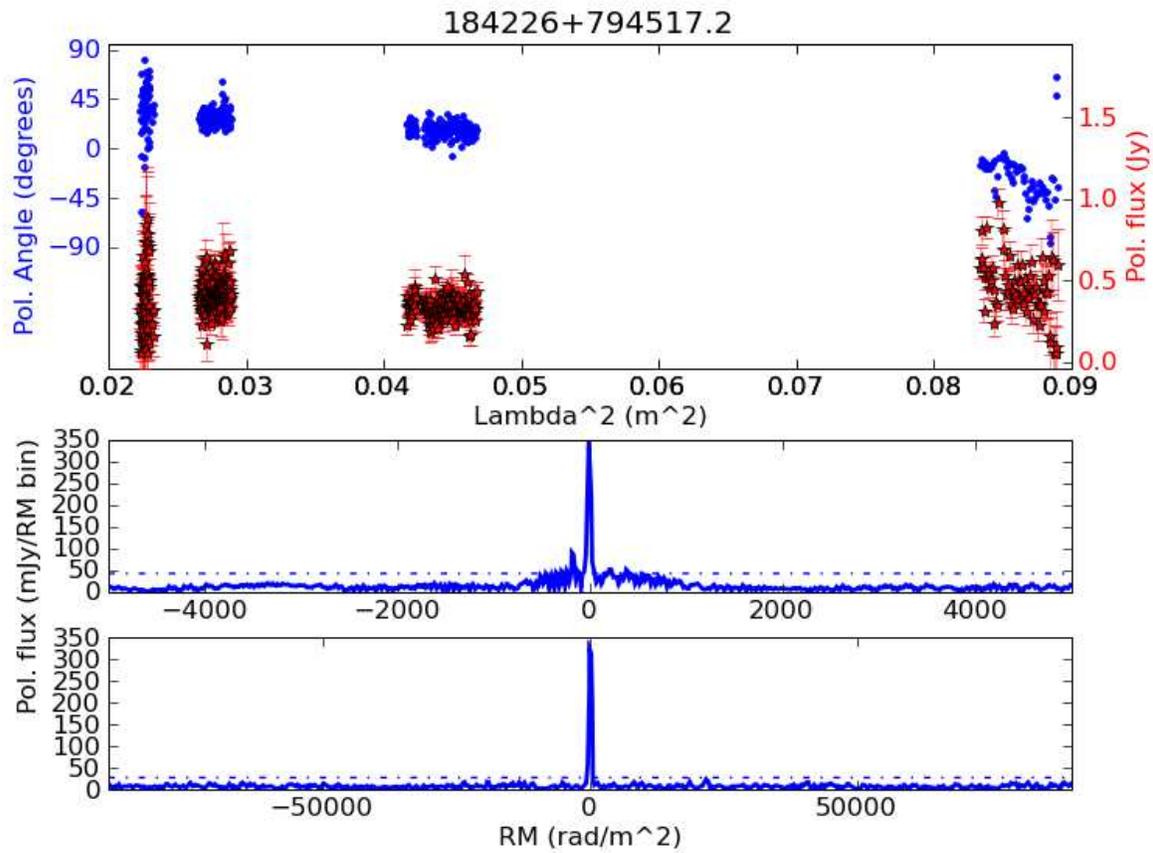}
\caption{RM synthesis summary plots for the secondary source in the field of J184226+794517 (J184150+794728), as in Figure \ref{rmplot1}.  Note that J184226+794517 and J184150+794728 are unresolved at 1.0 GHz, so up to a quarter of the polarized flux may be shared between this and Figure \ref{184226.1}.
\label{184226.2}}
\end{figure}

\begin{figure}[tbp]
\includegraphics[width=\textwidth]{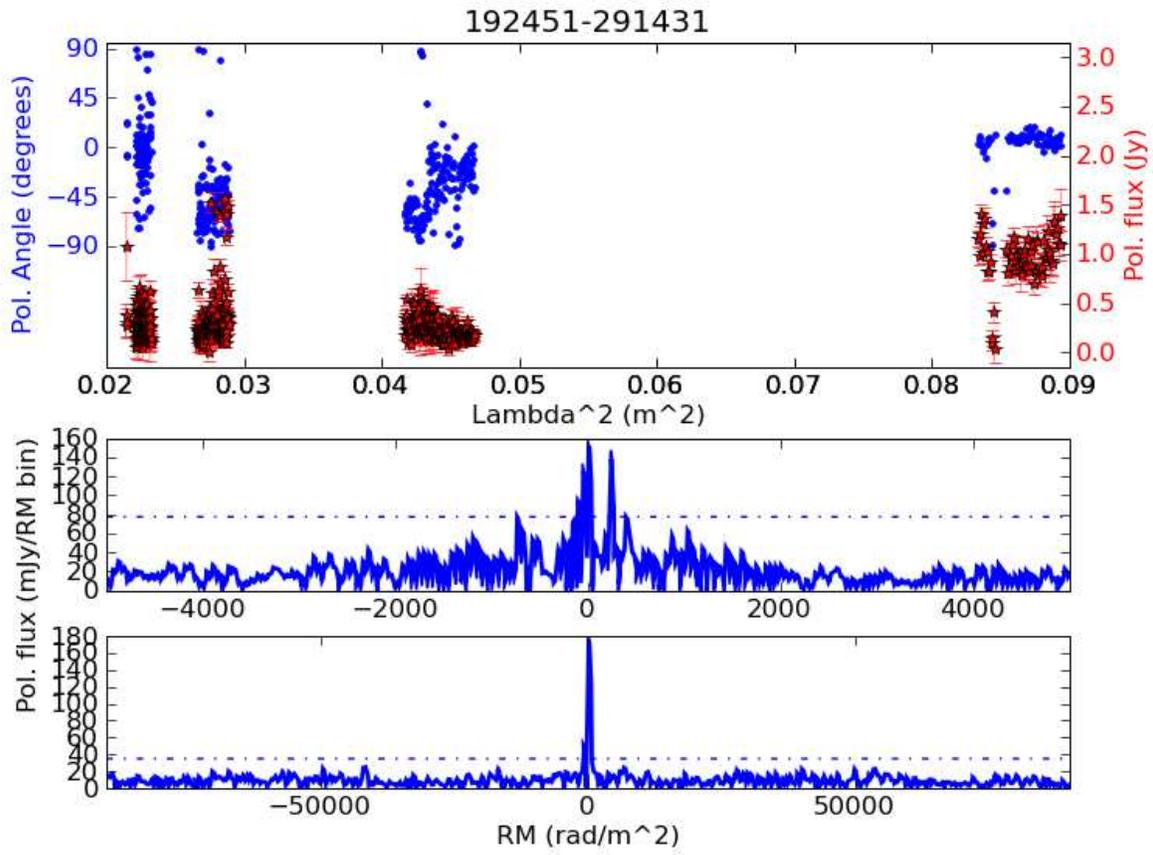}
\caption{RM synthesis summary plots for PKS1921-293 (J192451--291431), as in Figure \ref{rmplot1}. \label{rmplot4}}
\end{figure}

\begin{figure}[tbp]
\includegraphics[width=\textwidth]{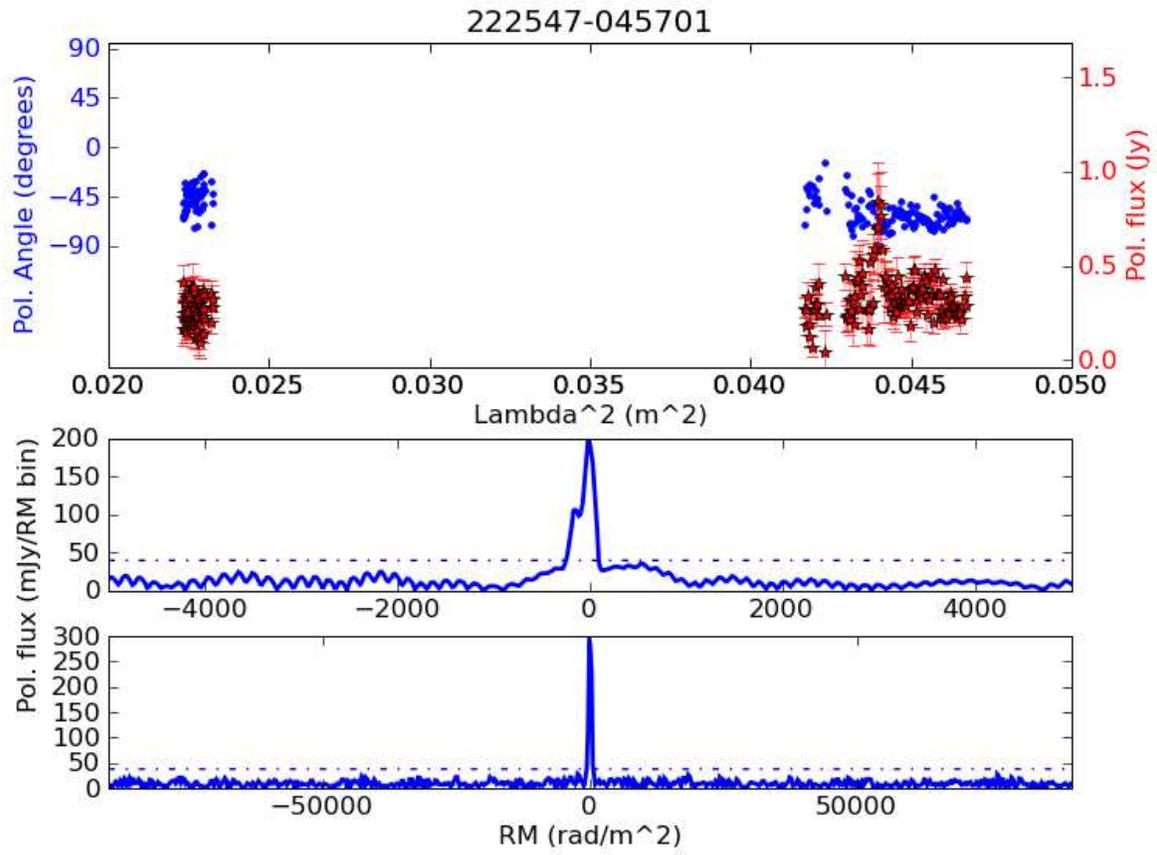}
\caption{RM synthesis summary plots for 3C 446 (J222547--045701), as in Figure \ref{rmplot1}. \label{rmplot5}}
\end{figure}

\begin{figure}[tbp]
\includegraphics[width=\textwidth]{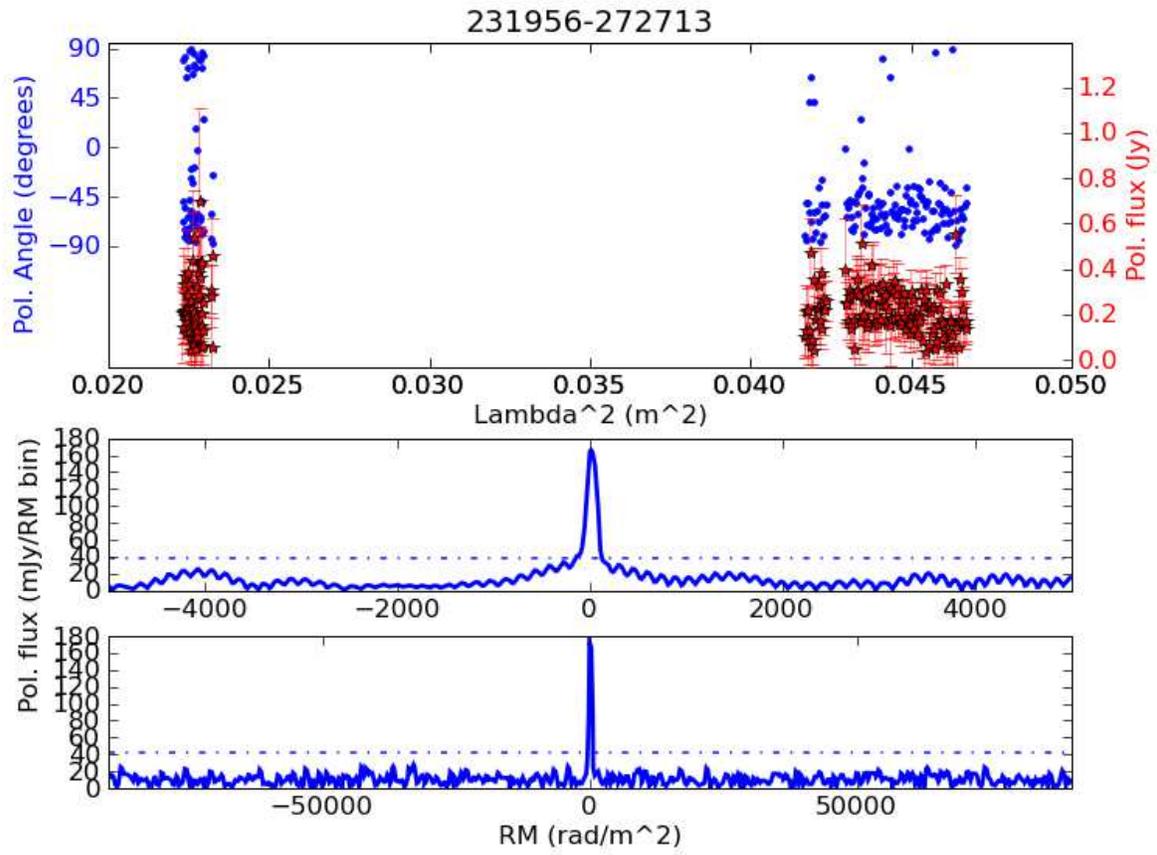}
\caption{RM synthesis summary plots for PKS 2317-27 (J231956--272713), as in Figure \ref{rmplot1}. \label{rmplotlast}}
\end{figure}

In total, we measured the Stokes Q and U spectra for 42 sources at a resolution of 0.5 MHz.  Applying the RM-CLEAN algorithm to the Stokes Q and U spectra produced cleaned RM spectra for each source.  The spectral resolution limits the RM spectra sensitivity beyond roughly $\pm90000$ rad m$^{-2}$.  Cleaned RM spectra for all sources are shown in Figures \ref{rmplot1} through \ref{rmplotlast}.

\notetoeditor{We show all plots from \ref{rmplot1} to \ref{rmplotlast} for completeness and to help the reader understand our technique.  However, they take up a lot of space and are often quite dull.  We would like to have five plots readily accessible to the reader:  \ref{rmplot2}, \ref{rmplot3}, \ref{rmplot4}, \ref{rmplot5}, and \ref{rmplot6}.  Perhaps the others can be hidden in the print version with a representative plot (perhaps \ref{rmplot1}) shown only?  Ideas welcome on how to do this...}

\clearpage

\begin{figure}[tbp]
\includegraphics[width=\textwidth]{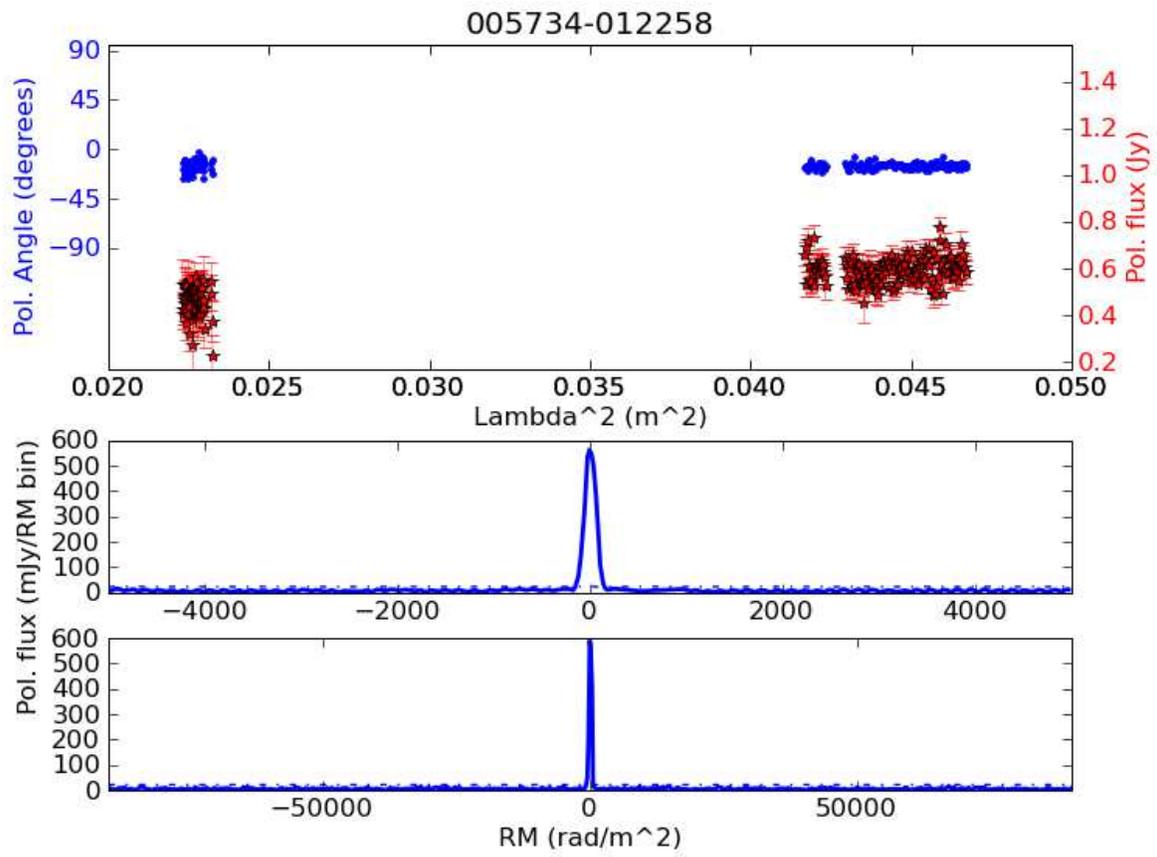}
\caption{RM synthesis summary plots for 3C 29 (J005734--012258), as in Figure \ref{rmplot1}.}
\end{figure}

\begin{figure}[tbp]
\includegraphics[width=\textwidth]{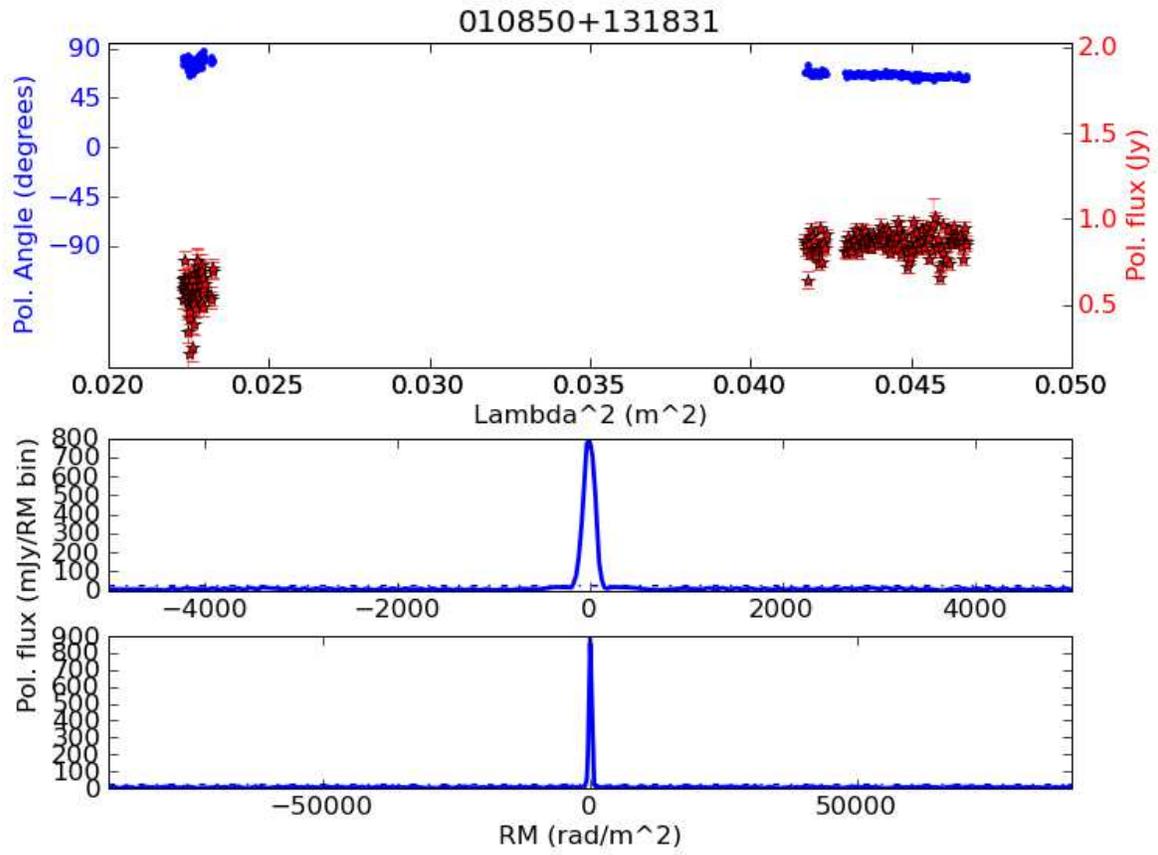}
\caption{RM synthesis summary plots for 3C 33 (J010850+131831), as in Figure \ref{rmplot1}.}
\end{figure}

\begin{figure}[tbp]
\includegraphics[width=\textwidth]{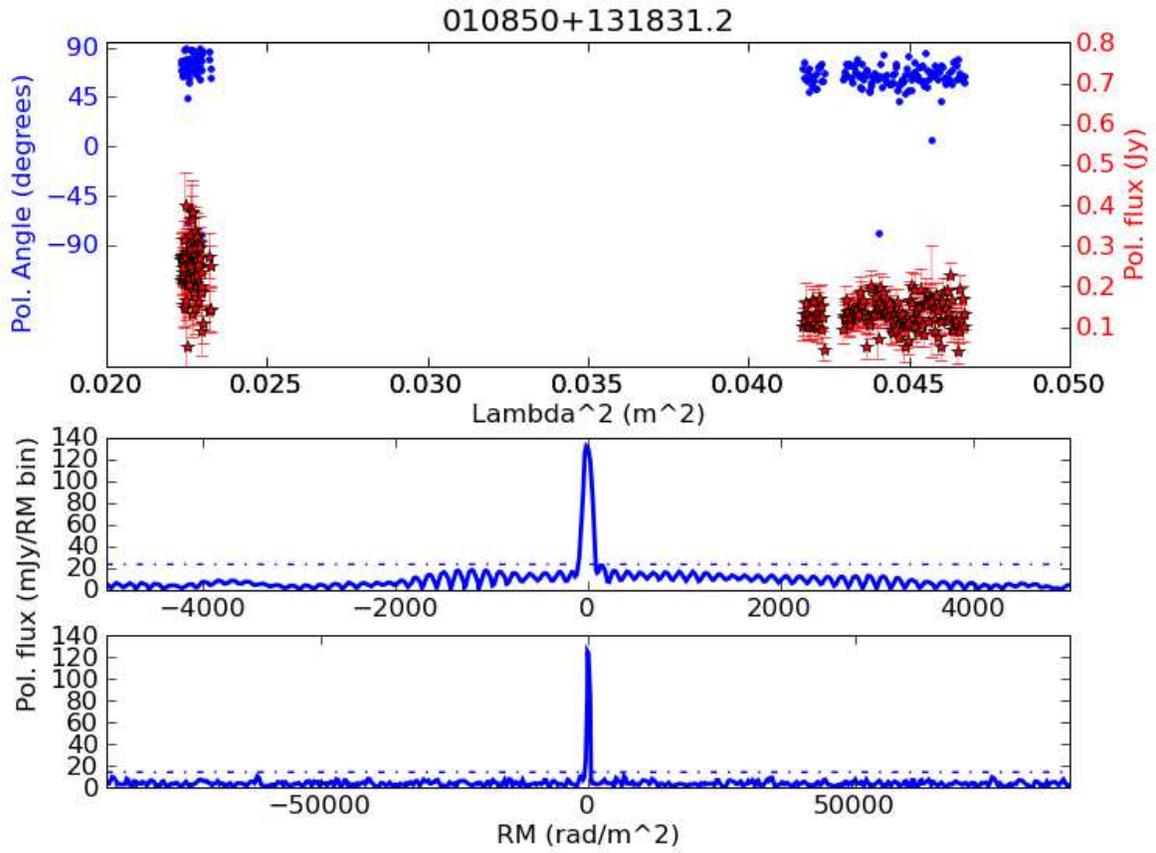}
\caption{RM synthesis summary plots for the secondary source in the field with 3C 33 (J010855+132214) as in Figure \ref{rmplot1}.}
\end{figure}

\begin{figure}[tbp]
\includegraphics[width=\textwidth]{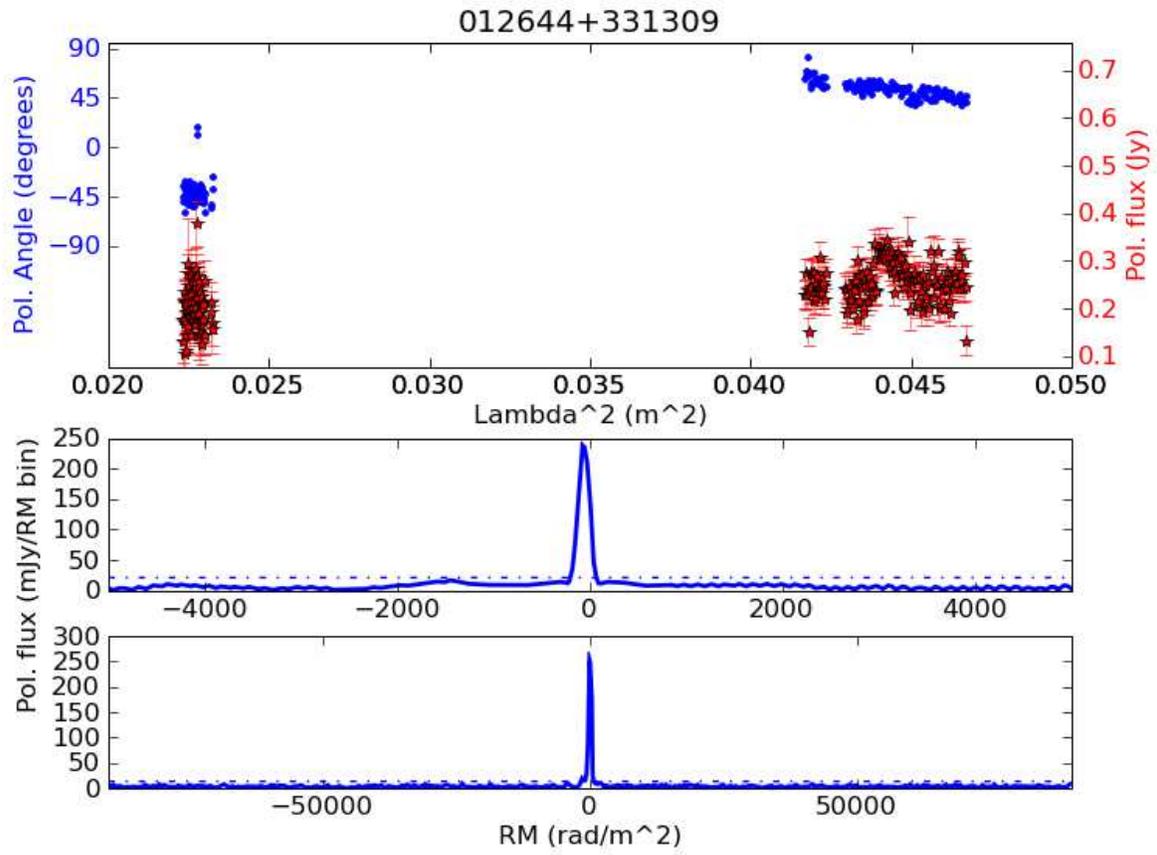}
\caption{RM synthesis summary plots for 3C 41 (J012644+331309), as in Figure \ref{rmplot1}.}
\end{figure}

\begin{figure}[tbp]
\includegraphics[width=\textwidth]{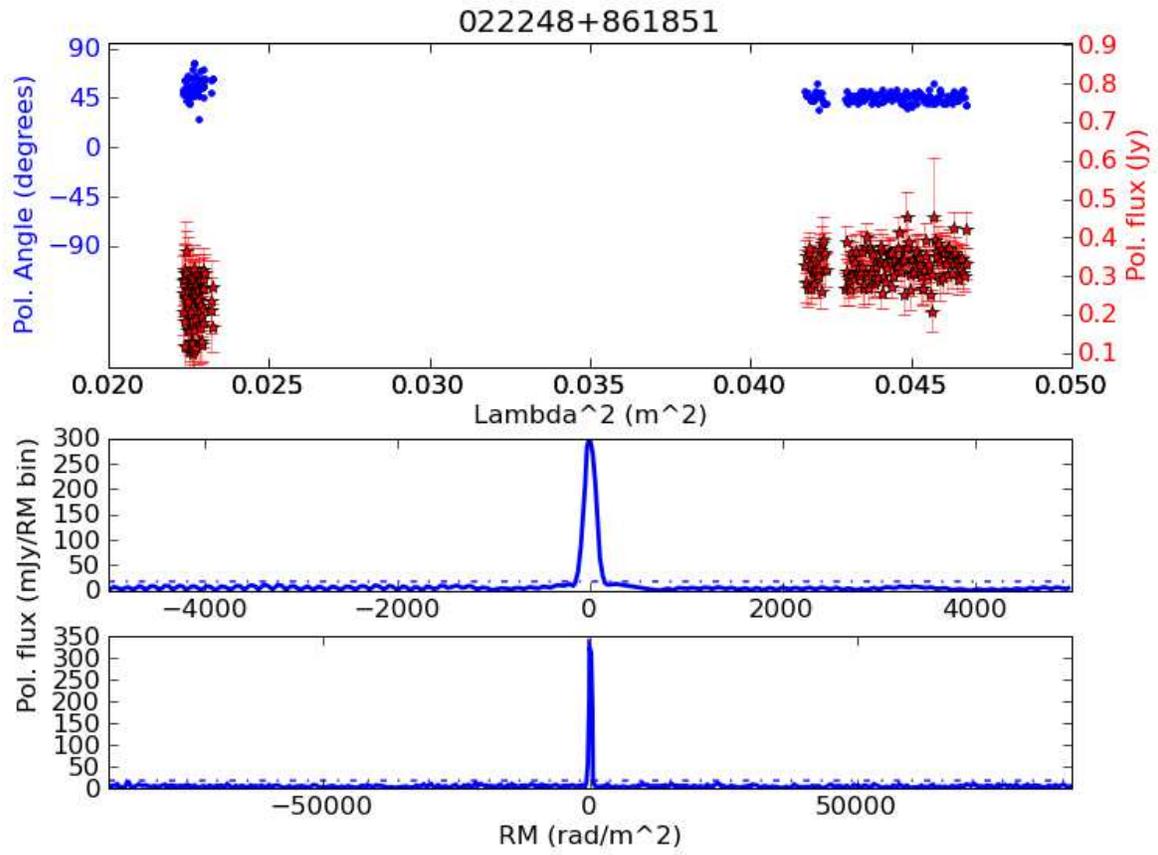}
\caption{RM synthesis summary plots for 3C 61.1 (J022248+861851), as in Figure \ref{rmplot1}.}
\end{figure}

\begin{figure}[tbp]
\includegraphics[width=\textwidth]{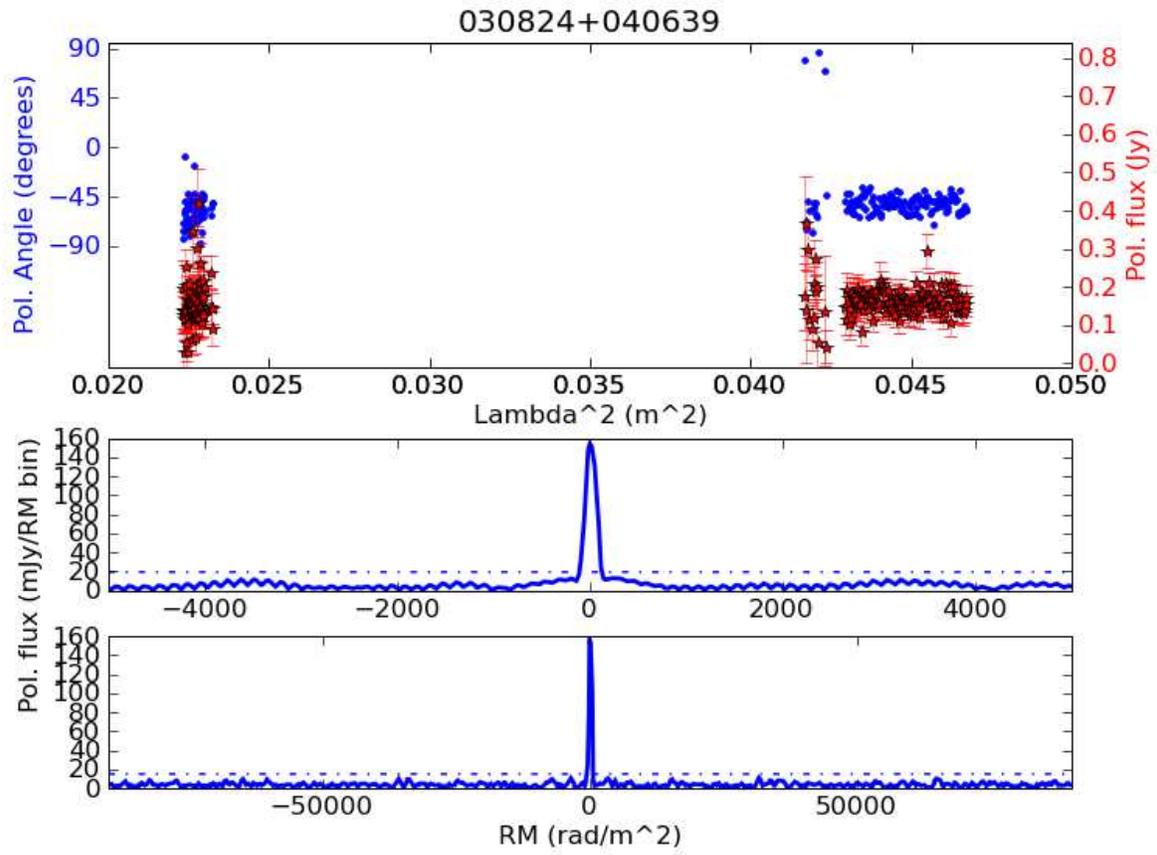}
\caption{RM synthesis summary plots for 3C 78 (J030824+040639), as in Figure \ref{rmplot1}.}
\end{figure}

\clearpage

\begin{figure}[tbp]
\includegraphics[width=\textwidth]{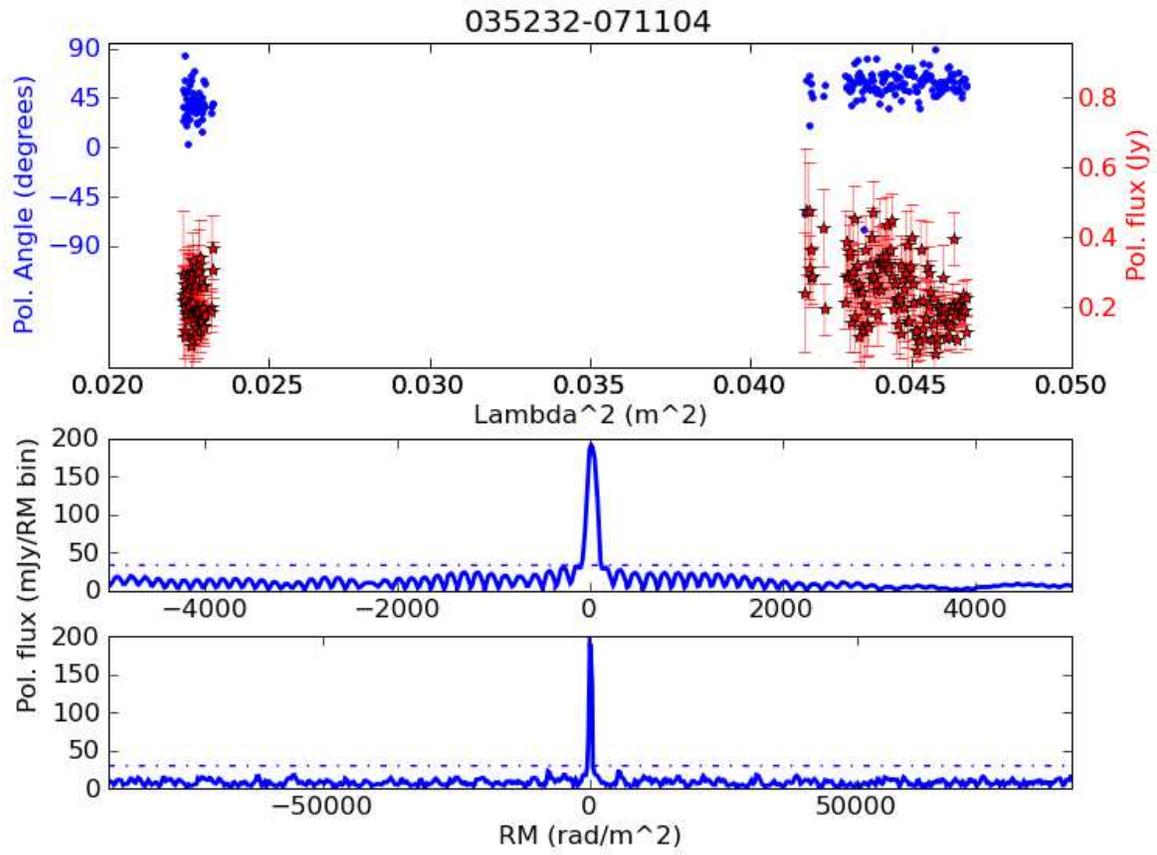}
\caption{RM synthesis summary plots for 3C 94 (J035232--071104), as in Figure \ref{rmplot1}.}
\end{figure}

\begin{figure}[tbp]
\includegraphics[width=\textwidth]{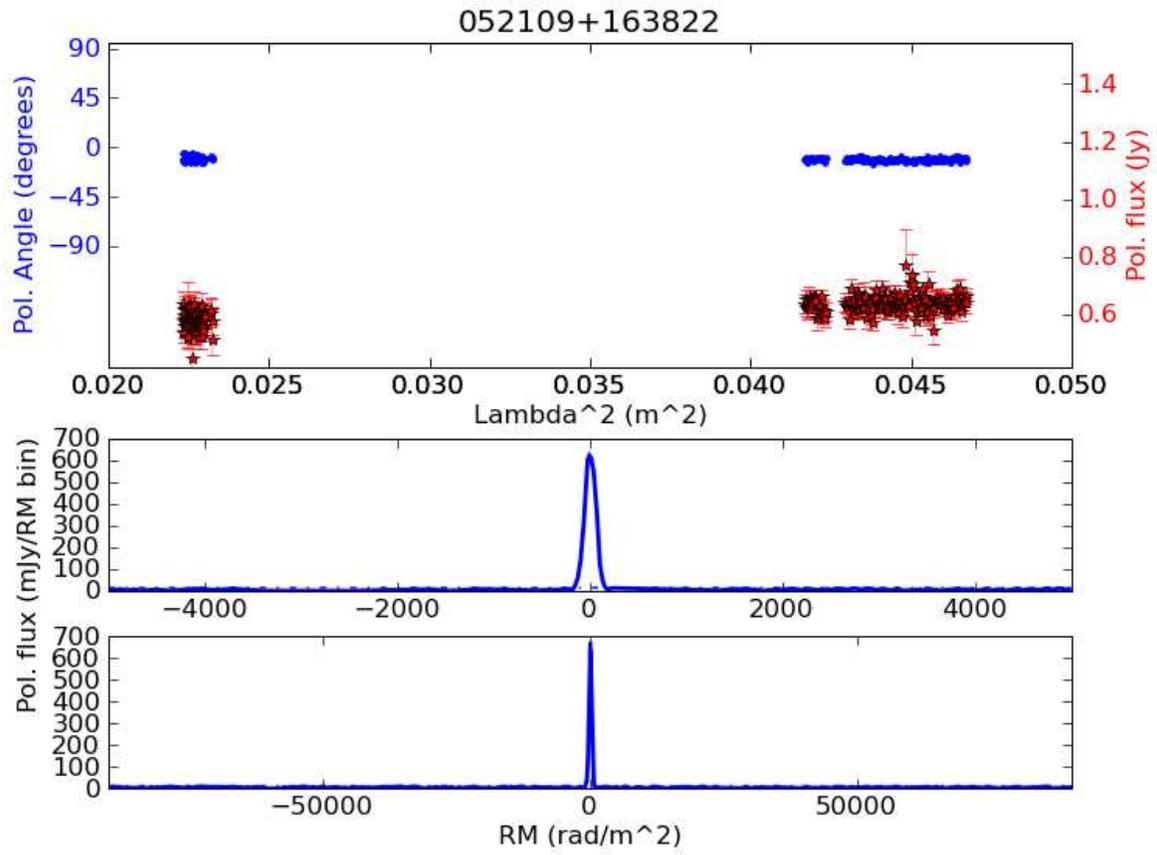}
\caption{RM synthesis summary plots for 3C 138 (J052109+163822), as in Figure \ref{rmplot1}.}
\end{figure}

\begin{figure}[tbp]
\includegraphics[width=\textwidth]{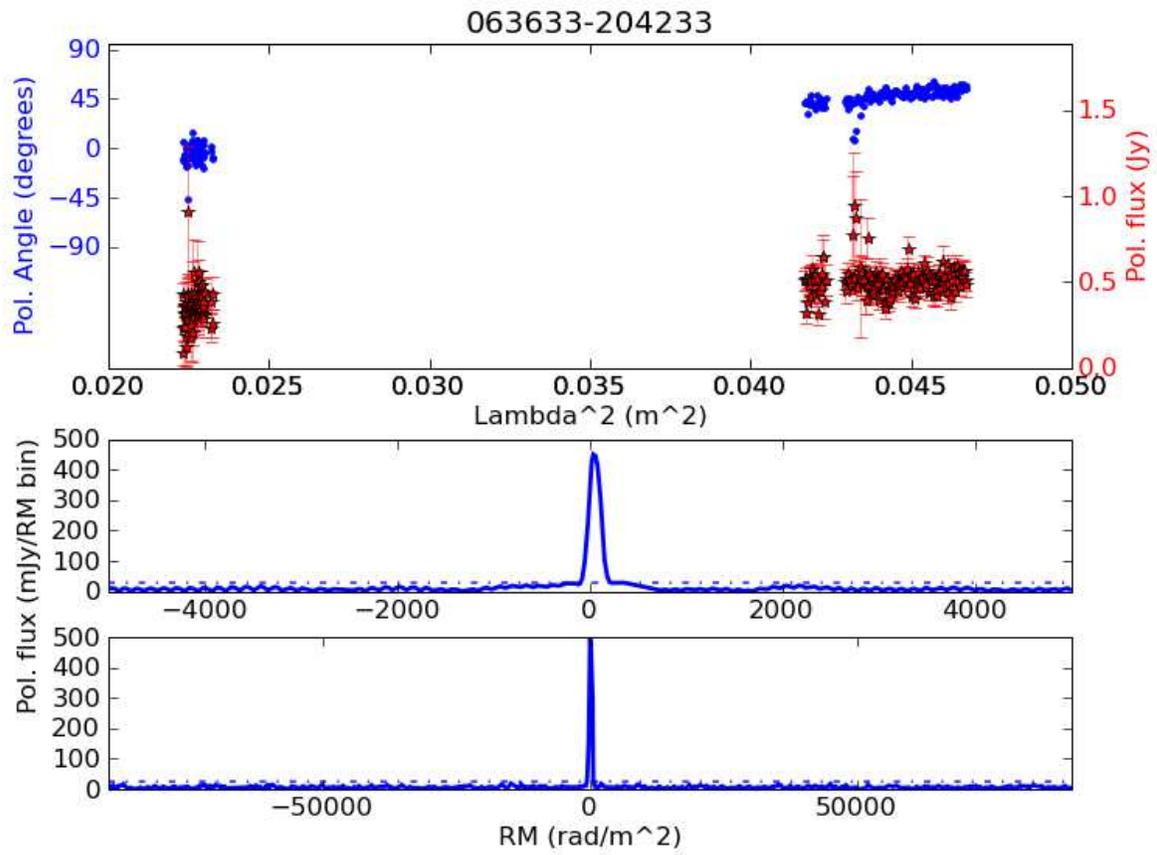}
\caption{RM synthesis summary plots for J063633--204233, as in Figure \ref{rmplot1}.}
\end{figure}

\begin{figure}[tbp]
\includegraphics[width=\textwidth]{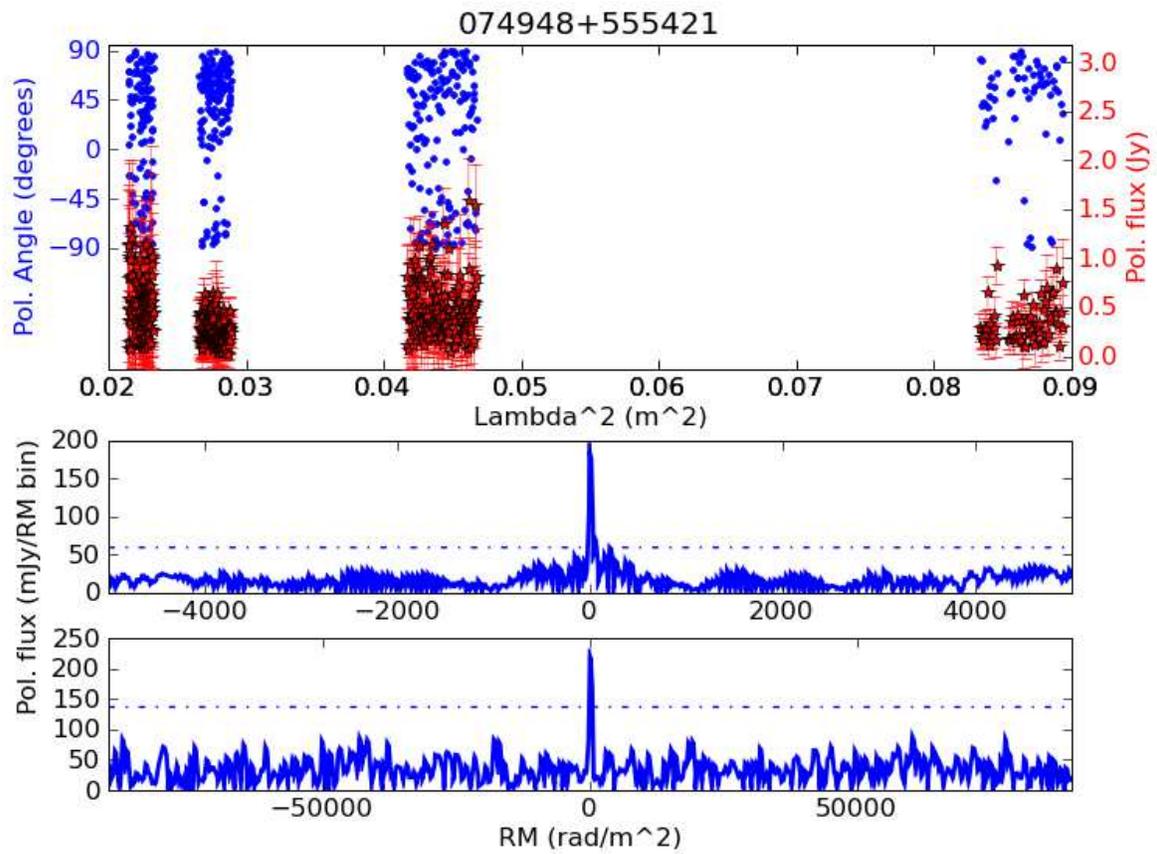}
\caption{RM synthesis summary plots for 4C 56.16 (J074948+555421), as in Figure \ref{rmplot1}.}
\end{figure}

\begin{figure}[tbp]
\includegraphics[width=\textwidth]{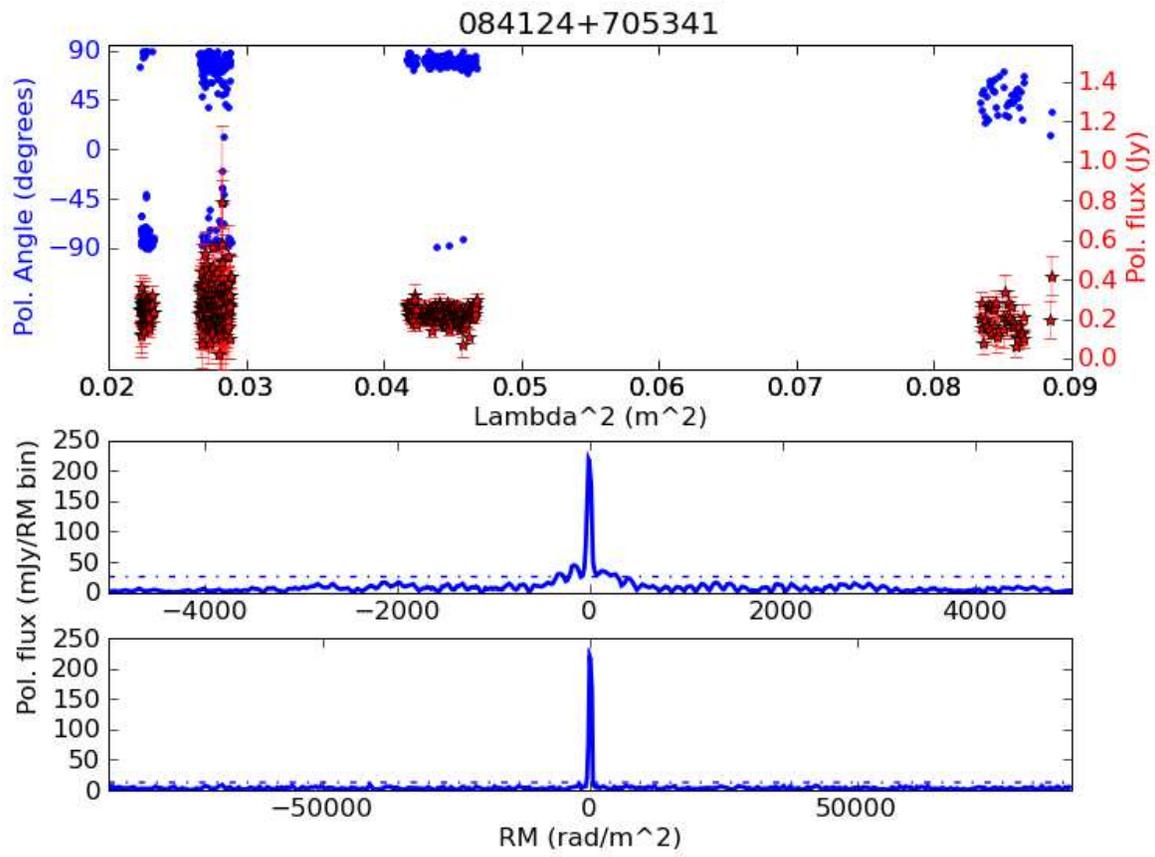}
\caption{RM synthesis summary plots for 4C 71.07 (J084124+705341), as in Figure \ref{rmplot1}.}
\end{figure}

\begin{figure}[tbp]
\includegraphics[width=\textwidth]{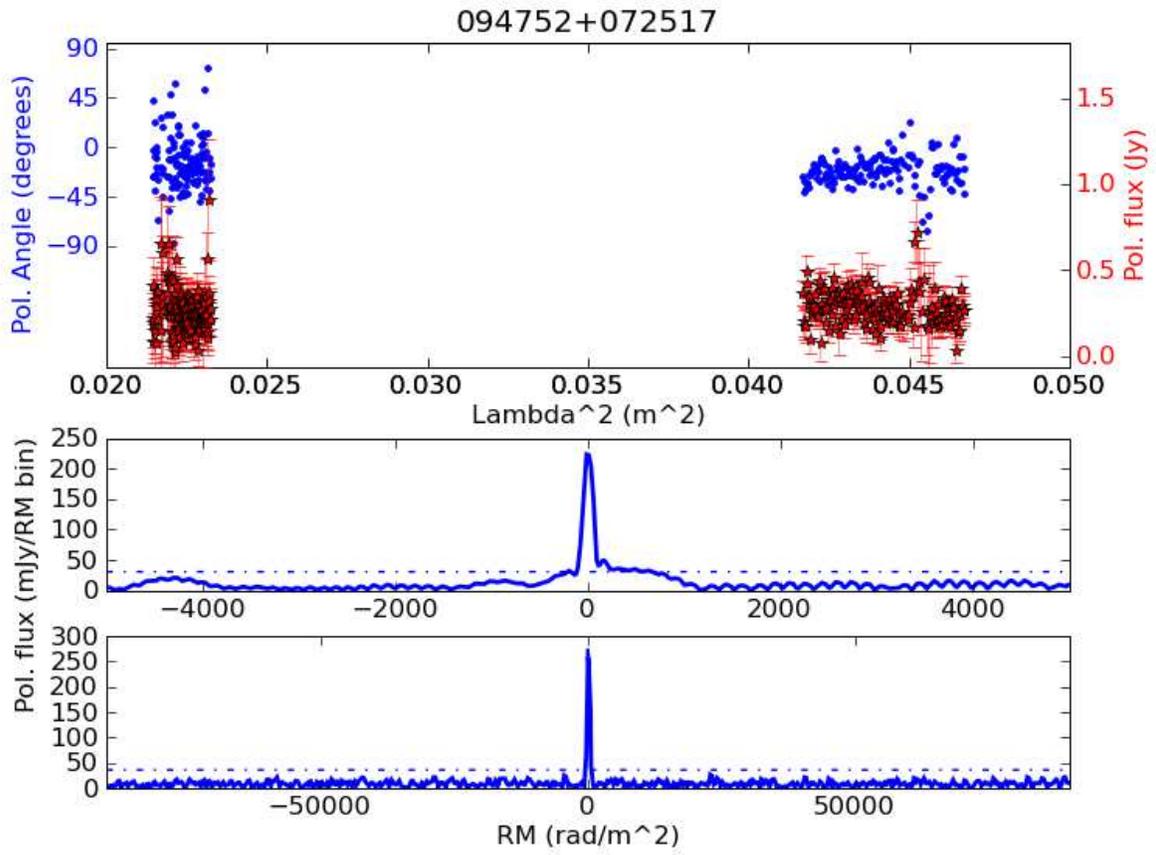}
\caption{RM synthesis summary plots for 3C 227 (J094752+072517), as in Figure \ref{rmplot1}.}
\end{figure}

\begin{figure}[tbp]
\includegraphics[width=\textwidth]{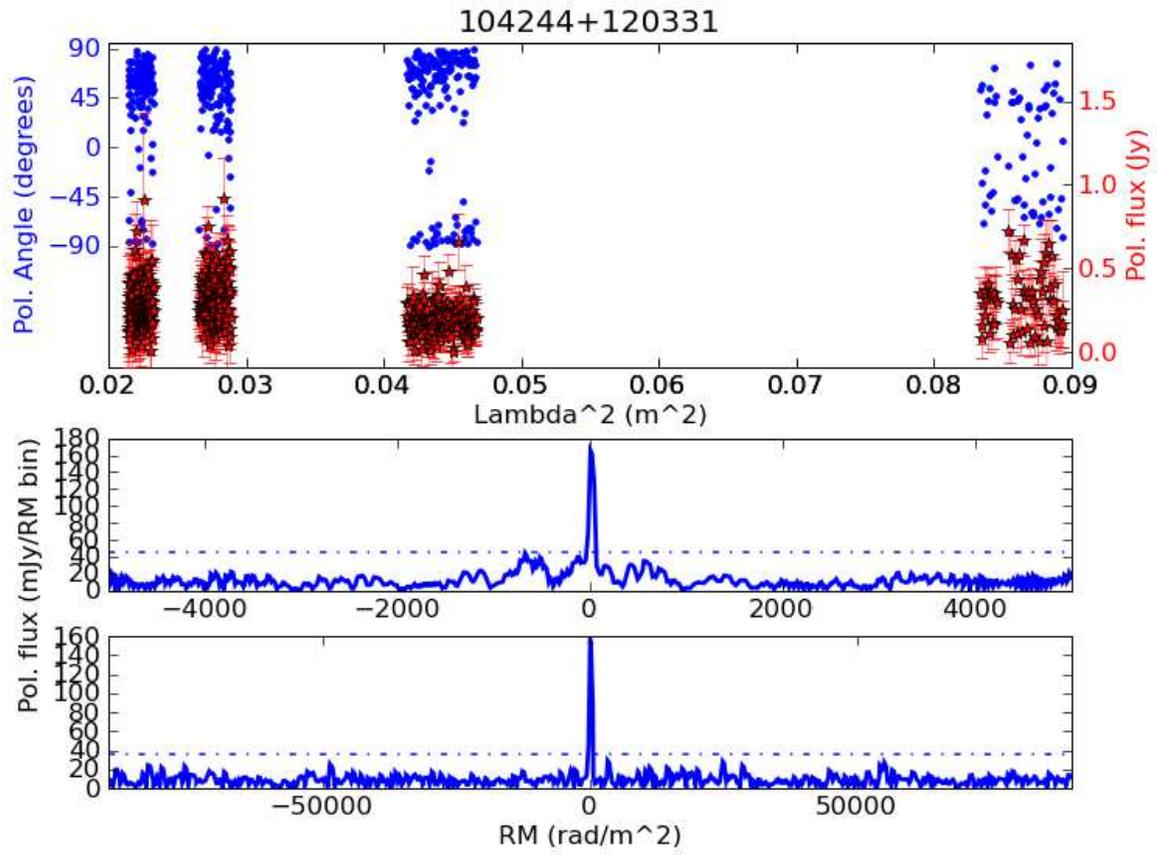}
\caption{RM synthesis summary plots for 3C 245 (J104244+120331), as in Figure \ref{rmplot1}.}
\end{figure}

\begin{figure}[tbp]
\includegraphics[width=\textwidth]{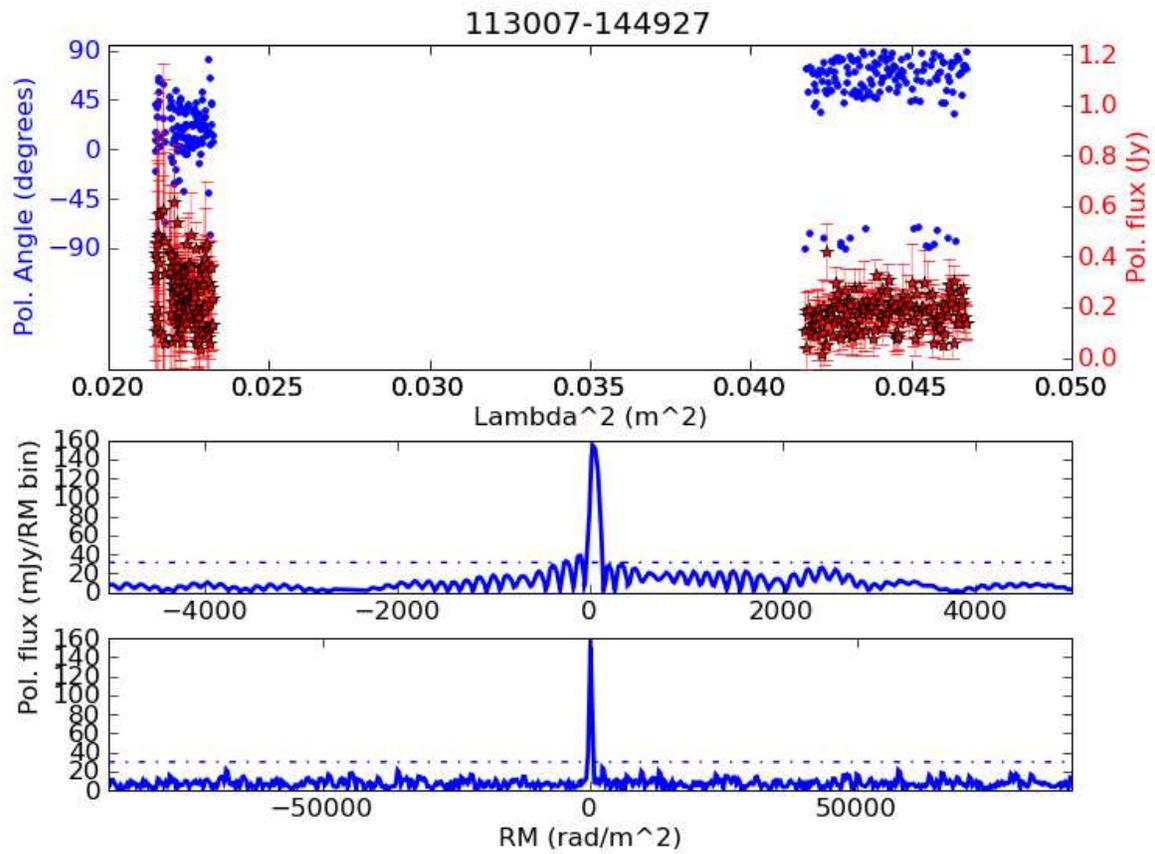}
\caption{RM synthesis summary plots for PKSJ1130-1449 (J113007--144927), as in Figure \ref{rmplot1}.}
\end{figure}

\clearpage

\begin{figure}[tbp]
\includegraphics[width=\textwidth]{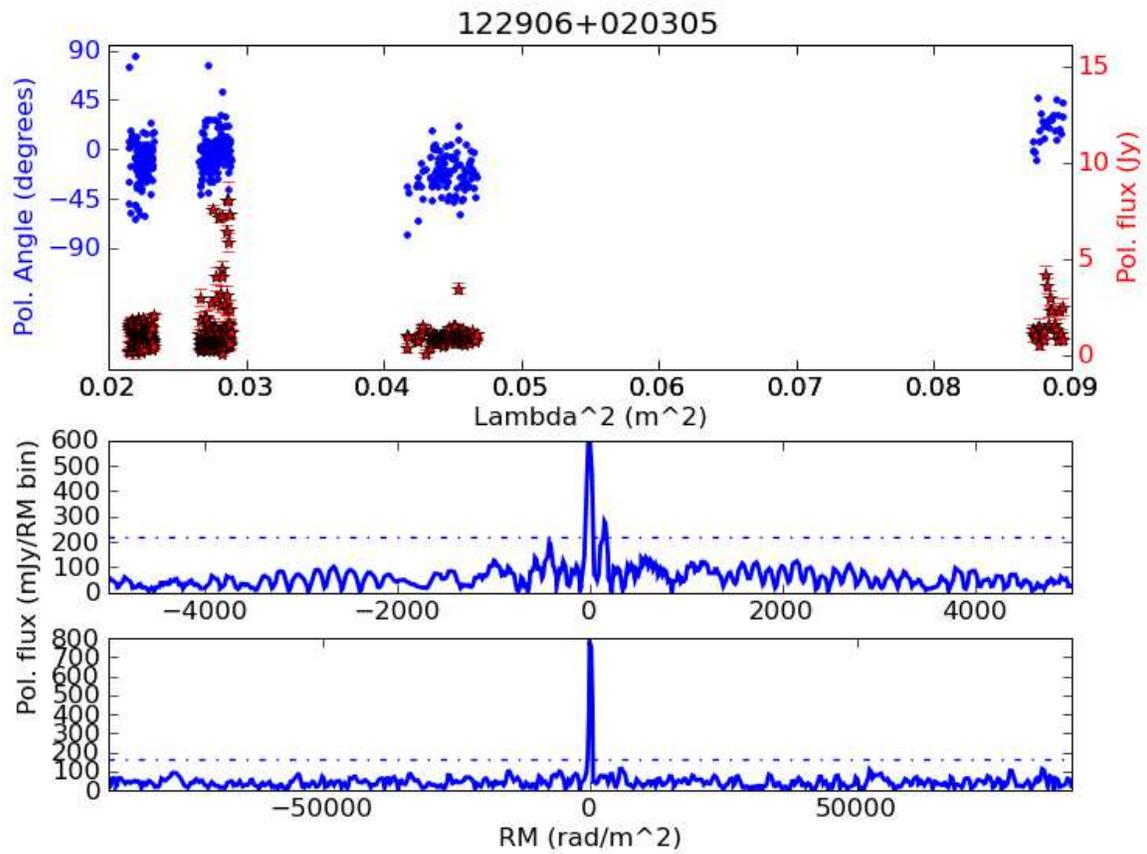}
\caption{RM synthesis summary plots for 3C 273 (J122906+020305), as in Figure \ref{rmplot1}.}
\end{figure}

\begin{figure}[tbp]
\includegraphics[width=\textwidth]{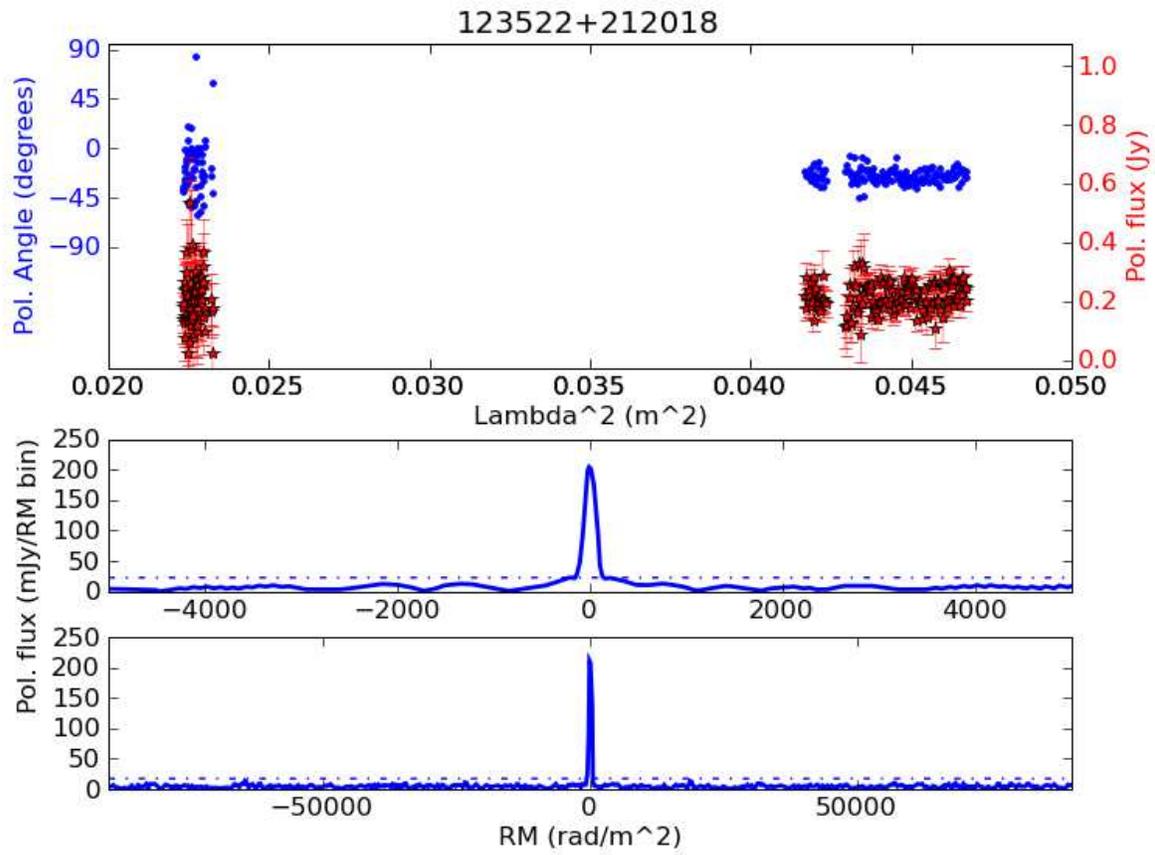}
\caption{RM synthesis summary plots for J123522+212018, as in Figure \ref{rmplot1}.}
\end{figure}

\begin{figure}[tbp]
\includegraphics[width=\textwidth]{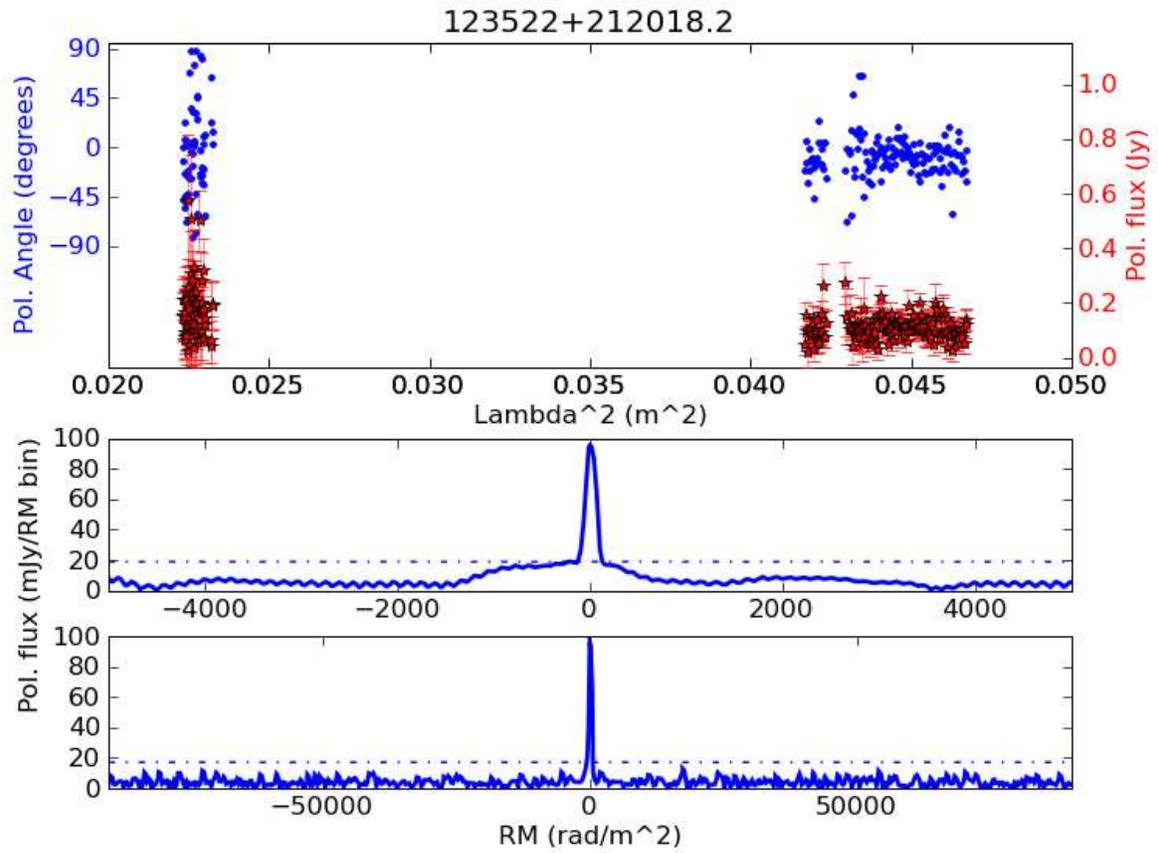}
\caption{RM synthesis summary plots for the secondary source in the field of J123522+212018, as in Figure \ref{rmplot1}.}
\end{figure}

\begin{figure}[tbp]
\includegraphics[width=\textwidth]{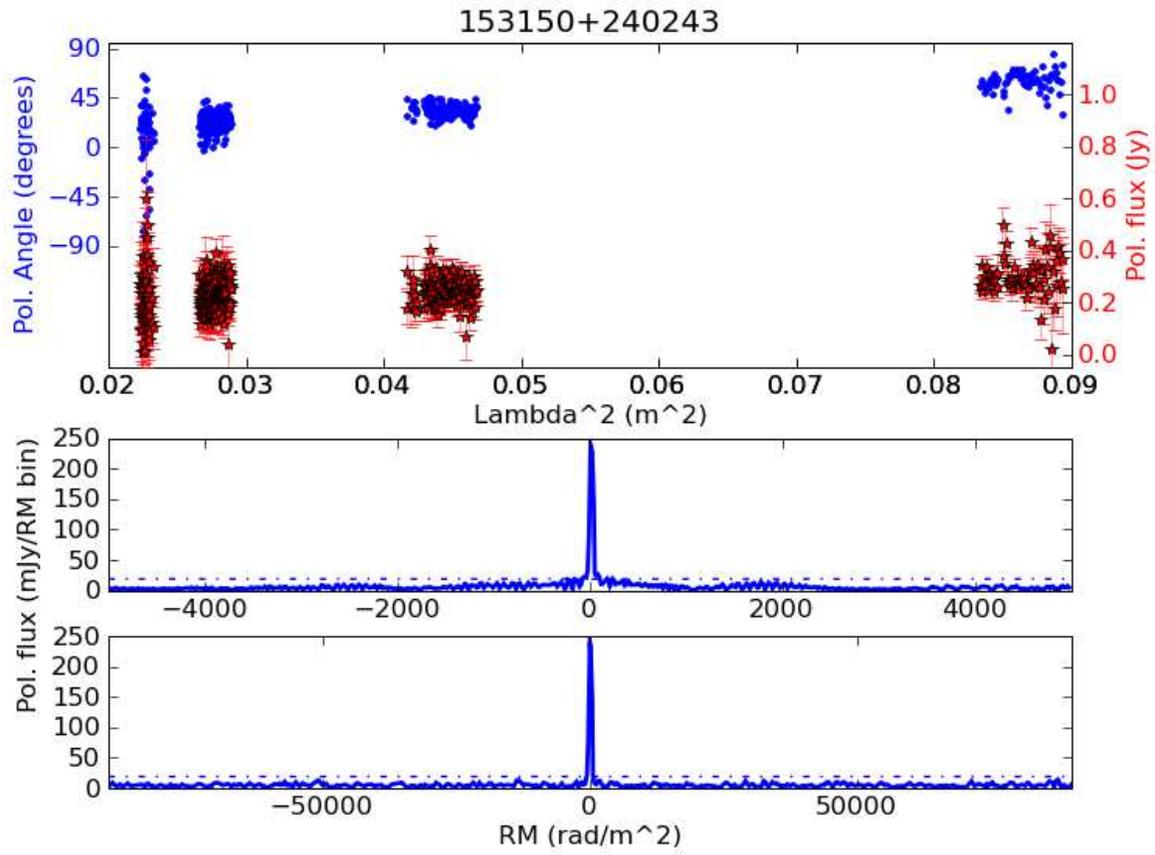}
\caption{RM synthesis summary plots for 3C 321 (J153150+240243), as in Figure \ref{rmplot1}.}
\end{figure}

\clearpage

\begin{figure}[tbp]
\includegraphics[width=\textwidth]{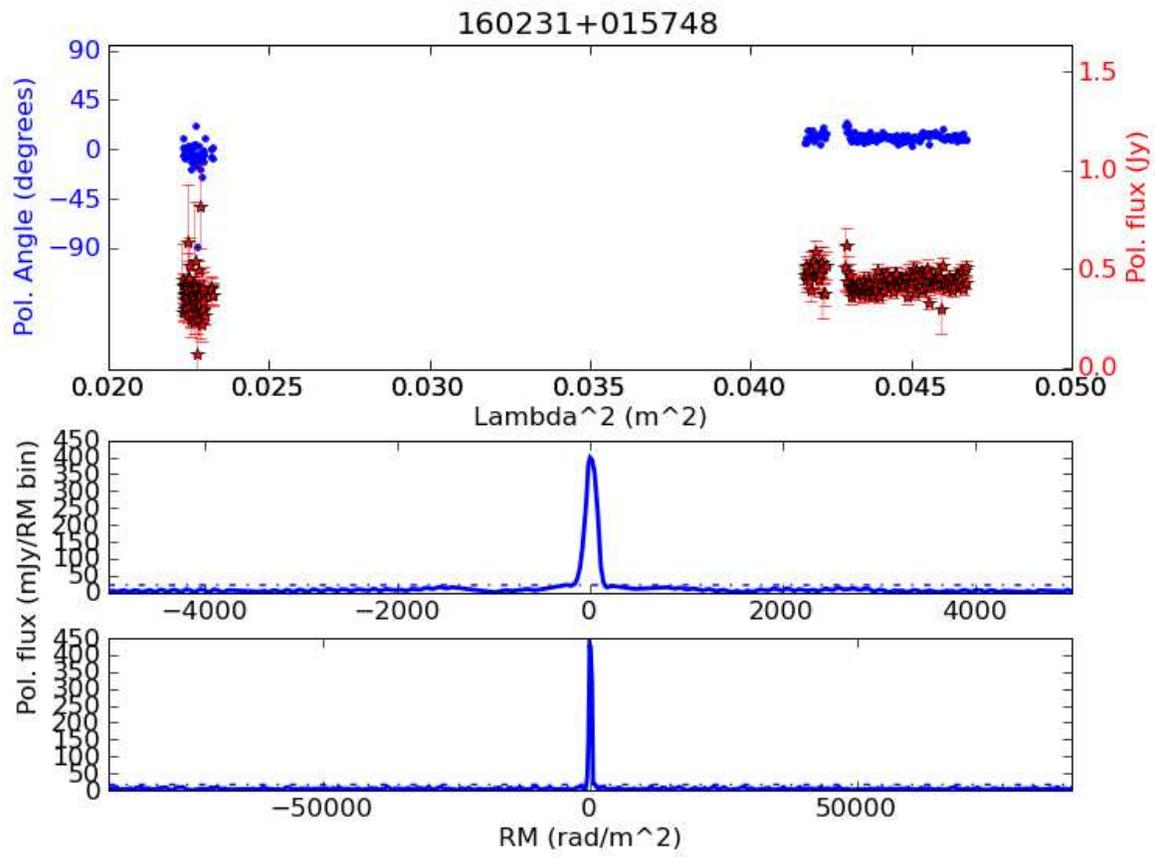}
\caption{RM synthesis summary plots for 3C 327 (J160231+015748), as in Figure \ref{rmplot1}.}
\end{figure}

\begin{figure}[tbp]
\includegraphics[width=\textwidth]{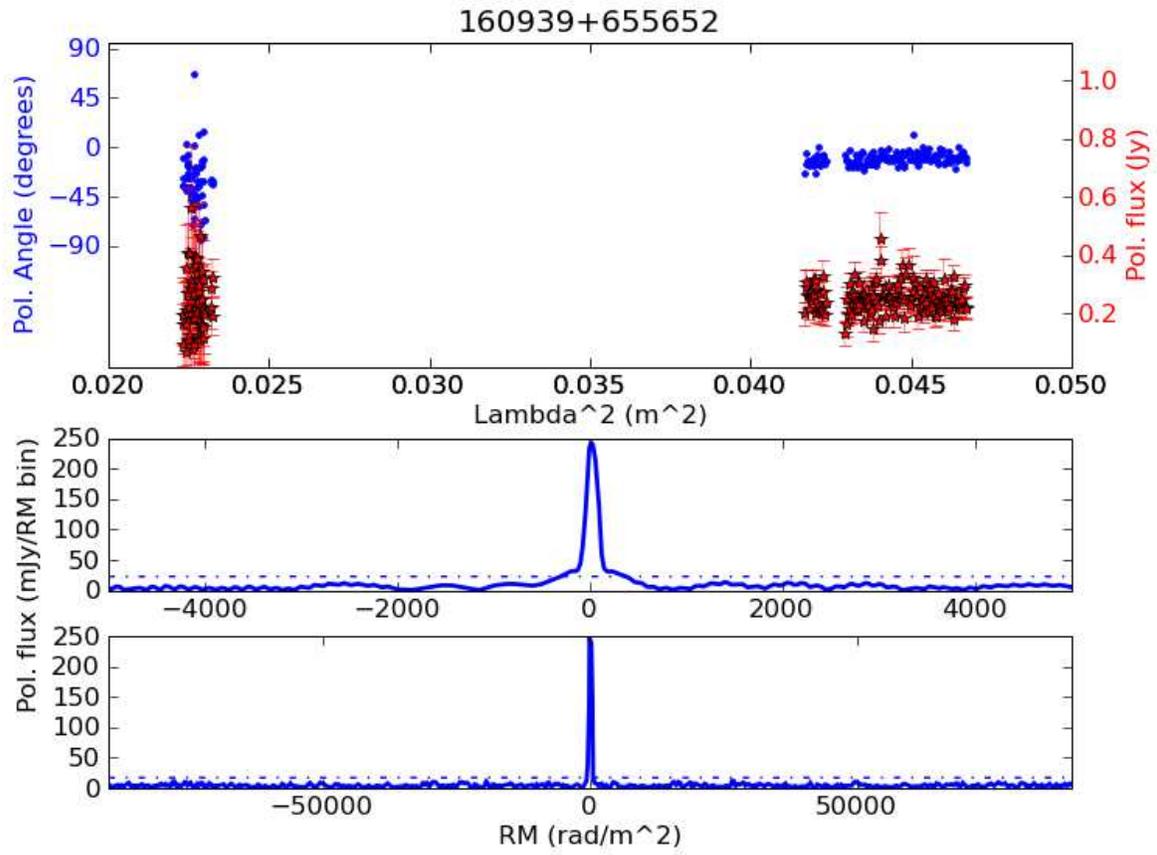}
\caption{RM synthesis summary plots for 3C 330 (J160939+655652), as in Figure \ref{rmplot1}.}
\end{figure}

\begin{figure}[tbp]
\includegraphics[width=\textwidth]{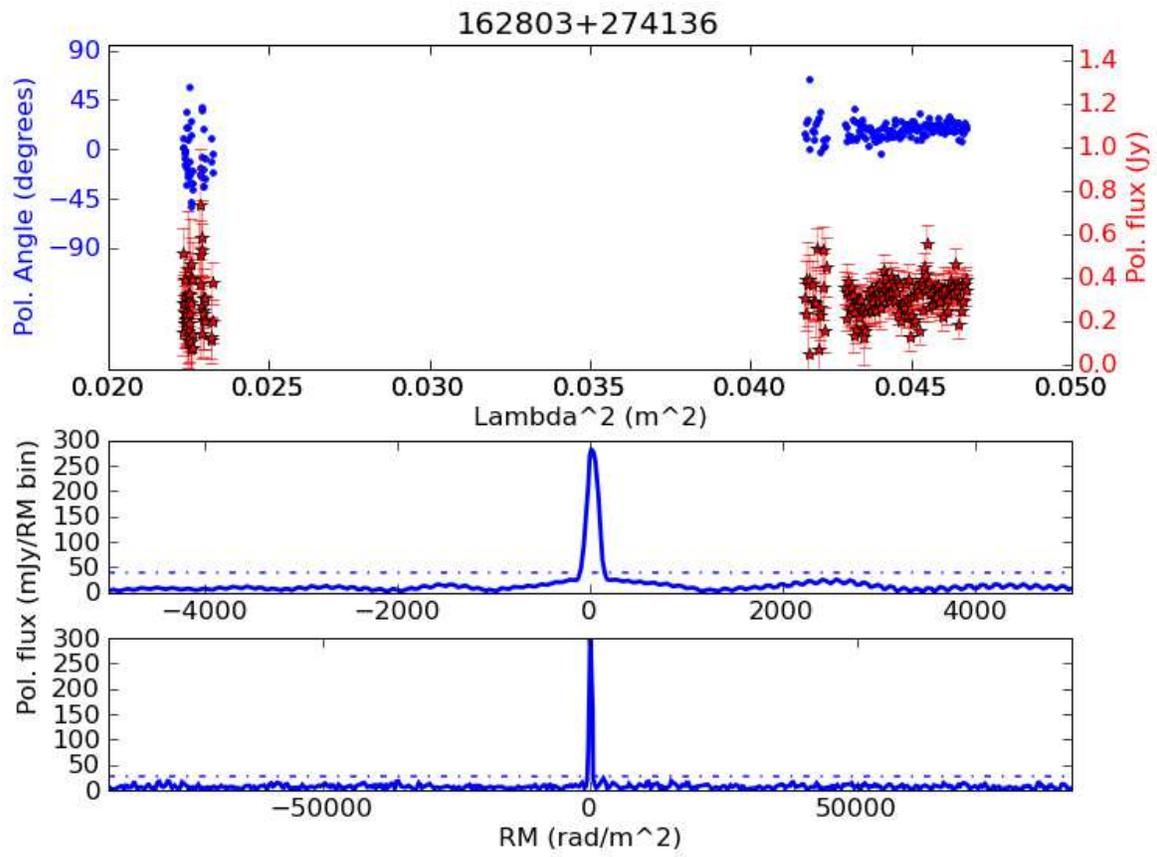}
\caption{RM synthesis summary plots for 3C 341 (J162803+274136), as in Figure \ref{rmplot1}.}
\end{figure}

\begin{figure}[tbp]
\includegraphics[width=\textwidth]{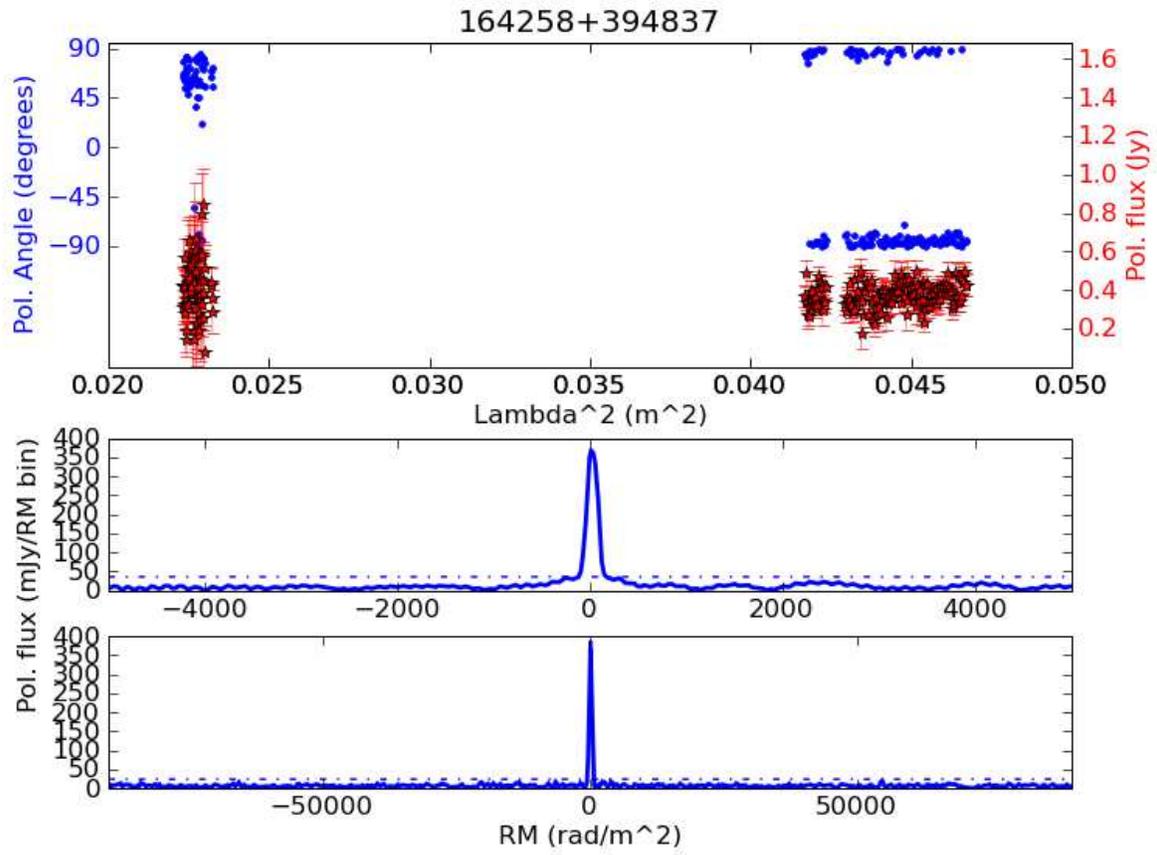}
\caption{RM synthesis summary plots for 3C 345 (J164258+394837), as in Figure \ref{rmplot1}.}
\end{figure}

\begin{figure}[tbp]
\includegraphics[width=\textwidth]{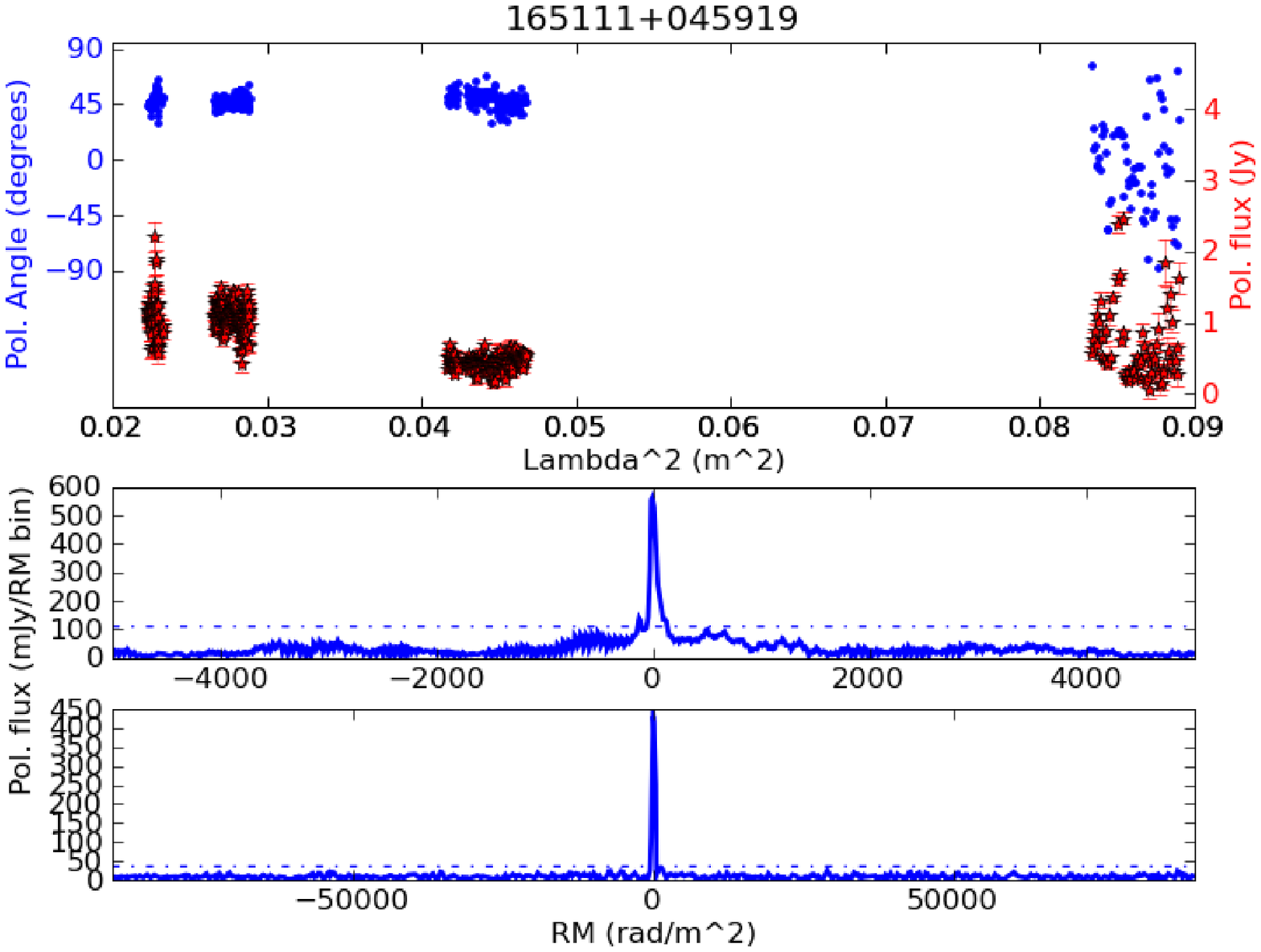}
\caption{RM synthesis summary plots for J165111+045919, as in Figure \ref{rmplot1}.}
\end{figure}

\begin{figure}[tbp]
\includegraphics[width=\textwidth]{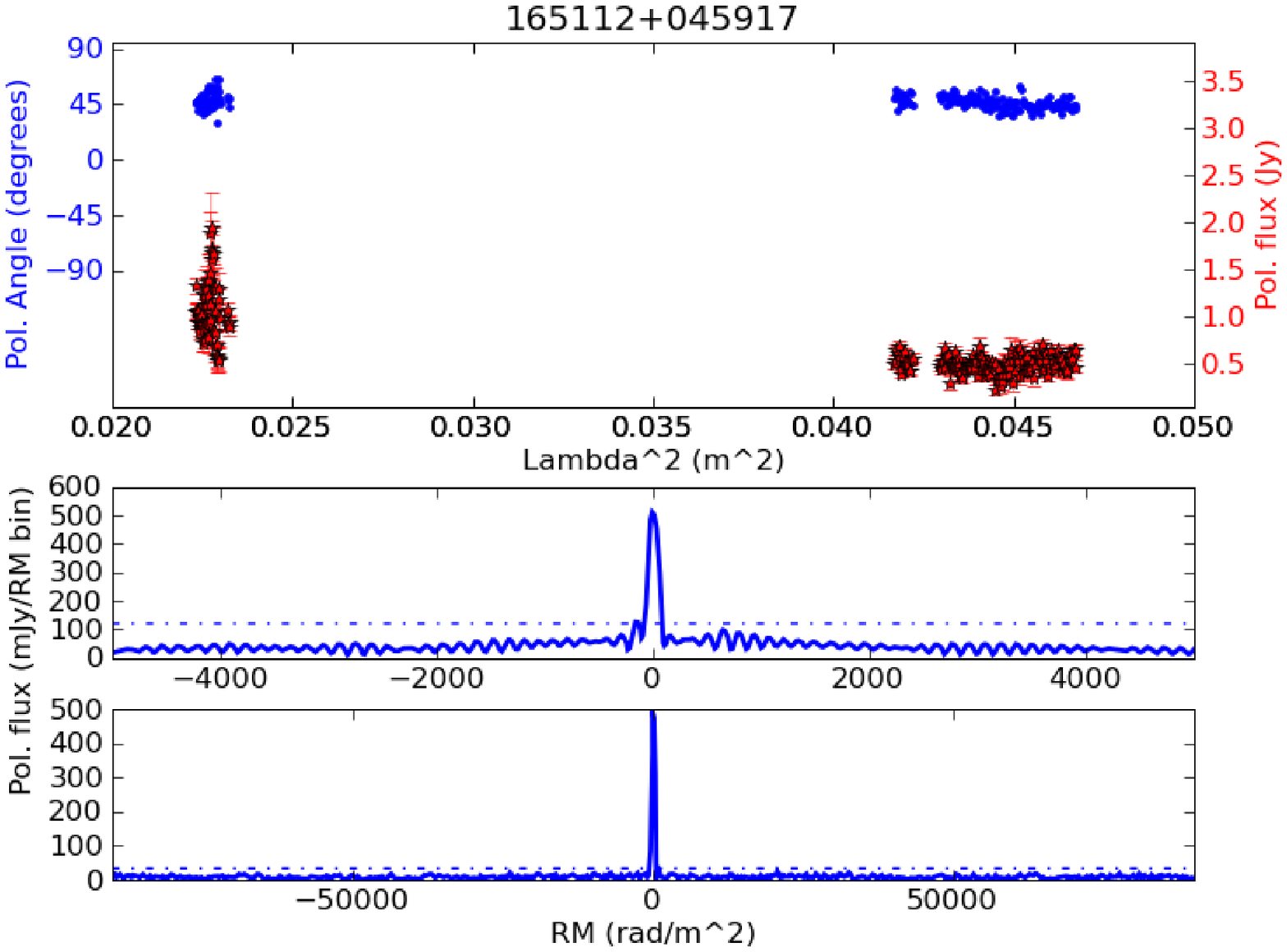}
\caption{RM synthesis summary plots for J165112+045917, as in Figure \ref{rmplot1}.}
\end{figure}

\clearpage

\clearpage

\begin{figure}[tbp]
\includegraphics[width=\textwidth]{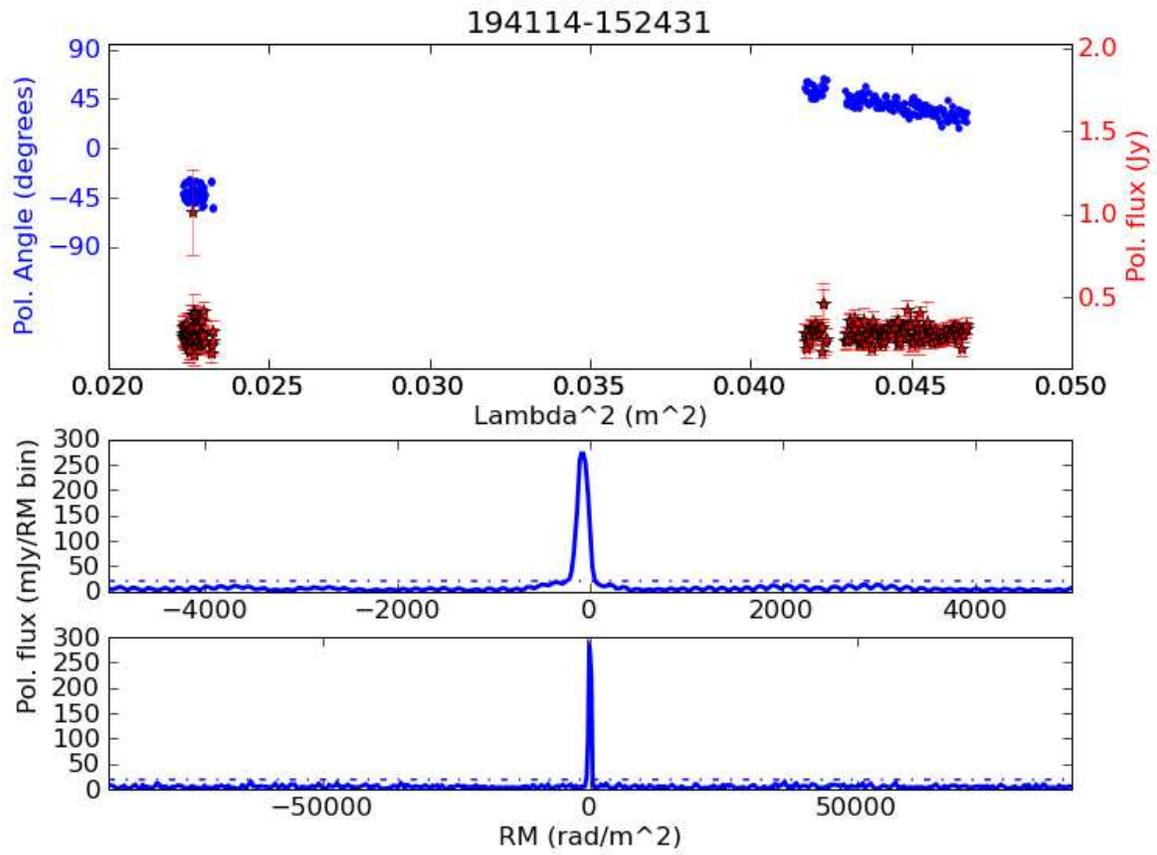}
\caption{RM synthesis summary plots for J194114-152431, as in Figure \ref{rmplot1}.}
\end{figure}

\begin{figure}[tbp]
\includegraphics[width=\textwidth]{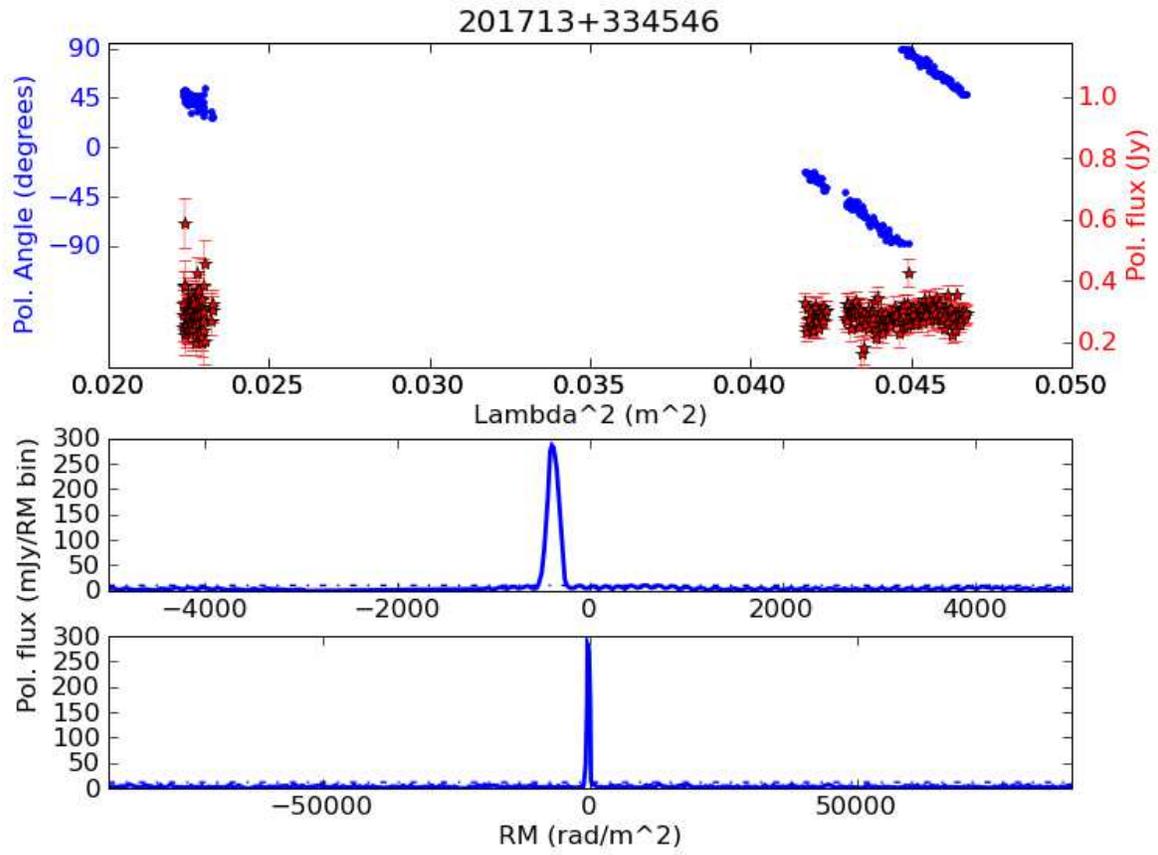}
\caption{RM synthesis summary plots for 4C 33.50 (J201713+334546), as in Figure \ref{rmplot1}.}
\end{figure}

\begin{figure}[tbp]
\includegraphics[width=\textwidth]{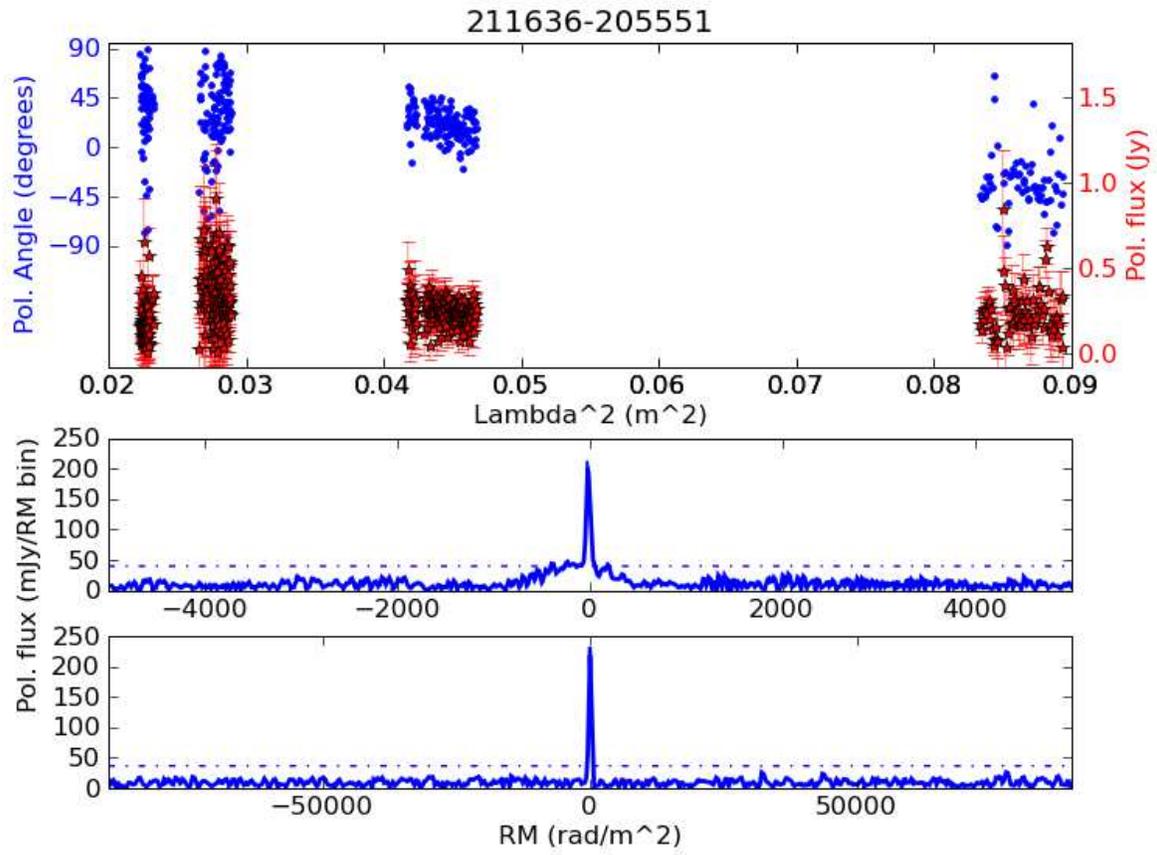}
\caption{RM synthesis summary plots for PKS J2116-2055 (J211636-205551), as in Figure \ref{rmplot1}.}
\end{figure}

\begin{figure}[tbp]
\includegraphics[width=\textwidth]{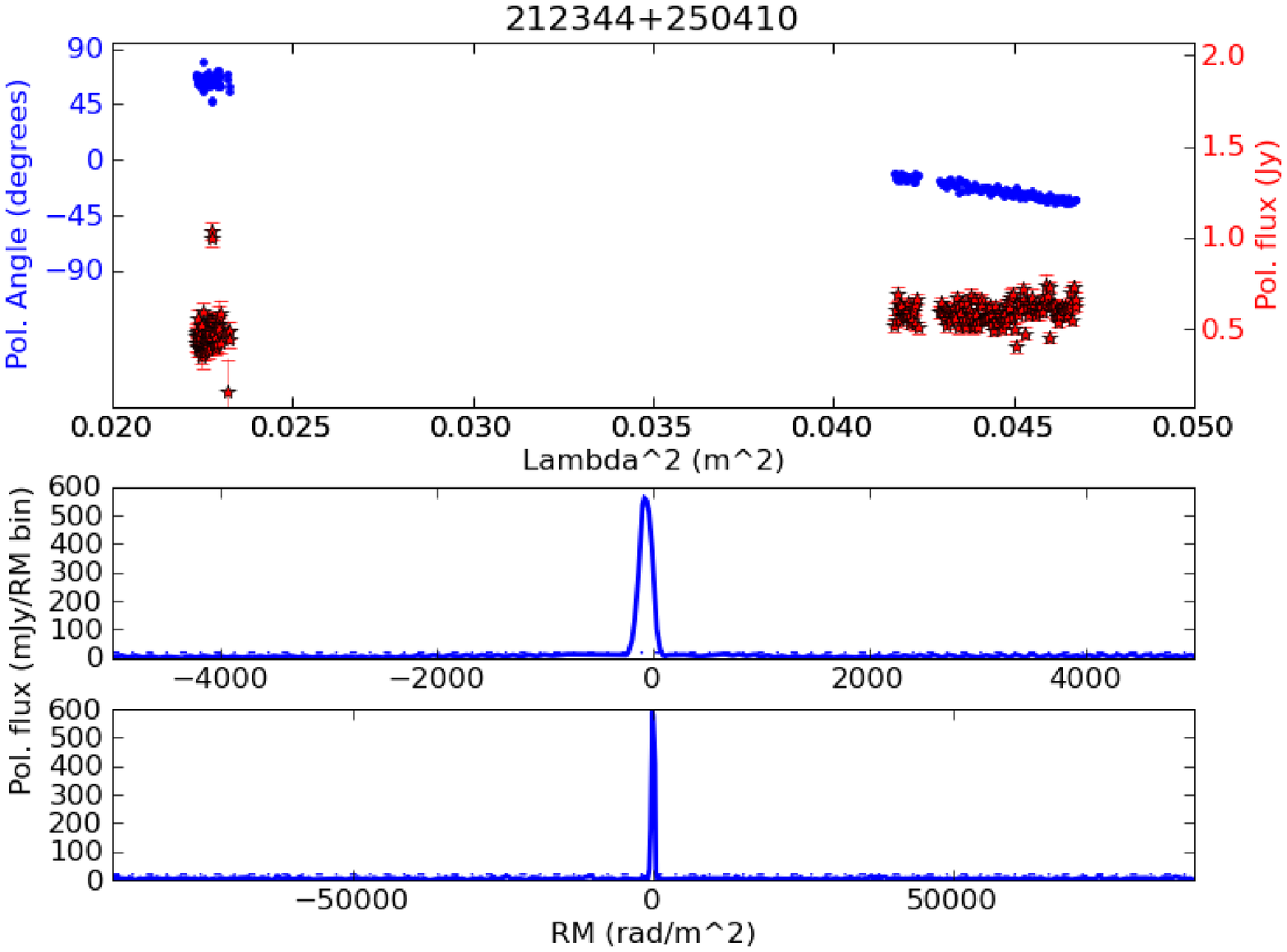}
\caption{RM synthesis summary plots for 3C 433 (J212344+250410), as in Figure \ref{rmplot1}.}
\end{figure}

\clearpage

\begin{figure}[tbp]
\includegraphics[width=\textwidth]{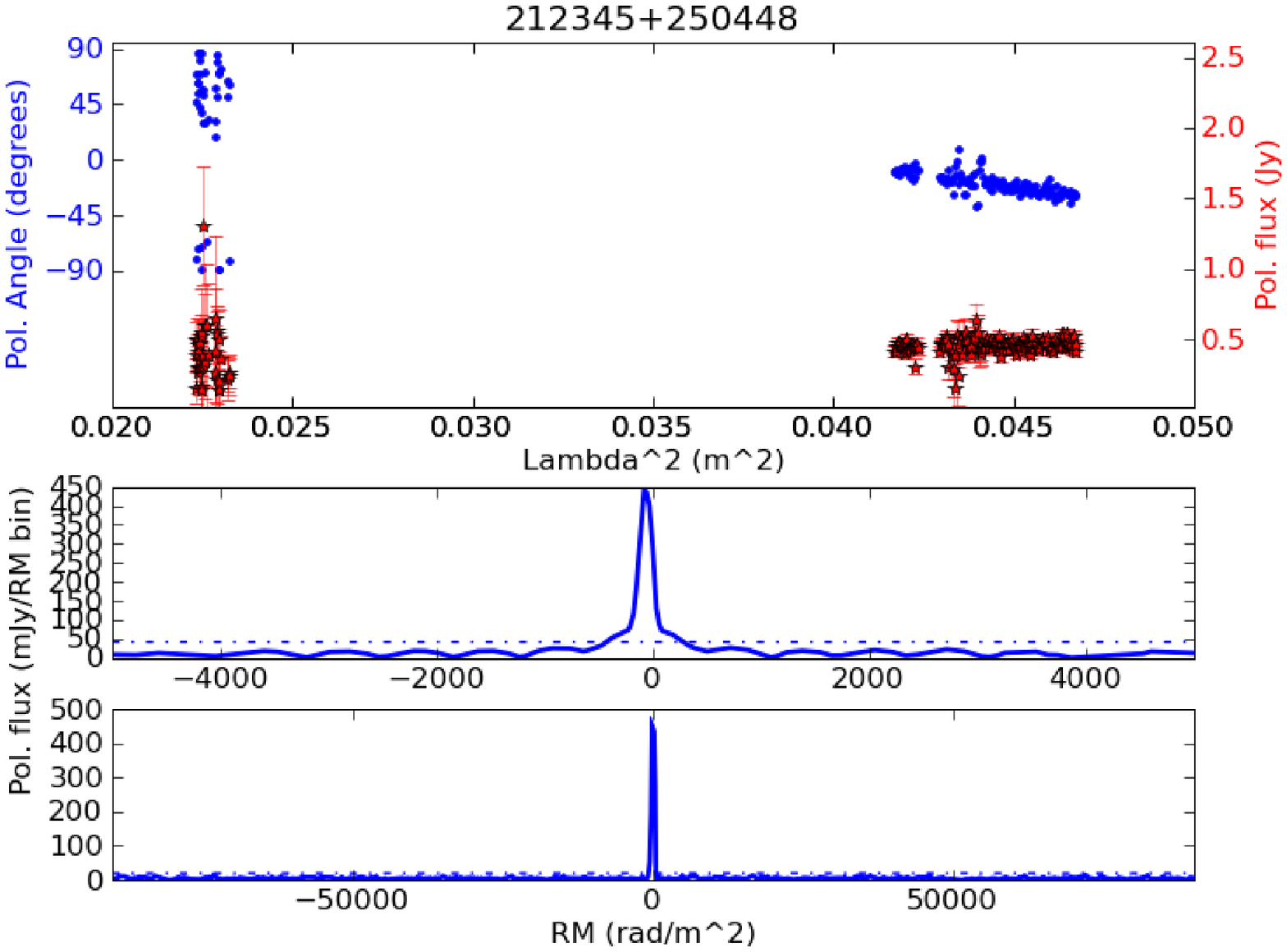}
\caption{RM synthesis summary plots for 3C 433 (J212345+250448), as in Figure \ref{rmplot1}.}
\end{figure}

\begin{figure}[tbp]
\includegraphics[width=\textwidth]{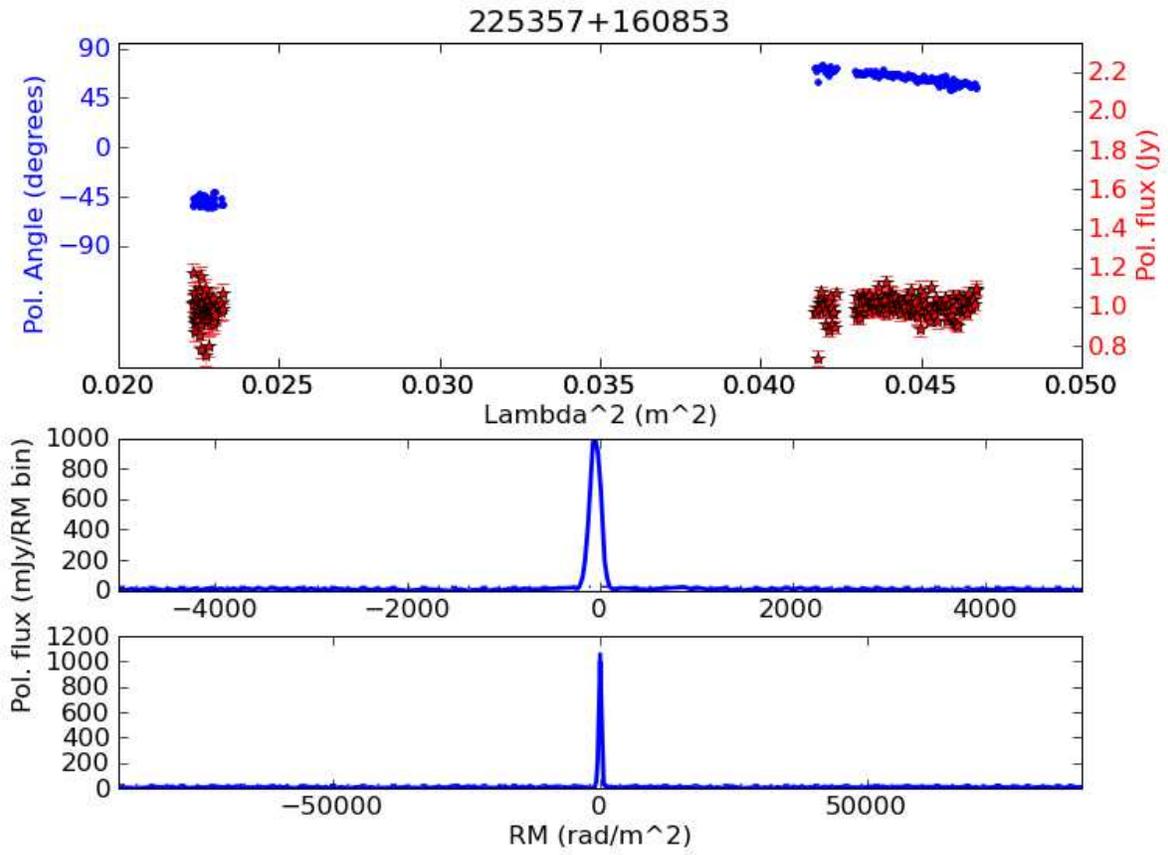}
\caption{RM synthesis summary plots for 3C 454.3 (J225357+160853), as in Figure \ref{rmplot1}.}
\end{figure}

\clearpage

One goal of the observations was to find evidence for multiple RM components in these sources.  To maximize our sensitivity to a wide range of RM components, we cleaned RM spectra in two ways.  First, a high-resolution RM spectrum was made to measure components at small RM very precisely.  Second, a high-sensitivity, low-resolution RM spectrum was created to search for large RM values.  While there are more data and resolution when using the entire bandwidth range, spectral index and sampling effects make the RMSF less smooth;  this is analogous to aperture synthesis with a sparse versus a dense array.  

For these reasons, we searched for components in RM spectra as shown in Figure \ref{rmplot1} through \ref{rmplotlast}.  First, the full frequency range (either 1.43--2.01 GHz or 1.0--2.01 GHz) was used to generate RM spectra for $-5000< RM <5000$\ rad m$^{-2}$ with a resolution of roughly 51 or 141 rad m$^{-2}$  \citep[RM beam $\rm{FWHM}=2 \sqrt{3} (\lambda_{\rm{max}}^2 - \lambda_{\rm{min}}^2)$; ][]{b05}.  Second, the 1.43 GHz data, which has the best quality and completeness, were used to generate RM spectra for $-90000< RM <90000$\ rad m$^{-2}$ with a resolution of roughly 688 rad m$^{-2}$.  Note that the largest Faraday thickness detectable is 150 rad m$^{-2}$ for the high resolution spectra and 75 rad m$^{-2}$ for the low resolution spectra.  Thus, most of the RM spectra should observe unresolved RM components;  the 1.0--2.01 GHz, high-resolution spectra are sensitive to mildly-resolved ($\sim2$\ RM beam widths) Faraday structures.

\subsection{RM Components}
\label{rmcompsec}
Table \ref{rmlimits} gives the $5\sigma$\ threshold and RM resolution of each of the 42 RM spectra over the two RM ranges.  As described in \S \ref{rmsynth}, the noise in Stokes (Q, U) spectra was measured in cleaned RM spectra for RM larger than 10 times the RM resolution.  The flux limit has units of mJy per synthesized beam per RM beam.  The typical limits on the polarized brightness are roughly 40 mJy per 51 rad m$^{-2}$\ beam and 25 mJy per 141 rad m$^{-2}$\ beam for $-5000 < RM < 5000$\ rad m$^{-2}$.  For $-90000 < RM < 90000$\ rad m$^{-2}$, the typical limit is 20 mJy per 688 rad m$^{-2}$\ beam.  The last column shows the fraction polarization limit at 1.43 GHz over the $-90000 < RM < 90000$ rad m$^{-2}$ range.

\begin{deluxetable}{lcccccc}
\tablecaption{Polarized Brightness Thresholds and Resolution of RM Spectra \label{rmlimits}}
\setlength{\tabcolsep}{0.002in}
\tablewidth{0pt}
\tablehead{
\colhead{Source} & \colhead{$P_{\pm5000}$} & \colhead{RM$^{FWHM}_{\pm5000}$} & \colhead{$P_{\pm90000}$} & \colhead{RM$^{FWHM}_{\pm90000}$} & \colhead{$f_{\pm90000}$} \\
  & \colhead{(mJy/beam/RM beam)} & \colhead{(rad m$^{-2}$)} & \colhead{(mJy/beam/RM beam)} & \colhead{(rad m$^{-2}$)} & \colhead{(\%)} \\
}
\startdata
J005558+682218	 & 18 & 141 & 17 & 688 &   0.3 \\ 
J005734--012258	 & 22 & 141 & 21 & 688 &   0.5 \\ 
J010850+131831	 & 24 & 141 & 20 & 688 &   0.2 \\ 
J010855+132214\tablenotemark{a}	 & 23 & 141 & 13 & 688 &   0.4 \\ 
J012644+331309	 & 21 & 141 & 13 & 688 &   0.3 \\ 
J022248+861851	 & 17 & 141 & 17 & 688 &   0.4 \\ 
J030824+040639	 & 19 & 141 & 14 & 688 &   0.2 \\ 
J035232--071104	 & 33 & 141 & 29 & 690 &   1.3 \\ 
J052109+163822	 & 12 & 141 & 11 & 688 &   0.1 \\ 
J063633--204233	 & 28 & 141 & 25 & 688 &   0.8 \\ 
J074948+555421	 & 59 & 51 & 136 & 688 &  10.5 \\ 
J084124+705341	 & 25 & 52 & 12 & 688 &   0.4 \\ 
J094752+072517	 & 30 & 136 & 36 & 688 &   0.5 \\ 
J104244+120331	 & 45 & 51 & 35 & 688 &   1.2 \\ 
J113007--144927	 & 31 & 136 & 29 & 688 &   0.6 \\ 
J122906+020305	 & 215 & 51 & 159 & 688 &   0.3 \\ 
J123039+121758	 & 117 & 51 & 122 & 688 &   2.0 \\ 
J123049+122323\tablenotemark{a}	 & 250 & 51 & 213 & 688 &   0.1 \\ 
J123522+212018	 & 22 & 141 & 17 & 688 &   1.1 \\ 
J123530+212048\tablenotemark{a}	 & 18 & 141 & 17 & 688 &   1.1 \\ 
J125611--054720	 & 87 & 51 & 101 & 1523\tablenotemark{b} &  1.1\tablenotemark{b} \\ 
J133108+303032	 & 25 & 51 & 24 & 688 &   0.2 \\ 
J153150+240243	 & 18 & 51 & 19 & 689 &   0.6 \\ 
J160231+015748	 & 21 & 141 & 16 & 688 &   0.3 \\ 
J160939+655652	 & 22 & 141 & 17 & 688 &   0.3 \\ 
J162803+274136	 & 39 & 141 & 27 & 688&   1.5 \\ 
J164258+394837	 & 35 & 141 & 23 & 688 &   0.3 \\ 
J165111+045919	 & 110 & 51 & 35 & 688 &   0.1 \\ 
J165112+045917	 & 120 & 141 & 33 & 688 &   0.1 \\ 
J172025--005852	 & 57 & 51 & 33 & 688 &   0.2 \\ 
J172034--005843\tablenotemark{a}	 & 55 & 51 & 44 & 688 &   0.1 \\ 
J184226+794517	 & 76 & 51 & 25 & 688 &   0.4 \\ 
J184150+794728\tablenotemark{a}	 & 41 & 51 & 27 & 688 &   0.7 \\ 
J192451--291431	 & 77 & 51 & 34 & 688 &   0.4 \\ 
J194114--152431	 & 20 & 141 & 18 & 688 &   0.3 \\ 
J201713+334546	 & 9 & 141 & 11 & 688 &   0.4 \\ 
J211636--205551	 & 39 & 51 & 36 & 688 &   1.4 \\ 
J212344+250410	 & 17 & 141 & 19 & 688 &   0.2 \\ 
J212345+250448	 & 41 & 141 & 19 & 688 &   0.2 \\ 
J222547--045701	 & 39 & 141 & 37 & 688 &   0.6 \\ 
J225357+160853	 & 24 & 141 & 25 & 688 &   0.2 \\ 
J231956--272713	 & 38 & 141 & 42 & 688 &   2.0 \\ 
\enddata
\tablenotetext{a}{Off-axis source.}
\tablenotetext{b}{No 1.43 GHz data available, so this limit measured at 1.8 GHz.}
\end{deluxetable}

\clearpage

Tables \ref{rmcomponents} and \ref{rmcomponentslow} show the brightness, RM, and signal-to-noise ratio for components detected in the high- and low-resolution RM spectra, respectively.  The component properties are measured by fitting a Gaussian to the cleaned RM spectra after subtracting the observed noise in quadrature.  All RM peaks with a best-fit peak brightness greater than $5\sigma$\ are shown.  While we find several plausible candidates at this threshold level (see \S \ref{dis}), past experience suggests that some $5\sigma$\ candidates are not real.  As such, the RM components in the tables should be considered candidates;  most analysis presented here uses a stricter threshold of 7$\sigma$.  Varying parameters of the cleaning process (e.g., depth of initial clean, RM grid size, maximum RM) showed a roughly 20\% variation in the brightness of components in a few of the most complicated RM spectra (noted in the tables).  Aside from this caveat, all components shown in the tables are significantly detected regardless of the parameters used during cleaning.  We also checked that RM spectra were robust to possible calibration errors in a few (Q, U) points (e.g., points with polarization angle $>45$\sdeg\ in Figure \ref{rmplot6}).  In all cases, the best-fit components were unchanged within their errors.

In addition to being filtered based on a simple noise threshold, RM candidates were also filtered according to a RM dynamic range limit.  The imaging and RM cleaning techniques described in \S \ref{imaging} and \S \ref{rmsynth} control for the effect of thermal noise and dynamic range limitations, but other systematic effects can create false RM components.  Figure \ref{rmsynthplot} shows how gain errors between the four bands create a wobble in the RM spectrum of 3C 286, when treated as a target \citep[absolute gain calibration accuracy $\approx3$\%;][]{w10}.  This wobble requires two $\sim6\sigma$\ RM components with RM $\approx-160$ and 90 rad m$^{-2}$\ and polarized brightness of about 35 mJy.  Furthermore, the spectral index of the source can create false RM components by effectively modifying the RMSF \citep{b05}.  

The gain error and spectral index effects are similar in that they act as unexpected scalings of the polarized flux on frequency scales larger than the 100-MHz bandwidth.  This kind of error will produce RM artifacts with polarized brightnesses that scale with the brightness of the source.  This effectively creates a RM dynamic range limit for all sources with similar spectral indices and calibration errors.  The RM spectrum of 3C 286 (treated as a target) allows us to estimate the RM dynamic range as the ratio of the true RM component to the largest false RM component.  This estimate gives a dynamic range limit of 38 or, equivalently, 2.6\% of the peak polarized brightness.  This limit only removes the extra RM components seen in 3C 286.  RM spectra with evidence for larger or more unusual spectral changes than 3C 286 (e.g., Figure \ref{rmplot3}) should be interpreted cautiously.

\begin{deluxetable}{lccc}
\vspace{-0.5in}
\tablecaption{Components Measured in High-resolution RM Spectra \label{rmcomponents}}
\setlength{\tabcolsep}{0.005in}
\tablewidth{0pt}
\tablehead{
\colhead{Source} & \colhead{P} & \colhead{RM} & \colhead{SNR} \\
  & \colhead{(mJy)} & \colhead{(rad m$^{-2}$)} & \\
}
\startdata
J005558+682218	& 305.6 & --99.0 $\pm$ 0.8 & 84 \\
J005734--012258	& 559.0 & 0.5 $\pm$ 0.5 & 123 \\
J010850+131831	& 789.7 & --12.3 $\pm$ 0.4 & 161 \\
J010855+132214\tablenotemark{a}	& 132.9 & --10.0 $\pm$ 1.9 & 28 \\
J012644+331309	& 238.5 & --68.0 $\pm$ 1.1 & 56 \\
J022248+861851	& 296.2 & --4.6 $\pm$ 0.8 & 83 \\
J030824+040639	& 153.9 & 5.4 $\pm$ 1.6 & 39 \\
J035232--071104	& 191.6 & 15.7 $\pm$ 2.0 & 28 \\
J052109+163822	& 622.4 & --0.7 $\pm$ 0.2 & 249 \\
J063633--204233	& 452.4 & 45.5 $\pm$ 0.8 & 78 \\
J074948+555421	& 186.2 & 1.5 $\pm$ 1.2 & 15 \\
                & 70.0 & 54.4 $\pm$ 3.2 & 5.9 \\
J084124+705341	& 218.2 & --12.7 $\pm$ 0.6 & 42 \\
                & 43.1 & --161.1 $\pm$ 8.9 & 8.3 \\
                & 30.2 & --313.3 $\pm$ 11.4 & 5.8 \\
		& 32.3 & 105.2 $\pm$ 39.2 & 6.2 \tablenotemark{b} \\
J094752+072517\tablenotemark{c} & 225.8 & --2.3 $\pm$ 1.6 & 36 \\
                & 47.8 & 158.4 $\pm$ 12.2 & 7.8 \\
                & 34.2 & 397.1 $\pm$ 81.0 & 5.6  \tablenotemark{b} \\
J104244+120331	& 165.0 & 13.5 $\pm$ 1.7 & 18 \\
J113007--144927	& 155.0 & 42.5 $\pm$ 2.1 & 24 \\
                & 39.5 & --109.8 $\pm$ 5.6 & 6.3 \\
		& 32.8 & --242.3 $\pm$ 7.5 & 5.2 \\
J122906+020305	& 604.7 & --10.7 $\pm$ 2.3 & 14 \\
                & 277.3 & 150.7 $\pm$ 4.1 & 6.4 \\
J123039+121758	& 306.4 & 83.5 $\pm$ 2.3 & 13 \\
		& 190.7 & 37.2 $\pm$ 3.1 & 8.1 \\
		& 137.0 & --118.1 $\pm$ 4.0 & 5.8 \\
		& 128.9 & --173.5 $\pm$ 3.9 & 5.7 \\
		& 209.5 & 241.7 $\pm$ 2.6 & 8.9 \\
		& 136.5 & 293.0 $\pm$ 4.3 & 5.8 \\
		& 129.5 & 454.1 $\pm$ 4.1 & 5.5 \\
		& 120.7 & 507.0 $\pm$ 4.6 & 5.1 \\
J123049+122323\tablenotemark{a}\tablenotemark{c} & 632.2 & 46.0 $\pm$ 1.4 & 12 \\
						 & 560.8 & 11.5 $\pm$ 2.2 & 11 \\
						 & 295.4 & --66.2 $\pm$ 3.0 & 5.9 \\
						 & 352.6 & 101.2 $\pm$ 2.5 & 7.1 \\
J123522+212018	& 204.2 & --3.1 $\pm$ 1.4 & 44 \\
J123530+212048\tablenotemark{a}	& 95.4 & 4.8 $\pm$ 2.6 & 25 \\
J125611--054720	& 283.5 & 15.3 $\pm$ 1.1 & 16 \\
                & 165.9 & 65.8 $\pm$ 1.7 & 9.4 \\
J133108+303032	& 1319.6 & 0.6 $\pm$ 0.1 & 261 \\
J153150+240243	& 239.3 & 11.4 $\pm$ 0.3 & 63 \\
		& 26.7 & 77.1 $\pm$ 4.0 & 7.1 \\
		& 22.8 & --62.0 $\pm$ 4.0 & 6.1 \\
J160231+015748	& 399.9 & 9.0 $\pm$ 0.7 & 92 \\
J160939+655652	& 242.3 & 17.8 $\pm$ 1.2 & 52 \\
J162803+274136	& 281.6 & 23.2 $\pm$ 1.8 & 36 \\
J164258+394837	& 368.2 & 20.8 $\pm$ 1.2 & 51 \\
J165111+045919\tablenotemark{c} & 543.1 & --6.0 $\pm$ 1.4 & 24 \\ 
				& 134.0 & --127.0 $\pm$ 4.9 & 6.1 \\
				& 137.5 & 107.5 $\pm$ 4.9 & 6.2 \\
J165112+045917	& 513.7 & --0.2 $\pm$ 2.5 & 21 \\
                & 125.9 & --152.2 $\pm$ 7.7 & 5.2 \\
J172025--005852	& 1065.4 & 39.8 $\pm$ 0.3 & 92 \\
                & 158.6 & --68.7 $\pm$ 2.3 & 13 \\
		& 85.9 & 134.4 $\pm$ 6.8 & 7.5 \\
                & 80.2 & 196.5 $\pm$ 4.9 & 7.0 \\
J172034-005843\tablenotemark{a}	& 288.4 & 107.5 $\pm$ 1.7 & 25 \\
				& 178.4 & --23.3 $\pm$ 2.9 & 15 \\
J184226+794517	& 354.6 & --8.4 $\pm$ 1.0 & 23 \\
		& 79.8 & --230.3 $\pm$ 4.0 & 5.2 \\
J184150+794728\tablenotemark{a}\tablenotemark{c} & 335.6 & --14.5 $\pm$ 0.6 & 40 \\
				& 91.8 & --179.6 $\pm$ 1.8 & 11 \\
				& 49.9 & --125.0 $\pm$ 2.9 & 6.0 \\
				& 48.1 & --234.3 $\pm$ 2.8 & 5.8 \\
				& 49.1 & --288.9 $\pm$ 2.8 & 5.9 \\
				& 46.4 & --343.4 $\pm$ 3.0 & 5.6 \\
				& 43.9 & 156.5 $\pm$ 4.6 & 5.3 \\
				& 52.9 & 210.6 $\pm$ 4.4 & 6.3 \\
				& 51.1 & 382.9 $\pm$ 4.0 & 6.1 \\
				& 43.0 & 549.7 $\pm$ 3.9 & 5.2 \\
J192451--291431\tablenotemark{c} & 152.4 & 12.4 $\pm$ 1.8 & 9.8 \\
				 & 139.0 & 240.4 $\pm$ 2.2 & 9.0 \\
				 & 125.0 & --42.4 $\pm$ 2.0 & 8.1 \\
				 & 94.7 & --96.0 $\pm$ 2.5 & 6.1 \\  
J194114--152431	& 274.9 & --79.1 $\pm$ 0.9 & 66 \\
J201713+334546	& 286.1 & --388.7 $\pm$ 0.4 & 144 \\
J211636--205551\tablenotemark{c} & 198.3 & --22.6 $\pm$ 1.1 & 24 \\  
				 & 42.5 & 185.4 $\pm$ 5.7 & 5.3 \\
				 & 45.0 & --387.8 $\pm$ 5.6 & 5.6 \\
				 & 46.7 & --236.9 $\pm$ 7.8 & 5.8 \\
				 & 43.6 & --173.3 $\pm$ 10.0 & 5.5 \\
				 & 41.6 & --104.0 $\pm$ 11.3 & 5.2 \\
J212344+250410	& 561.1 & --71.4 $\pm$ 0.4 & 157 \\
J212345+250448	& 440.1 & --71.7 $\pm$ 1.3 & 52 \\
J222547--045701	& 190.7 & --19.3 $\pm$ 3.0 & 24 \\
		& 105.9 & --149.9 $\pm$ 5.5 & 13 \\
J225357+160853	& 989.9 & --53.9 $\pm$ 0.3 & 205 \\
J231956--272713	& 165.7 & 14.0 $\pm$ 2.9 & 21 \\
\enddata
\tablenotetext{a}{Off-axis source.}
\tablenotetext{b}{Component is wider than the RM beam.  Significance is underestimated.}
\tablenotetext{c}{RM component brightnesses have extra 20\% uncertainty introduced by RM cleaning algorithm.}
\end{deluxetable}

\clearpage

\begin{deluxetable}{lccc}
\vspace{-0.5in}
\tablecaption{Components Measured in Low-resolution RM Spectra \label{rmcomponentslow}}
\setlength{\tabcolsep}{0.005in}
\tablewidth{0pt}
\tablehead{
\colhead{Source} & \colhead{P} & \colhead{RM} & \colhead{SNR} \\
  & \colhead{(mJy)} & \colhead{(rad m$^{-2}$)} & \\
}
\startdata
J005558+682218	& 341.3 & --104.7 $\pm$ 3.1 & 96 \\
J005734--012258	& 593.0 & 0.9 $\pm$ 2.2 & 137 \\
J010850+131831	& 861.5 & --14.6 $\pm$ 1.5 & 205 \\
J010855+132214\tablenotemark{a}	& 126.1 & --3.3 $\pm$ 6.8 & 45 \\
J012644+331309	& 253.9 & --66.6 $\pm$ 3. & 95 \\
                & 18.8 & -1464.4 $\pm$ 55.4 & 7.1 \\
J022248+861851	& 326.1 & --0.3 $\pm$ 3.2 & 92 \\
J030824+040639	& 157.2 & 9.5 $\pm$ 5.8 & 52 \\
J035232--071104	& 191.0 & 7.6 $\pm$ 9.5 & 32 \\
J052109+163822	& 637.8 & --0.9 $\pm$ 1.0 & 289 \\
J063633--204233	& 496.9 & 51.4 $\pm$ 3.0 & 99 \\
J074948+555421	& 217.4 & 54.6 $\pm$ 41.5 & 8.0 \\
J084124+705341	& 221.4 & --15.4 $\pm$ 3.4 & 86 \\
J094752+072517	& 257.7 & 21.4 $\pm$ 8.6 & 35 \\
J104244+120331	& 156.2 & 57.1 $\pm$ 14.0 & 21 \\
J113007--144927	& 151.0 & 22.2 $\pm$ 12.2 & 25 \\
J122906+020305	& 778.0 & 10.8 $\pm$ 12.7 & 24 \\
J123039+121758	& 575.5 & 75.4 $\pm$ 14.1 & 23 \\
J123049+122323\tablenotemark{a}	& 218.8 & --29.5 $\pm$ 59.0 & 5.1 \\
J123522+212018	& 213.2 & --4.9 $\pm$ 4.8 & 61 \\
J123530+212048\tablenotemark{a}	& 96.2 & --2.2 $\pm$ 11.4 & 28 \\
J125611--054720	& 314.6 & 232.7 $\pm$ 41.3 & 15 \\
J133108+303032	& 1357.4 & --3.4 $\pm$ 1.1 & 273 \\
J153150+240243	& 240.2 & --12.3 $\pm$ 4.9 & 63 \\
J160231+015748	& 433.1 & --0.7 $\pm$ 2.2 & 131 \\
J160939+655652	& 247.8 & 17.6 $\pm$ 4.2 & 71 \\
J162803+274136	& 297.9 & 26.9 $\pm$ 5.5 & 54 \\
J164258+394837	& 368.6 & 20.4 $\pm$ 3.8 & 78 \\
J165111+045919	& 435.2 & --29.5 $\pm$ 4.7 & 61 \\
J165112+045917	& 494.7 & --24.7 $\pm$ 4.0 & 72 \\
J172025--005852	& 1553.3 & 37.5 $\pm$ 1.3 & 234 \\
J172034--005843\tablenotemark{a}	& 737.8 & 49.1 $\pm$ 3.7 & 82 \\
J184226+794517	& 316.5 & --7.3 $\pm$ 4.9 & 61 \\
J184150+794728\tablenotemark{a}	& 326.0 & --12.0 $\pm$ 5.0 & 60 \\
J192451--291431	& 175.5 & 207.8 $\pm$ 11.7 & 25 \\
                & 49.1 & --738.7 $\pm$ 39.2 & 7.0 \\
J194114--152431	& 281.1 & --87.2 $\pm$ 4.0 & 74 \\
J201713+334546	& 285.3 & --388.3 $\pm$ 2.5 & 122 \\
J211636--205551	& 220.7 & --60.6 $\pm$ 10.1 & 30 \\
J212344+250410	& 593.9 & --73.1 $\pm$ 1.9 & 152 \\
J212345+250448	& 453.4 & --72.0 $\pm$ 2.6 & 114 \\
J222547--045701	& 290.4 & --59.4 $\pm$ 8.5 & 38 \\
J225357+160853	& 1002.3 & --55.7 $\pm$ 1.6 & 193 \\
J231956--272713	& 173.5 & --1.0 $\pm$ 16.1 & 20 \\
\enddata
\tablenotetext{a}{Off-axis source.}
\tablenotetext{b}{Component detected in low resolution RM spectrum.}
\end{deluxetable}

\clearpage

With high confidence ($7\sigma\approx40$\ mJy), we find multiple RM components toward 12 out of 42 sources, for a total of 61 components.  For a lower significance threshold ($5\sigma\approx30$\ mJy), we find multiple RM components in 19 out of 42 sources, for a total of 92 components.  The brightest component of each multi-component source ranges from roughly 30\% to 90\% of the total polarized flux in all components, with only a weak dependence on the significance threshold.

As expected for our frequency coverage, most RM components are unresolved.  The multi-component RM spectra have components separated by the RM beam size.  These multi-component spectra are likely caused by how our observing configuration resolves flux from Faraday-thick sources.  In general, the low-resolution RM spectra have best-fit peaks consistent with the flux-weighted mean RM of the components detected at high resolution.  This suggests that the low- and high-resolution spectra are seeing the same components.  This is consistent with the fact that most components are thinner than the 50 rad m$^{-2}$, high-resolution beam and the low-resolution spectra are sensitive to Faraday thicknesses less than 75 rad m$^{-2}$.

\section{Discussion}
\label{dis}

\subsection{Peak RM vs. Mean RM}
\label{resrm}
This survey provides a sample of low- and high-resolution RM spectra to study the effect of resolution on measuring RM.  Comparing Tables \ref{rmcomponents} and \ref{rmcomponentslow} shows that the mean, low-resolution RM is similar to the flux-weighted mean of the components seen at high resolution.  In other words, the low-resolution spectra are usually smoothed versions of the high-resolution spectra.  However, this means that the peak RM for a source is resolution dependent.

\begin{figure}[tbp]
\includegraphics[width=\textwidth]{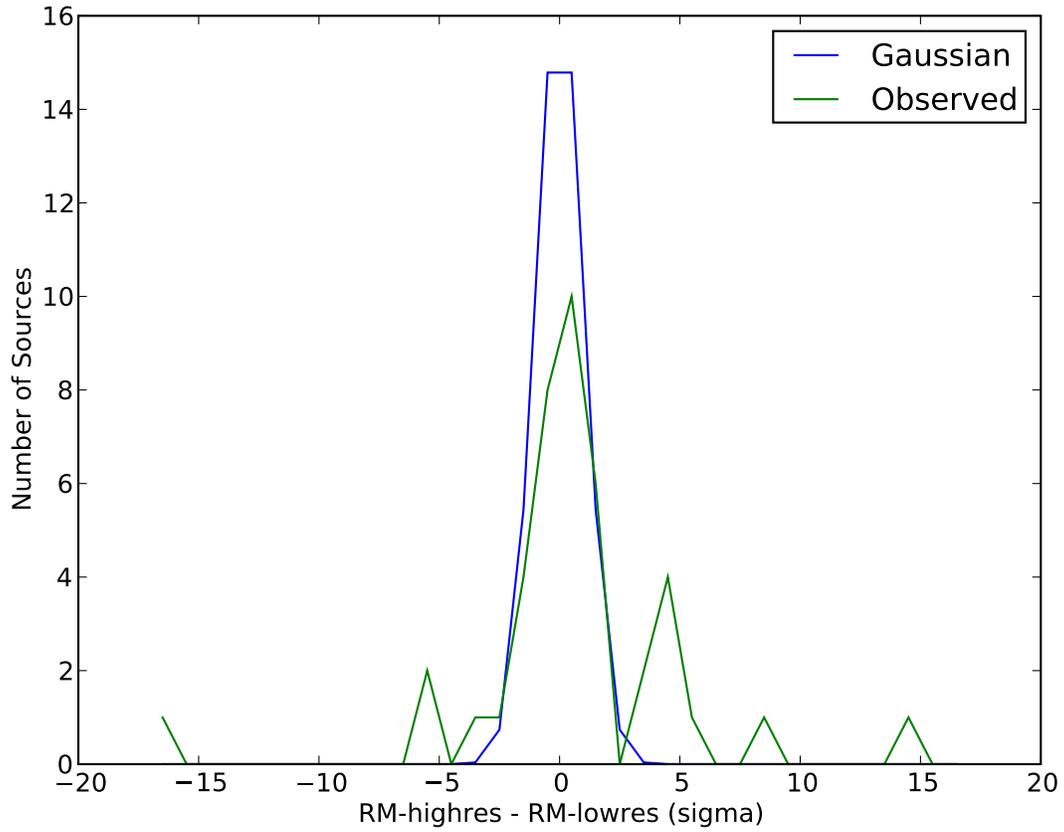}
\caption{Difference in peak RM measured in high- and low-resolution RM spectra in terms of the significance of the difference.  The error in the difference is the quadrature sum of the individual RM measurements.  The green line shows the observed RM difference and the blue line shows the difference expected under Gaussian statistics. \label{rmdiff}}
\end{figure}

Figure \ref{rmdiff} shows a histogram of the shift in peak RM between the high- and low-resolution spectra in units of significance.  An idealized distribution is shown as a Gaussian containing 42 sources with width $1\sigma$.  The observed distribution of RM values has a wider spread than the idealized distribution, which reflects the change in peak RM with resolution.

While the observed distribution of peak RM change is somewhat irregular, it is similar to a Gaussian of width $2-3\sigma$.  For a typical RM difference error of 5 rad m$^{-2}$, this suggests an extra RM uncertainty of 10--15 rad m$^{-2}$ for RM spectra with resolution less than 600 rad m$^{-2}$.  This extra uncertainty represents the precision with which one can specify the peak RM of a source in our low resolution spectra.  

Generally speaking, any individual RM component cannot be specified more accurately than the RM beam width.  This is analogous the uncertainty in measuring any centroid of an unresolved cluster of sources.  For some applications \citep[e.g., measuring mean line-of-sight properties; ][]{g05,t09}, the mean RM is adequate, since it averages over all RM components.  However, any individual component cannot be specified more precisely than the beam size due to the uncertainty in the distribution of components within the RM beam.

\subsection{Comparison with T09 Results}
\begin{figure}[tbp]
\includegraphics[width=0.5\textwidth]{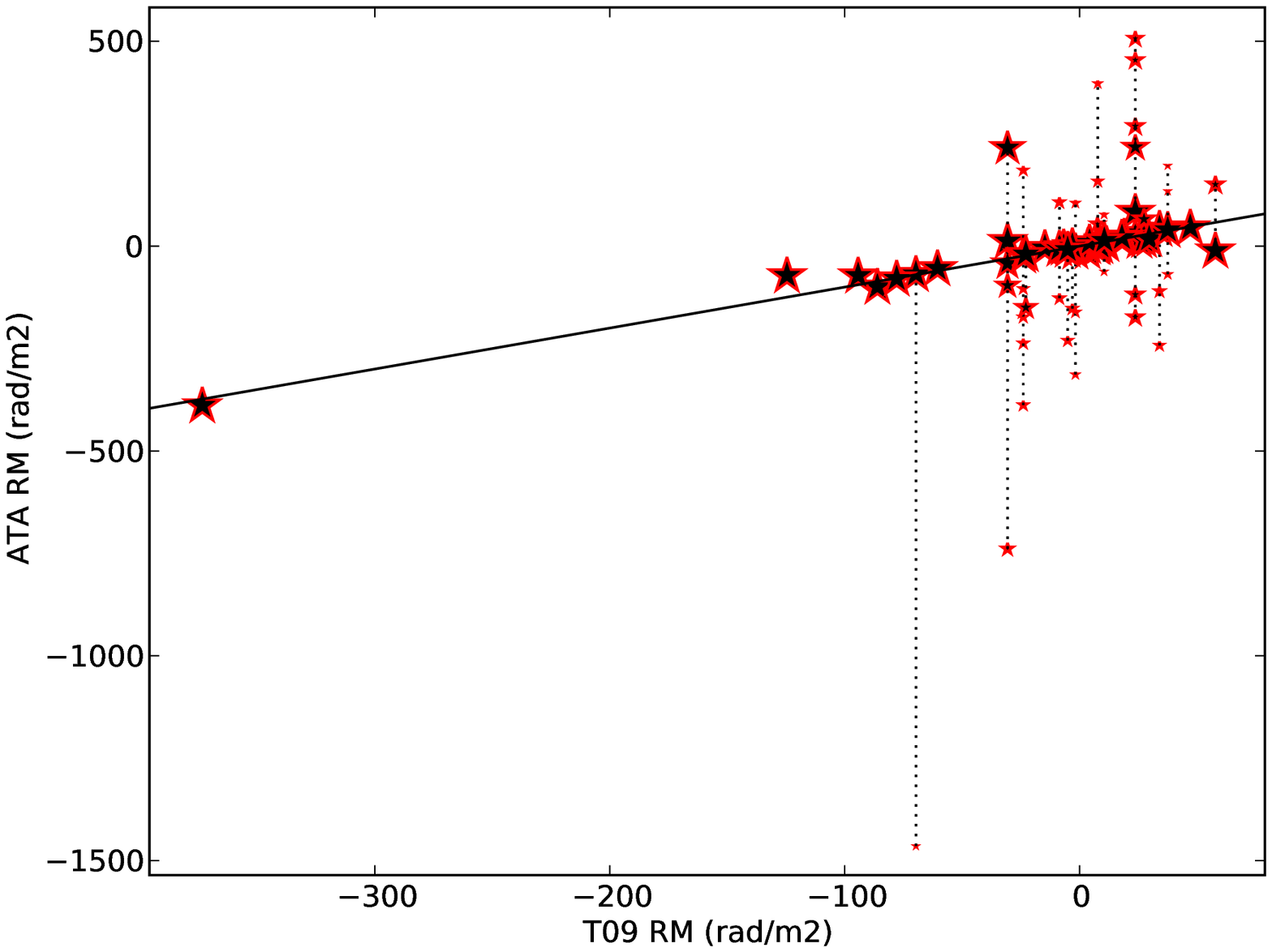}
\includegraphics[width=0.5\textwidth]{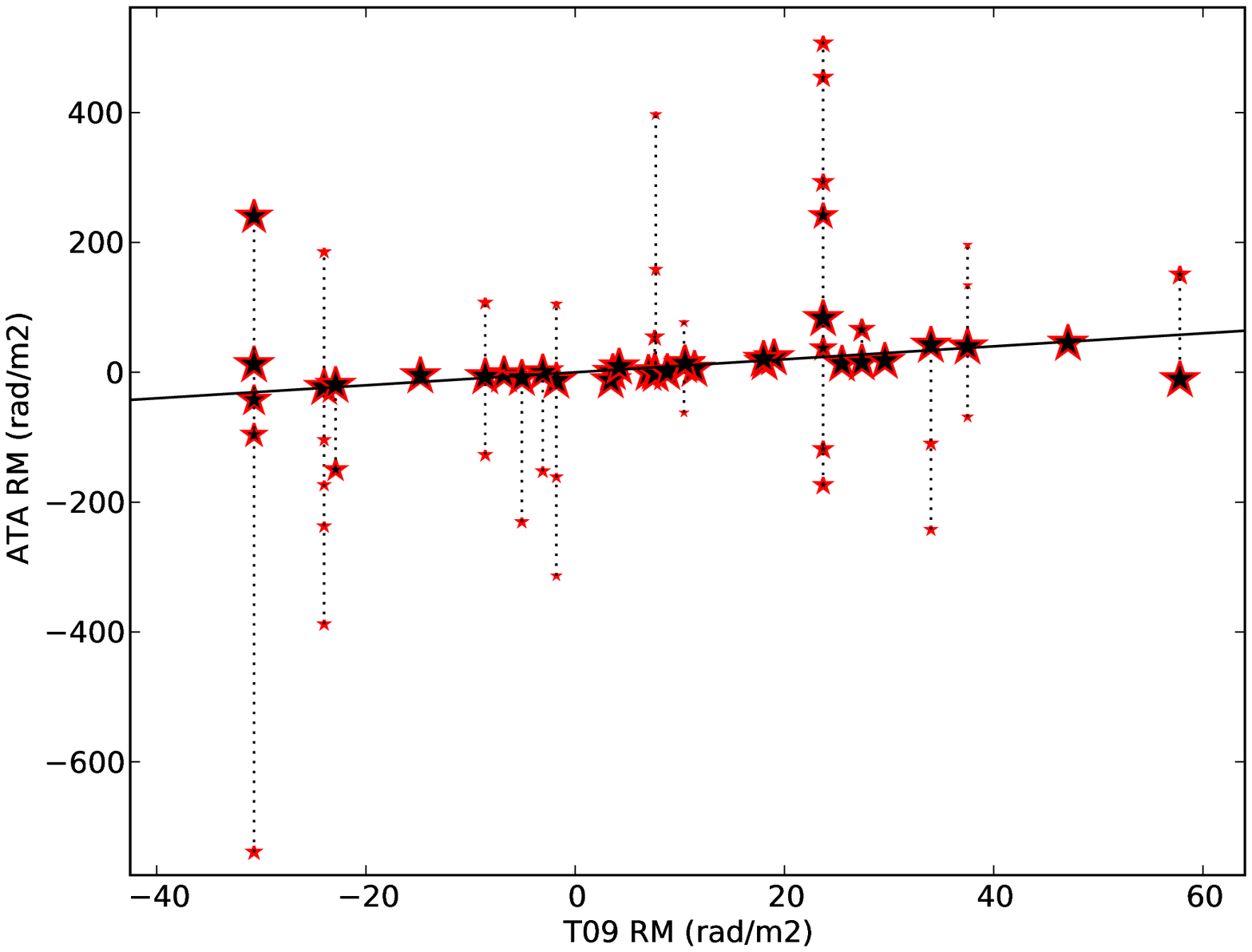}
\caption{\emph{Left:} All RM components measured by the ATA from 1--2 GHz compared to \citetalias{t09} at 1.4 GHz for the 37 central sources.  The solid line shows where the ATA and \citetalias{t09} RMs are equal.  \citetalias{t09} sources with multiple ATA RM components are connected with a dotted line.  The symbol size is proportional to the flux density of the component relative to the brightest component in the source. \emph{Right:}  A detail of the plot above for small Faraday depths. \label{compare}}
\end{figure}

Figure \ref{compare} compares the RM components observed here with the results of \citetalias{t09}.  Since the ATA is able to measure multiple significant components per source, the plot shows the relative flux in multi-component sources.  We find that the flux-weighted mean RM of all high-resolution ATA components is similar to the low-resolution, \citetalias{t09} RM value.  The most significant potential difference between the ATA and \citetalias{t09} values appears in the double source J212344+250410/J212345+250448, as discussed in \S \ref{dbl}.  This confirms that sources with polarized flux densities greater than 200 mJy in \citetalias{t09} do not suffer from $n\pi$\ ambiguity problems and that their RM values are at least 95\% reliable (more than 35 out of 37 sources).  However, as noted in \S \ref{resrm}, low RM resolution spectra are most appropriate for measuring the mean RM along a line of sight, not the peak RM of a source.

\begin{figure}[tbp]
\includegraphics[width=\textwidth]{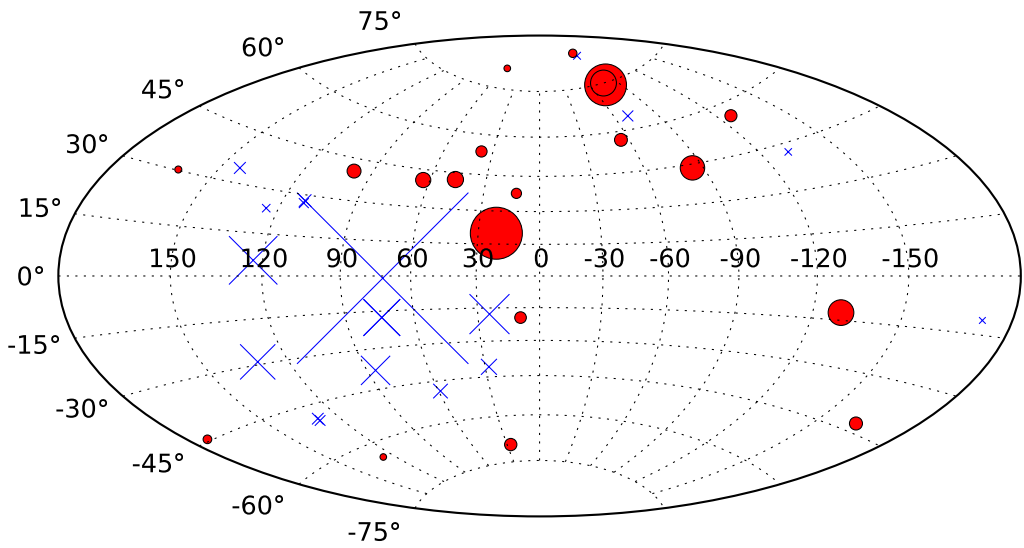}
\caption{Galactic distribution of the peak RM measured in the high-resolution spectra (Table \ref{rmcomponents}) shown with a Hammer-Aitoff projection.  Positive RM is shown with a circle, negative RM is shown with a cross.  The symbol width scales linearly with the magnitude of the RM, with a minimum width similar to the smallest symbols shown. \label{skydistribution}}
\end{figure}

Figure \ref{skydistribution} shows the peak RM measured in high-resolution RM spectra (see Table \ref{rmcomponents}) projected onto the Galactic coordinate system.  The RM values show large-scale structure similar to that seen in other RM surveys \citep{s81,t09}.  Since this map shows the peak RM, it is more affected by intrinsic source structure than in low RM resolution observations.

\subsection{Unresolved Double Sources}
\label{dbl}
As noted in Table \ref{complexfields}, two pairs of fields in our survey are unresolved by the ATA:  (J165111+045919, J165112+045917) and (J212344+250410, J212345+250448).  These four fields were each observed independently, so they provide a unique test of the quality of ATA RM synthesis analysis.

The RM spectra toward J165111+045919 and J165112+045917 are consistent with each other.  Nominally the peak RM is different, but this is a resolution effect, similar to that discussed in \S \ref{resrm}.  The peak RM measured in two spectra is roughly $-$5 rad m$^{-2}$ \citepalias{t09}.  This is consistent with the flux-weighted mean RM of the individual sources reported by \citetalias{t09}, since they are unresolved spatially and spectrally.  The sum of the \citetalias{t09} polarized flux in the two components is about twice that detected here, so the ATA is beam depolarized by a factor of 2.

The RM spectra toward two, unresolved parts of 3C 433 (J212344+250410 and J212345+250448) are also consistent with each other, showing a single RM component at roughly $-$72 rad m$^{-2}$.  This value differs significantly from the RM measured by \citetalias{t09} toward the two sources individually, roughly $-$125 and $-$94 rad m$^{-2}$ \citepalias{t09}.  The value seen by the ATA is similar to the value of $-73\pm1$ rad m$^{-2}$\ seen in single-dish, wide-bandwidth observations \citep{s81}.  The low-resolution RM spectra presented here have sensitivity to similar Faraday thicknesses as \citetalias{t09}, so their difference (and the consistency of the ATA and single-dish observations) may indicate an error in \citetalias{t09}.  Alternatively, it has been suggested that spatially unresolved pairs of sources can confuse the RM synthesis technique in some cases (L. Rudnick, private communication).  We cannot distinguish between these possibilities with our current analysis technique.

\subsection{Comparison with VLBA RM Images}
\label{vlba}
Several previous studies have presented milliarcsecond, Very Long Baseline Array (VLBA), polarimetric maps of dozens of AGN \citep{a02,z03,z04,z05,m10}.  These observations were generally done at higher frequencies, ranging from 5 to 22 GHz.  As the VLBA observations are at higher frequency and have better spatial resolution, they are less subject to depth and beam depolarization and show a relatively unambiguous view of the polarized emission.  Below, we compare the properties observed at high spatial resolution with those of our RM spectra.

\subsubsection{3C 273}
RM maps of 3C 273 (J122906+020305) from 5 to 22 GHz find time-variable RM with polarized flux density of a few hundred mJy, or a few percent of total intensity \citep{a02,z05}.  The RM distribution for the core changes from $\pm3000$\ rad m$^{-2}$ over year timescales, while the jet (within 10 mas) is predominantly from 0 to 1000 rad m$^{-2}$.  From 1 to 2 GHz, we find two components at $-$11 and 151 rad m$^{-2}$ with polarization fractions of 1 and 0.5\%, respectively.  While 3C 273 has variable RM and brightness distribution, the 1--2 GHz observations generally have a lower polarization fraction and RM range than seen by the VLBA.  Another source showing this kind of frequency dependence is 3C 446 \citep[see Figure \ref{rmplotlast}; ][]{z03}.  

\subsubsection{3C 279}
Our observations of 3C 279 (J125611--054720) find two RM components at 15 and 66 rad m$^{-2}$ with polarized brightnesses of 270 and 160 mJy ($\sim2.7$ and 1.6\% of total intensity).  The 1--2 GHz RM is similar to the that seen at 8 GHz with the VLBA in June 2000 \citep{z03}, which is much less than that seen at earlier dates \citep{z01}.  The observations of \citet{z03} show two sources at milliarcsecond-resolution with RM of $-$91 and 13 rad m$^{-2}$ and polarized brightnesses of about 1.4 Jy ($\sim9$\%).  Intrinsic variability makes it hard to make a strong conclusion, but this similarity suggests that a similar part of 3C 279 is seen from 1 to 8 GHz.  

\subsubsection{3C 454.3}
Source J225357+160853 is 3C 454.3 (a.k.a.\ B2251+158), which has a low polarization fraction of $\sim0.7$\% at 8 GHz, as compared to 9\% in our observations \citep{z03}.  The RM distribution is irregular with a value at peak brightness of $-$263 rad m$^{-2}$ at 8 GHz; we measure a single component of $-$54 rad m$^{-2}$ from 1--2 GHz.  This source is a blazar and is highly variable at radio wavelengths, so it is difficult to make strong conclusions in this comparison.  However, monitoring at 43 GHz shows that its polarization fraction rarely reaches 9\% \citep{j10}.  If so, this source may have a higher polarization fraction from 1--2 GHz than at higher frequencies, which differs from the trend observed toward 3C 273, 3C 446, and 3C 279.

\subsubsection{M87}
\label{m87}
Field J123039+121758 is particularly complicated as it is host to the starburst galaxy M87 and its jet (see Figures \ref{rmplot2} and \ref{rmplot3}).  Comparing the ATA observations to previous work \citep[e.g.,][]{r96} shows a similar bright core and faint jet in total intensity.  In polarized intensity the situation is reversed and the brightest source the ATA sees from 1--2 GHz is the termination of the southwest jet;  in fact, this polarized source is what was identified by our survey selection criteria.  In general, the core and other parts of the jet have varying levels of polarized flux that change rapidly with frequency.  

The RM spectra of the core and jet of M87 are amongst the most complicated in this survey.  This is probably not a coincidence, since M87 is one of the few sources that has a core-jet structure resolved by the ATA.  The spatial separation allows us to resolve multiple Faraday components, which otherwise cause depth depolarization.  Still, the dramatic increase in polarized flux in the 1.0 GHz band (especially in the core) suggests that there are strong spectral index effects that may confuse the RM cleaning process.  Also, the RM values measured for the M87 core in the low-resolution RM spectra are very different from those at high-resolution, which is likely related to strong Faraday thickness effects.  In principle an algorithm could be developed to disentangle these effects, similar to the ``multi-frequency synthesis'' technique used in synthesis imaging.

For the M87 jet, we find RM up to $\sim300$\ rad m$^{-2}$, while in the M87 core the RM ranges up to 100 rad m$^{-2}$.  Despite the relatively large RM we find from 1--2 GHz, M87 has an even larger range of known RM values measured by VLBA imaging \citep[$|RM|<1000$\ rad m$^{-2}$; ][]{z03}.

\subsection{Limits on Large RM Values}
Limits on RM up to $\pm9\times10^{4}$ rad m$^{-2}$ are typically 20 mJy.  At the maximum RM, this is equivalent to an intrinsic fractional polarization of order 0.5\% for our sample.  Since this is more than an order of magnitude less than the polarization fraction of typical synchrotron sources, these components are not likely to be Faraday-thin synchrotron sources.  The low-resolution RM spectra have a sensitivity to sources with RM thickness up to 75 rad m$^{-2}$.  If the Faraday thickness is related to a physical thickness, Equation \ref{phi} requires a physical thickness of $93 (n_{e}/1\ \rm{cm}^{-3})^{-1} (B/1\ \mu \rm{G})^{-1}$\ pc.  However, it is equally likely that the Faraday thickness of the source is caused by multiple RM components that are spatially unresolved.

In general, the 1--2 GHz RM spectra do not detect the large polarization fractions and RM values seen by the VLBA from 5--22 GHz.  Since the spectral indices for these sources are relatively flat, they should be bright enough to detect from 1--2 GHz.  That requires that either (1) the Faraday thickness be larger than 75 rad m$^{-2}$ (in the low-resolution RM spectra), or (2) the polarized source structure change between 1 and 10 GHz.

High-spatial-resolution polarimetry of quasars has found a decreasing polarization fraction toward lower frequencies \citep{a02,m10}, which suggests that Faraday-thickness depolarization occurs even at high spatial resolution.  If this is the sole cause of frequency-dependence of the polarization fraction, the lack of large RM requires that Faraday thickness increases with Faraday depth.  If so, the typical thickness is larger than 75 rad m$^{-2}$ for depths larger than $|RM|>1300$\ rad m$^{-2}$, the largest, high-significance RM detected here.

It is also likely that wide-bandwidth RM synthesis analysis can be confused by relative spectral index changes within a source.  \citet{z03} have shown that the spectral index of the 3C 454.3 jet is steeper than that of the core, which suggests that 1--2 GHz observations will tend to see more of its flux.  The inner core and jet are hosts to the strongest and most organized magnetic fields, where one expects large RM and high fractional polarization.  The outer jet is likely to be interacting with ambient gas, and is thus less organized and has lower RM.  The RM synthesis does not formally account for these relative changes, so some care is required when interpreting RM results from data with large fractional bandwidths.

\section{Conclusions}
\label{con}
We have conducted a survey of 37 bright, polarized radio sources from 1--2 GHz with the ATA.  Applying polarimetric calibration allows us to measure the full Stokes information over this frequency range.  RM synthesis uses these wide-bandwidth spectra to create high-resolution RM spectra, sensitive to RM up to 90000 rad m$^{-2}$ with a resolution as good as 50 rad m$^{-2}$.  Observations of 3C 286 show that these data can measure RM with a precision up to 0.6 rad m$^{-2}$.

In the 37 fields observed, 42 sources were brightly polarized enough to perform RM synthesis.  Twelve of the 42 sources have multiple RM components with greater than 7$\sigma$\ significance ($\approx40$\ mJy), for a total of 61 components.  Comparing the RM values measured by the ATA at high and low RM resolution show that multi-component RM spectra can bias the measurement of the peak RM of a source.  We show that the peak RM of our sources cannot be known to better than 10--15 rad m$^{-2}$ if observed at RM resolution larger than 600 rad m$^{-2}$.  More generally, the peak RM can be measured no more accurately than the width of the RM beam.  However, for typical applications the mean RM measured by low-resolution RM spectra is not affected by this bias.  

The RM spectra measured by the ATA confirm the RM results presented by \citetalias{t09} for sources with polarized flux densities greater than 200 mJy.  There is no evidence for $n\pi$\ ambiguity problems and the narrow-bandwidth VLA observations are at least 95\% reliable for this sample of sources.

VLBA maps with milliarcsecond resolution spatially resolve some components that we identify in RM spectra.  This correspondence suggests that RM synthesis can detect blobs of gas in massive AGN jets, which ordinarily must be spatially resolved to be studied.  Observations of the time and frequency dependence of Faraday thickness and polarization fraction in AGN can be compared to simulations to constrain physical models.

We have shown that RM synthesis is useful because it quantifies the full complexity of the Faraday rotation process.  Future work will exploit the relative ease of the technique to resolve the time-dependent RM structure in AGN.  Improving calibration models will also enable wide-field, wide-bandwidth RM surveys with the ATA and other radio interferometers.  Large, accurate RM samples will probe magnetic fields on Galactic and extragalactic scales.

\acknowledgements{We dedicate this paper to the memory of Don Backer, whose curiosity and generosity inspired us all.  The authors would like to acknowledge the generous support of the Paul G. Allen Family Foundation that has provided major support for design, construction and operations of the ATA. The U.S. Naval Observatory provided significant funds for ATA construction. Contributions from Nathan Myhrvold, Xilinx Corporation, Sun Microsystems and other private donors have been instrumental in supporting the ATA. The US National Science Foundation grants AST-0321309, AST-0540690 and AST-0838268 have contributed to the ATA project. 

We thank Bob Sault, Jeroen Stil, and Mattieu de Villiers for useful discussions.  We thank the many people that helped build the ATA, including Rob Ackermann, Shannon Atkinson, Peter Backus, William C. Barott, Leo Blitz, Douglas Bock, Tucker Bradford, Calvin Cheng, Chris Cork, Mike Davis, Dave DeBoer, Matt Dexter, John Dreher, Greg Engargiola, Ed Fields, Matt Fleming, Gerry Harp, Tamara Helfer, Jane Jordan, Susanne Jorgensen, Tom Kilsdonk, Joeri van Leeuwen, John Lugten, Peter McMahon, Oren Milgrome, Tom Pierson, Karen Randall, John Ross, Seth Shostak, Andrew Siemion, Ken Smolek, Jill Tarter, Douglas Thornton, Lynn Urry, Artyom Vitouchkine, Niklas Wadefalk, Jack Welch, and Dan Werthimer.  B.M.G. and L.H.S. acknowledge the support of the Australian Research Council through grant DP0986386.  This research has made use of NASA's Astrophysics Data System Bibliographic Services.}

{\it Facilities:} \facility{ATA ()}

\end{document}